\documentclass[twocolumn,superscriptaddress,floatfix]{revtex4-1}

\newcommand{\br}{\boldsymbol{\textbf{r}}}

\newcommand{\bA}{\boldsymbol{A}}
\newcommand{\bB}{\boldsymbol{\textbf{B}}}

\newcommand{\bC}{\boldsymbol{C}}
\newcommand{\bE}{\boldsymbol{E}}

\newcommand{\Hopsig}{\hat{\mathcal{{H}}}_{\sigma}}

\newcommand{\bPhi}{\boldsymbol{\Phi}}

\newcommand{\epsilonksig}{\epsilon_{k,\sigma}}

\newcommand{\dr}{\,d\br}

\newcommand{\vxc}{v_{\text{xc}}}

\newcommand{\vxcsig}{v_{\text{xc},\sigma}}
\newcommand{\rhod}{\rho^{\text{data}}}

\newcommand{\rhodsig}{\rho^{\text{data}}_{\sigma}}

\newcommand{\rhosig}{\rho_{\sigma}}
\newcommand{\bPsiksig}{\boldsymbol{\Psi}_{k,\sigma}}
\newcommand{\bPsitildeksig}{\boldsymbol{\widetilde{\Psi}}_{k,\sigma}}

\newcommand{\bPksig}{\mathbf{P}_{k,\sigma}}
\newcommand{\bPtildeksig}{\mathbf{\widetilde{P}}_{k,\sigma}}

\newcommand{\bDksig}{\mathbf{D}_{k,\sigma}}

\newcommand{\bQksig}{\textbf{Q}_{k,\sigma}}
\newcommand{\bpsiksig}{\psi_{k,\sigma}}

\newcommand{\bEksig}{\mathbf{E}_{k,\sigma}}

\newcommand{\vH}{v_{\text{H}}}
\newcommand{\vN}{v_{\text{N}}}
\newcommand{\bI}{\textbf{I}}
\newcommand{\vFA}{v_{\text{FA}}}

\newcommand{\trans}{^{\text{T}}}

\newcommand{\tr}{\textbf{Tr}}
\newcommand{\bDtildeksig}{\mathbf{\widetilde{D}}_{k,\sigma}}
\newcommand{\Eksigij}{{E}_{k,\sigma}^{ij}}
\newcommand{\Eij}{E_{ij}}
\newcommand{\lag}{\mathcal{L}}
\newcommand{\bLambdaij}{\boldsymbol{\Lambda}^{ij}}
\newcommand{\Lambdaijkl}{\left(\Lambda^{ij}\right)_{kl}}
\newcommand{\Lambdaijab}{\Lambda^{ij}_{\alpha\beta}}
\newcommand{\Lambdaijaa}{\Lambda^{ij}_{\alpha\alpha}}

\newcommand{\ket}[1]{\left| #1 \right>} 
\newcommand{\bra}[1]{\left< #1 \right|} 

\usepackage{mathtools}
\usepackage{subfiles}
\usepackage{graphicx}
\usepackage{graphics}
\usepackage{multirow}
\usepackage{fancyhdr, ifpdf}
\usepackage{amsxtra}
\usepackage{stmaryrd}
\usepackage{mathrsfs}
\usepackage{psfrag}
\usepackage{subfigure}
\usepackage{epsfig}
\usepackage{rotating}
\usepackage{setspace}
\usepackage{float}
\usepackage{amsfonts}
\usepackage{amsbsy}
\usepackage{amscd}
\usepackage{amsthm}
\usepackage{color}
\usepackage{tabularx}
\usepackage{sidecap}
\usepackage{dcolumn}
\usepackage{footnote}
\usepackage{algorithm}
\usepackage{algorithmic}
\usepackage[normalem]{ulem}
\usepackage[titletoc,title]{appendix}

\definecolor{hellgruen}{rgb}{0.2,0.7,0.2}

\newcommand{\tcb}{\textcolor{black}}
\newcommand{\tcm}{\textcolor{black}}
\newcolumntype{M}[1]{>{\centering\arraybackslash}m{#1}}
\newcolumntype{N}{@{}m{0pt}@{}}

\newcommand{\norm}[1]{\left\lVert#1\right\rVert}
\newcommand{\modulus}[1]{\left\vert#1\right\vert}

\title{Exact and Model Exchange-Correlation Potentials for Open-Shell Systems}

\begin{document}
\title{Exact and model exchange-correlation potentials for open-shell systems}
\author{Bikash Kanungo}
\affiliation{Department of Mechanical Engineering, University of Michigan, Ann Arbor, Michigan 48109, USA}
\author{Jeffrey Hatch}
\affiliation{Department of Chemistry, University of Michigan, Ann Arbor, Michigan 48109, USA}
\author{Paul M. Zimmerman}
\affiliation{Department of Chemistry, University of Michigan, Ann Arbor, Michigan 48109, USA}
\author{Vikram Gavini}
\affiliation{Department of Mechanical Engineering, University of Michigan, Ann Arbor, Michigan 48109, USA}
\affiliation{Department of Materials Science and Engineering, University of Michigan, Ann Arbor, Michigan 48109, USA}


\begin{abstract}
The conventional approaches to the inverse density functional theory problem typically assume non-degeneracy of the Kohn-Sham (KS) eigenvalues, greatly hindering their use in open-shell systems. We present a generalization of the inverse density functional theory problem that can seamlessly admit degenerate KS eigenvalues. Additionally, we allow for fractional occupancy of the Kohn-Sham orbitals to also handle non-interacting ensemble-v-representable densities, as opposed to just non-interacting pure-v-representable densities. We present the exact exchange-correlation (XC) potentials for six open-shell systems---four atoms (Li, C, N, and O) and two molecules (CN and $\text{CH}_2$)---using accurate ground-state densities from configuration interaction calculations. We compare these exact XC potentials with model XC potentials obtained using non-local (B3LYP, SCAN0) and local/semi-local (SCAN, PBE, PW92) XC functionals. Although the relative errors in the densities obtained from these DFT functionals are of $\mathcal{O}(10^{-3}-10^{-2})$, the relative errors in the model XC potentials remain substantially large---$\mathcal{O}(10^{-1}-10^0)$. 
\end{abstract}
\maketitle

Density functional theory~\cite{Hohenberg1964, Kohn1965} (DFT) has become the most used electronic structure method due to its superior balance of speed and accuracy. Approximations to the exchange-correlation (XC) functional, however, limit DFT’s accuracy even though the theory itself is formally exact. A wide range of theoretical and empirical approaches have been proposed to improve the XC functional, but truly systematic means to doing so are wholly unknown. A promising alternative is to study the XC potential ($\vxc(\br)$), which comes from the derivative of the XC functional. This information can be gained through the inverse DFT problem, which transforms the ground-state density (e.g. from an accurate wavefunction) into the XC potential.~\cite{Zhao1994, Leeuwen1994, Peirs2003, Wu2003, Jensen2018, Kanungo2019, Shi2021} This provides not only critical input data for conventional and machine-learning based XC approximations,~\cite{Schmidt2019, Zhou2019} but also the ability to probe deficiencies of existing model XC functionals.~\cite{Nam2020, Kanungo2021}

Inverse DFT has been applied by our group~\cite{Kanungo2019,Kanungo2021} and others~\cite{Gorling1992, Wang1993, Zhao1994, Leeuwen1994, Tozer1996, Wu2003, Peirs2003, Jacob2011, Gould2014, Jensen2018, Shi2021, Shi2022, Gould2023, Aouina2023} to electronic densities corresponding to spin singlet, spin-restricted open-shell, or predominantly closed-shell states. The widespread interest in strongly correlated materials~\cite{Dagotto2005}, including magnetic materials~\cite{Malrieu2014}, superconductors~\cite{Orenstein2000}, and transition metal based catalysts~\cite{Paier2013}, however, puts a great distance between existing studies of XC potentials and materials systems of contemporary interest. As one key step towards bridging this divide, XC functionals must be modeled to handle open-shell systems accurately, where unpaired electrons are crucial. Although there have been a few attempts to compute the XC potentials for open-shell systems~\cite{Gritsenko2004, Boguslawski2013, Gould2023}, the accuracy and the robustness afforded by these approaches remain a concern. Thus, while significant progress has been made in inverse DFT methods for closed-shell electronic states, systematically convergent approaches are desired for open-shell systems.

The inverse DFT problem is solved as an iterative procedure ~\cite{Gorling1992, Wang1993, Leeuwen1994, Peirs2003, Ryabinkin2012} or constrained optimization ~\cite{Zhao1994, Wu2003, Jacob2011, Kanungo2019, Kumar2020}. While most of these approaches suffer from numerical instabilities ~\cite{Burgess2007, Bulat2007,Jacob2011} and/or are based on electron densities with incorrect asymptotic behavior ~\cite{Mura1997, Schipper1997, Gaiduk2013, Kanungo2019}, recent efforts have worked to address these challenges. In particular, one strategy uses the two-electron density matrix instead of just the density~\cite{Ryabinkin2015, Cuevas2015, Ospadov2017, Tribedi2023} and another employs constrained optimization in a complete finite-element (FE) basis~\cite{Kanungo2019, Kanungo2021}. The latter strategy finds XC potentials from densities containing correct asymptotics, giving highly accurate potentials that can be used in learning XC functionals.

Most of the above approaches have an implicit assumption of non-degeneracy in all or frontier Kohn-Sham (KS) eigenvalues, and hence, their applicability and robustness for systems with degenerate KS eigenvalues remain unclear. To elaborate, any inverse DFT algorithm relies on iterative updates to $\vxc(\br)$ of the form  $\vxc^{(n+1)}(\br) =\vxc^{(n)}(\br) + u^{(n)}(\br)$, where $u^{(n)}(\br)$ is defined in terms of the KS system and other auxiliary quantities (e.g., difference between KS and target density, adjoint functions, etc.) obtained using $\vxc^{(n)}(\br)$. Given the non-uniqueness of the KS orbitals in case of degeneracy (i.e., one can choose any orthogonal transformation of the degenerate KS orbitals), there is non-uniqueness in the update $u^{(n)}$, leading to either non-convergence or ``sloshing" of the XC potential. \tcb{Additionally, any robust inverse DFT approach for open-shell system should be able to handle non-interacting ensemble-v-representable (e-$v_s$) density (i.e., density corresponding to an ensemble of KS determinant) as opposed to only non-interacting pure-v-representable (pure-$v_s$ ) density (i.e., density corresponding to a single KS determinant), which remains non-trivial in the conventional approaches. Among the past efforts at inverse DFT for open-shell systems, the Lieb-response based approach by Gould~\cite{Gould2023} elegantly handles both degeneracy as well as any e-$v_s$ density. However, the handling of densities with continuous symmetry (e.g., open-shell atoms) remains a challenge. Lastly, beyond the above conceptual challenges, inverse DFT for open-shell systems inherit the same numerical challenges as closed-shell systems---ill-posedness and/or spurious oscillations due to the incompleteness of the basis and the incorrect asymptotics in the target densities. Thus, overall, a robust approach to inverse DFT that can simultaneously resolve the conceptual and numerical challenges in inverse DFT for open-shell systems is desired.} 



\tcb{For an open-shell system, the inverse problem can be posed in two different ways: (i) using spin-restricted KS-DFT, where a single XC potential $\vxc(\br)$ that yields the target total density $\rhod(\br)$ is to be computed; or (ii) using spin-unrestricted KS-DFT where two different XC potentials $\vxcsig(\br)$, $\sigma=1,2$ being the spin index, which yields the target spin-densities $\rhodsig(\br)$ are sought. Since our eventual objective is to use accurate XC potentials from inverse DFT to improve the XC approximation, including its spin dependence, we use the more generic spin-unrestricted KS-DFT formalism. However, the key ideas presented can be easily extended to the spin-restricted KS-DFT case.} The main idea in this work is to reformulate the partial differential equation (PDE) constrained optimization approach to the inverse DFT problem~\cite{Kanungo2019, Shi2021, Shi2022} such that it guarantees a unique update to $\vxcsig(\br)$ at each iteration, even in the case of degenerate KS eigenvalues. Given the target spin densities $\rhodsig(\br)$, the inverse DFT problem of finding the $\vxcsig(\br)$ that yields the target densities can be posed as the following PDE-constrained optimization: 
\tcb{
\begin{equation}\label{eq:objective}
    \min_{\{\vxcsig(\br)\}}\sum_{\sigma=1}^2\int w_\sigma(\br)\left(\rhodsig(\br)-\rhosig(\br)\right)^2\dr\,,  
\end{equation}
where  $\rhosig(\br)$ are the KS spin-densities and $w_\sigma(\br)$ is a positive weight that expedites the convergence, especially in the low density region.} The  $\rhosig(\br)$ are obtained from the solutions of the KS eigenvalue problem, which, for a non-periodic system (e.g., atoms, molecules), is given by
\begin{equation}\label{eq:KSSig}
   \Hopsig\bPsiksig(\br)=\bPsiksig(\br)\bEksig\,,\quad k=1,2,\ldots,M_{\sigma}\,.
\end{equation}
In the above equation, $\Hopsig=-\frac{1}{2}\nabla^2 + \vH(\br) + \vN(\br) +\vxcsig(\br)$ is the KS Hamiltonian for spin index $\sigma$; $k$ indexes the distinct eigenvalues, with the $k^{\rm{th}}$ eigenvalue having a multiplicity of $m_{k,\sigma}$. $\bEksig=\epsilonksig\bI_{m_{k,\sigma}}$ is the diagonal eigenvalue matrix, with $\epsilonksig$ as the $k^{\rm{th}}$ distinct eigenvalue. $\bPsiksig(\br)=\left[\bpsiksig^{(1)}(\br) ~\middle| ~\bpsiksig^{(2)}(\br)~ \middle| ~ \cdots ~\middle| ~ \bpsiksig^{(m_{k,\sigma})}(\br)\right]$ comprises of the $m_{k,\sigma}$ real-valued degenerate eigenfunctions.
Although the formulation presented here is for non-periodic systems, the main ideas can also be extended to periodic systems. Typically, one is interested in the canonical eigenfunction, which are orthonormal. The orthogonality between eigenfunctions belonging to different eigenvalues is guaranteed by the Hermiticity of the KS Hamiltonian. Thus, to obtain the canonical eigenfunctions, we enforce orthonormality among degenerate eigenfunctions, 
which can be expressed as 
\begin{equation} \label{eq:orthonormal}
\int\bPsiksig\trans(\br)\bPsiksig(\br)\dr=\bI_{m_{k,\sigma}}\,.
\end{equation}
Given the canonical eigenfunctions $\{\bPsiksig\}\rvert_{k=1}^{M_\sigma}$, its spin density can be defined as
\begin{equation}\label{eq:rhosig}
    \rhosig(\br) = \sum_{k=1}^{M_{\sigma}}\tr\left(f(\bEksig)\bPsiksig\trans(\br)\bPsiksig(\br)\right)\,,
\end{equation}
where $f(\bEksig)=\left(\bI_{m_{k,\sigma}}+e^{-(\bEksig-\mu_{\sigma}\bI_{m_{k,\sigma}})/k_BT}\right)^{-1}$ is Fermi-Dirac occupancy matrix with 
\tcb{$\mu_\sigma$ being the chemical potential for the $\sigma$ spin, given through the conservation of the number of electrons ($N_\sigma$),
\begin{equation} \label{eq:sumf}
    \sum_{k=1}^{M_{\sigma}}\tr\left(f(\bEksig)\right)=\int \rhodsig(\br) \dr = N_\sigma\,.
\end{equation}
}
We emphasize that the use of occupancy is crucial to seamlessly handle both pure-$v_s$ and e-$v_s$ densities. In general it is difficult to \textit{a priori} ascertain if a target density is pure-$v_s$ or e-$v_s$. To this end, the use of occupancy allows a unified means to admit both kinds of densities. 
Using the various constraints (Eqs.~\ref{eq:KSSig},~\ref{eq:orthonormal}, and ~\ref{eq:sumf}), the optimization in Eq.~\ref{eq:objective} can be recast as an unconstrained optimization of the Lagrangian,
%
\begin{widetext}
\tcb{
\begin{equation}\label{eq:L}
\begin{split}
    \lag = & \sum_{\sigma}\int w_\sigma(\br)\left(\rhodsig(\br)-\rhosig(\br)\right)^2\dr + \sum_{\sigma}\sum_{k=1}^{M_{\sigma}}\tr\left(\int\bPksig\trans(\br)\left(\Hopsig\bPsiksig(\br)-\bPsiksig(\br)\bEksig\right)\dr\right) + \\
    &\sum_\sigma\eta_\sigma\left(\sum_{k=1}^{M_{\sigma}}\tr\left(f(\bEksig)\right)-N_\sigma\right) 
    + \sum_{\sigma}\sum_{k=1}^{M_{\sigma}}\tr\left(\bDksig\left(\int\bPsiksig\trans(\br)\bPsiksig(\br)\dr-\bI_{m_{k,\sigma}}\right)\right)\,,
\end{split}
\end{equation}
}
\end{widetext}
where $\bPksig(\br)=\left[p^{(1)}_{k,\sigma}(\br) ~\middle| p^{(2)}_{k,\sigma}(\br)~\middle| \cdots ~\middle|p^{(m_{k,\sigma})}_{k,\sigma}(\br)\right]$ comprises of the adjoint functions that enforce the constraints corresponding to the KS eigenvalue problem (Eq.~\ref{eq:KSSig}); $\bDksig \in \mathbb{R}^{m_{k,\sigma}\times m_{k,\sigma}}$ is the Lagrange multiplier matrix enforcing the orthonormality constraints in Eq.~\ref{eq:orthonormal}; \tcb{$\eta_\sigma\in \mathbb{R}$} is the Lagrange multiplier enforcing Eq.~\ref{eq:sumf}. Optimizing $\lag$ with respect to $\bPksig$, $\bDksig$ and \tcb{$\eta_\sigma$}, yields the constraints equations Eq.~\ref{eq:KSSig}, Eq.~\ref{eq:orthonormal}, and Eq.~\ref{eq:sumf}, respectively. Optimizing $\lag$ with respect to $\bPsiksig$, $\bEksig$, and $\mu_\sigma$ results in:
\begin{widetext}
\tcb{
\begin{equation}\label{eq:adjoint}
    \Hopsig(\br)\bPksig-\bPksig\bEksig = 4 w_\sigma(\br)\left(\rhodsig(\br)-\rhosig(\br)\right)\bPsiksig(\br)f(\bEksig)
    - \bPsiksig(\br)\left(\bDksig+\bDksig\trans\right)\,, 
\end{equation}
\begin{equation}\label{eq:adjointOverlap}
    \int \bPsiksig\trans(\br)\bPksig(\br)\dr = \frac{\partial f_{k,\sigma}^{\mu_\sigma}}{\partial\epsilonksig}\left[-2 \int w_\sigma(\br)\left(\rhodsig(\br)-\rhosig(\br)\right)\bPsiksig\trans(\br)\bPsiksig(\br)\dr+\eta_\sigma\bI_{m_{k,\sigma}}\right]\,,
\end{equation}
\begin{equation}\label{eq:eta}
    \eta_\sigma\sum_{k=1}^{M_{\sigma}}m_{k,\sigma}\frac{\partial f_{k,\sigma}^{\mu_\sigma}}{\partial \mu_\sigma} = 2\sum_{k=1}^{M_{\sigma}}\frac{\partial f_{k,\sigma}^{\mu_\sigma}}{\partial \mu_\sigma}\int w_\sigma(\br)\left(\rhodsig(\br)-\rhosig(\br)\right)\tr\left(\bPsiksig\trans(\br)\bPsiksig(\br)\right)\dr\,,
\end{equation}
}
\end{widetext}
%
where $f_{k,\sigma}^{\mu_\sigma} = \left(1+e^{-(\epsilonksig-\mu_\sigma)/k_BT}\right)^{-1}$. Having solved Eqs.~\ref{eq:KSSig},~\ref{eq:orthonormal},~\ref{eq:sumf},~\ref{eq:adjoint},~\ref{eq:adjointOverlap}, and ~\ref{eq:eta}, we can write 
\begin{equation}\label{eq:vxc}
    \frac{\delta \lag}{\delta \vxcsig(\br)} = \sum_{k=1}^{M_{\sigma}}\tr\left(\bPksig\trans(\br) \bPsiksig(\br)\right)\,.
\end{equation}
The above provides the update to $\vxcsig(\br)$ to be used in any gradient-based optimization method. In the case of degeneracy, $\bPsiksig$ cannot be determined uniquely. That is any $\bPsitildeksig=\bPsiksig\bQksig$ (with $\bQksig$ being any $m_{k,\sigma} \times m_{k,\sigma}$ orthogonal matrix) will satisfy Eqs.~\ref{eq:KSSig}, ~\ref{eq:orthonormal} as well as preserve the density (Eq.~\ref{eq:rhosig}). However, using $\bPsitildeksig$ in the adjoint equation (Eq.~\ref{eq:adjoint}), the corresponding adjoint is given by $\bPtildeksig=\bPksig\bQksig$. Thus for any orthogonal transformation of $\bPsiksig$, its corresponding adjoint functions also are transformed similarly. 
Finally, rewriting Eq.~\ref{eq:vxc} in terms of $\bPsitildeksig$ and $\bPtildeksig$, we have
\begin{equation}\label{eq:LVxcInvariance}
\begin{split}
        \frac{\delta \lag}{\delta \vxcsig(\br)} &= \sum_{k=1}^{M_{\sigma}}\tr\left(\bPtildeksig\trans(\br) \bPsitildeksig(\br)\right) \\
        &=  \sum_{k=1}^{M_{\sigma}}\tr\left(\bQksig\trans\bPksig\trans(\br) \bPsiksig(\br)\bQksig\right) \\
        &= \sum_{k=1}^{M_{\sigma}}\tr\left(\bPksig\trans(\br) \bPsiksig(\br)\right)\,,
\end{split}
\end{equation}
where we used the invariance of the trace of products of matrices with respect to cyclic permutations. This shows the uniqueness of $\delta \lag/\delta \vxcsig(\br)$ for a given $\vxcsig(\br)$. We refer to the Supporting Information (SI) for a detailed derivation of Eqs.~\ref{eq:adjoint}--\ref{eq:LVxcInvariance}. 

In order to numerically solve the above equations, we discretize the $\bPsiksig$ and $\vxcsig$  using an adaptively refined finite-element basis~\cite{MOTAMARRI2013308,MOTAMARRI2020106853,das2022dft} that provides systematic convergence for all-electron DFT calculations, and is essential for an accurate solution of the inverse DFT problem. In particular, we use an adaptive discretization based on a fourth-order spectral finite-element (FE) basis for discretizing the KS eigenfunctions and the corresponding adjoint functions. A discretization based on linear finite-element basis is sufficient for $\vxcsig$, which is a smoother field in comparison to the KS eigenfunctions. \tcb{To mitigate the unphysical artifacts in $\vxcsig$ from the lack of cusp on nuclei in the Gaussian target densities we add a small cusp-correction $\Delta\rhodsig$ to $\rhodsig(\br)$ near the nuclei, given as
\begin{equation} \label{eq:deltaRho}
    \Delta \rhodsig(\br) = \rho_{\sigma,\text{FE}}^{\text{DFT}}(\br) - \rho_{\sigma,\text{G}}^{\text{DFT}}(\br)\,,
\end{equation}
where $\rho_{\sigma,\text{FE}}^{\text{DFT}}(\br)$ denotes the groundstate density of a chosen XC approximation (say LDA or GGA) solved using a finite element basis, and $\rho_{\sigma,\text{G}}^{\text{DFT}}(\br)$ denotes the same, albeit solved using the Gaussian basis used in the generation of the target density $\rhodsig(\br)$. The key idea here is that $\rho_{\sigma,\text{FE}}^{\text{DFT}}(\br)$, obtained using the finite element basis, contains the cusp at nuclei, and hence, one can expect $\Delta\rhodsig(\br)$ to reasonably capture the Gaussian basis-set error near the nuclei. We refer to~\cite{Kanungo2019} to illustrate the efficacy and robustness of this cusp-correction towards mitigating any spurious oscillation in the XC potentials. Beyond the missing cusp, the Gaussian densities also have wrong far-field decay---Gaussian decay instead of exponential decay---which can induce spurious oscillations in the XC potentials}. To mitigate this, we enforce appropriate boundary condition on $\vxcsig(\br)$ in the low density region ($\rhodsig(\br) < 10^{-7}$). 
\tcb{This is done by using an initial guess for the $\vxcsig(\br)$ that is consistent with the expected decay and then applying homogeneous Dirichlet boundary conditions on the adjoint functions while solving Eq.~\ref{eq:adjoint}. In effect, this fixes $\vxcsig$ to its initial value in the low density region. In particular, we use a scaled Fermi-Amaldi potential, $\vFA=-\frac{\alpha}{N}\int\frac{\rhod(\br')}{|\br-\br'|}\dr'$, where $\rhod$ and $N$ are the total density and total number of electrons, respectively. We choose $\alpha=1$ for the exact XC potentials corresponding to configuration interaction (CI) densities, so as to ensure the expected $-1/r$ decay. For the model potentials corresponding to the hybrid functionals (B3LYP and SCAN0), $\alpha$ is set to the fraction of the Hartree–Fock exchange used in the hybrid XC functional, ensuring the consistent far-field decay of the model XC potentials. In our calculations involving densities obtained from SCAN functionals, we use the Slater exchange potential, $v_\text{S}(\br) = -(3/\pi)^{1/3}\rhod(\br)^{1/3}$, as the boundary condition to ensure the expected exponential decay in the $\vxcsig(\br)$.}

In our numerical studies, we consider four atoms ---Li (doublet), C (triplet, $\nu=2$), N (quartet, $\nu=3$), and O (triplet, $\nu=2$), where $\nu$ denotes the difference in the number of majority (up) and minority (down) spin electrons. Additionally, we also consider two molecules---CN (doublet) and CH$_2$ (triplet, $\nu=2$). All our groundstate CI and DFT calculations to obtain $\rhodsig$ are done using the QChem software package~\cite{QChem4}. \tcm{In our inverse calculations, we use weights of the form $w_\sigma(\br)=1/(\rhodsig(\br)+\tau)^t$, with $t=\{0,1,2\}$, used in sequence, and $\tau=10^{-6}$ to avoid any singularity (see Eq.~\ref{eq:objective} for definition). The values $t=1$ and $t=2$ are crucial to attaining good agreement between the Kohn-Sham HOMO level corresponding to the exact XC potential and the negative of the ionization potential (Koopmans' theorem~\cite{Perdew1997}).} For all inverse DFT calculations, we use a temperature $T=100$~K for the Fermi-Dirac distribution. The $L_2$ error in the density---$\norm{\rhodsig-\rhosig}_{L_2}$---is driven below $10^{-4}$ (except for B3LYP, where it is driven below $3\times10^{-4}$) at convergence. We remark that while the use of a non-zero temperature formally results in an e-$v_s$ density, practically, for systems with non-degenerate frontier orbitals or with finite HOMO-LUMO gap, it results in a pure-$v_s$ density. 

\tcb{We first examine the accuracy of the proposed approach with an LDA density generated using finite element basis as the target density. This allows for a direct comparison of the XC potential from inverse DFT with the known LDA potential.} \tcm{Using N as a benchmark system, we accurately reproduce the LDA potential.} \tcb{We note that the accuracy of the potential is crucially dependent on the adequacy of the Gaussian basis (while using Gaussian densities) used to generate the target densities as well as the refinement of the finite element basis in the inverse DFT calculations. To this end, we examine the sensitivity of the XC potentials to the choice of Gaussian basis and finite element discretization,} \tcm{by using the groundstate density of N from heat-bath configuration interactions (HBCI) calculations~\cite{Dang2023}}. \tcb{We use two increasingly larger polarized Gaussian basis with tight cores~\cite{Pritchard2019}, namely, cc-pCVQZ and cc-pCV5Z, for computing the groundstate density, and two increasingly refined finite element basis---one with fourth-order finite elements and the other with fifth-order finite elements---for inverse DFT calculation. We observe negligible differences in the XC potentials obtained from different combination of Gaussian and finite element basis.} \tcm{Notably, with different choices of Gaussian and finite element basis, we observe a difference of $2-3$ mHa in the KS eigenvalues and the correlation part of kinetic energy ($T_\text{c}$), which is close to chemical accuracy. For all subsequent calculations reported in this work, unless stated otherwise, we use the combination of cc-pCVQZ and fourth-order finite elements.} \tcb{We, next, evaluate the exact XC potentials for all the six benchmark systems (Li, C, N, O, CN, CH2) using  their HBCI based groundstate densities as the target densities. For all the exact XC potentials obtained, we attain $\sim$10 mHa agreement between the highest KS eigenvalue and the negative of the ionization potential (Koopmans' theorem). Further, for all the atoms considered, we find good agreement of $\sim$4 mHa between the virial of the XC potential and $E_{\text{xc}} + T_{\text{c}}$~\cite{Levy1985}.  We present the details of all the above accuracy tests---verification study with LDA densities, sensitivity to Gaussian and finite element basis, test using Hartree-Fock density, agreement with Koopmans' theorem, and the virial relation test---in the SI.}     

Having established the accuracy of the proposed approach to inverse DFT for open-shell systems, we present a comparison of the exact and model XC potentials for all the benchmark systems considered. The model XC potentials are obtained using DFT-based groundstate densities of widely used approximate XC functionals, which includes two hybrid (B3LYP~\cite{Becke1993b,Lee1988} and SCAN0~\cite{Hui2016}), one meta-GGA (SCAN~\cite{Sun2015}), one GGA (PBE~\cite{Perdew1996}), and one LDA (PW92~\cite{Perdew1992}) functional. \tcm{For the hybrid and the meta-GGA functionals, the groundstate densities are evaluated within the generalized Kohn-Sham (GKS) formalism~\cite{Seidl1996}.} \tcm{While most of the approximate XC functionals are modeled to yield accurate groundstate energies or energy differences, accurate potentials are crucial for any response calculation.} We quantify the difference between the exact and the model XC potentials using two $\rhodsig$ weighted error metrics
\begin{equation}\label{eq:vxcErrors}
e_{1,\sigma} = \frac{\norm{\rhodsig\delta\vxcsig}_{L_2}}{\norm{\rhodsig\vxcsig^{\text{exact}}}_{L_2}}\,,
\quad e_{2,\sigma} = \frac{\norm{\rhodsig\modulus{\nabla\delta\vxcsig}}_{L_2}}{\norm{\rhodsig\modulus{\nabla\vxcsig^{\text{exact}}}}_{L_2}}\,,
\end{equation}
where $\delta\vxcsig=\vxcsig^{\text{exact}}-\vxcsig^{\text{model}}$ (cf. SI for error metrics without the $\rhodsig$ weight and additional error metrics with the chemical potentials aligned). For all the open-shell systems considered, the target densities (from HBCI and DFT) turn out to be pure-$v_s$ densities. However, to test the efficacy of the proposed approach in handling e-$v_s$ density, we used a SCAN0 based density for the B atom obtained using a Fermi-Dirac smearing in the ground state calculation. Upon inversion, the density leads to an ensemble of KS single Slater determinants for the majority-spin---three degenerate orbitals near the Fermi level with 1/3 occupancy (see SI for details).  

\begin{figure}[htbp!]
  \centering
  \subfigure{\includegraphics[scale=0.75]{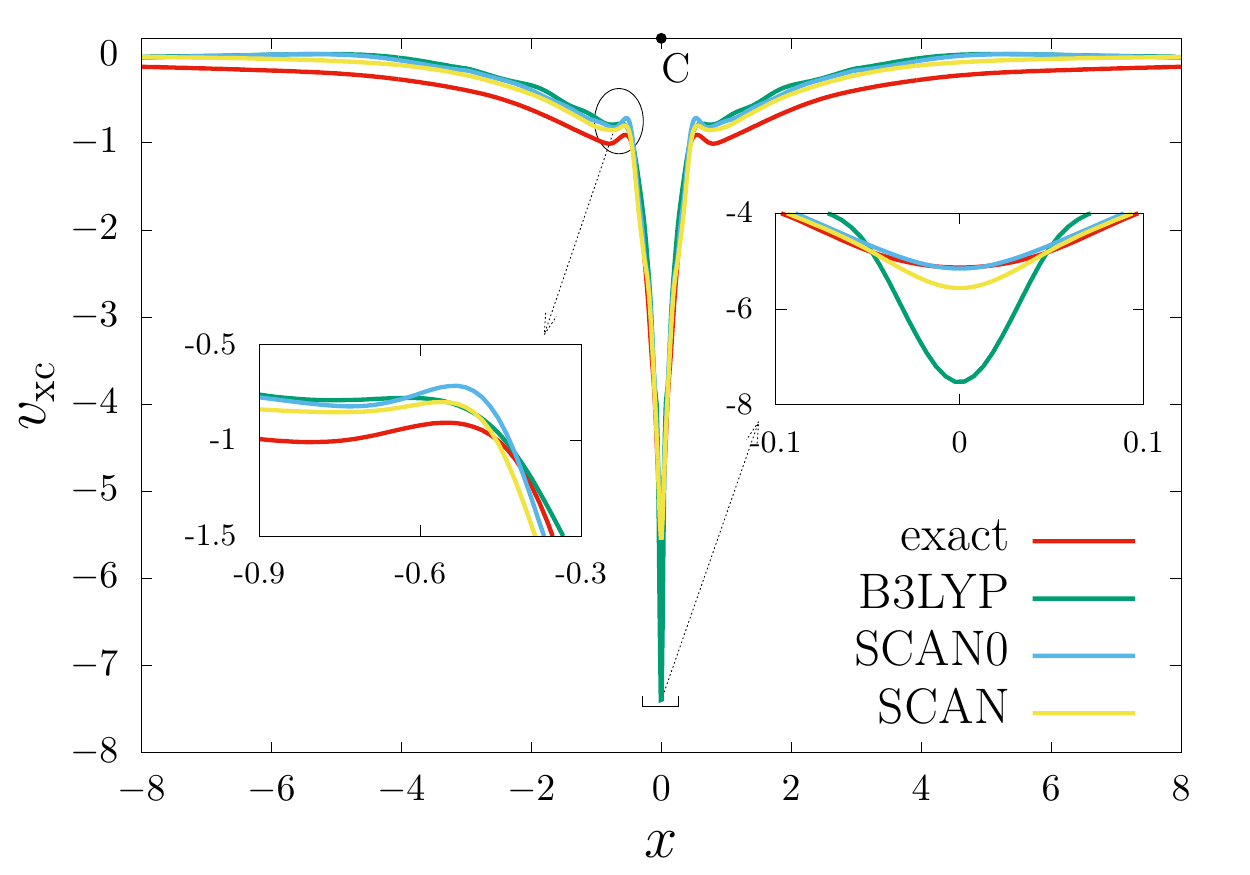}}
  \subfigure{\includegraphics[scale=0.75]{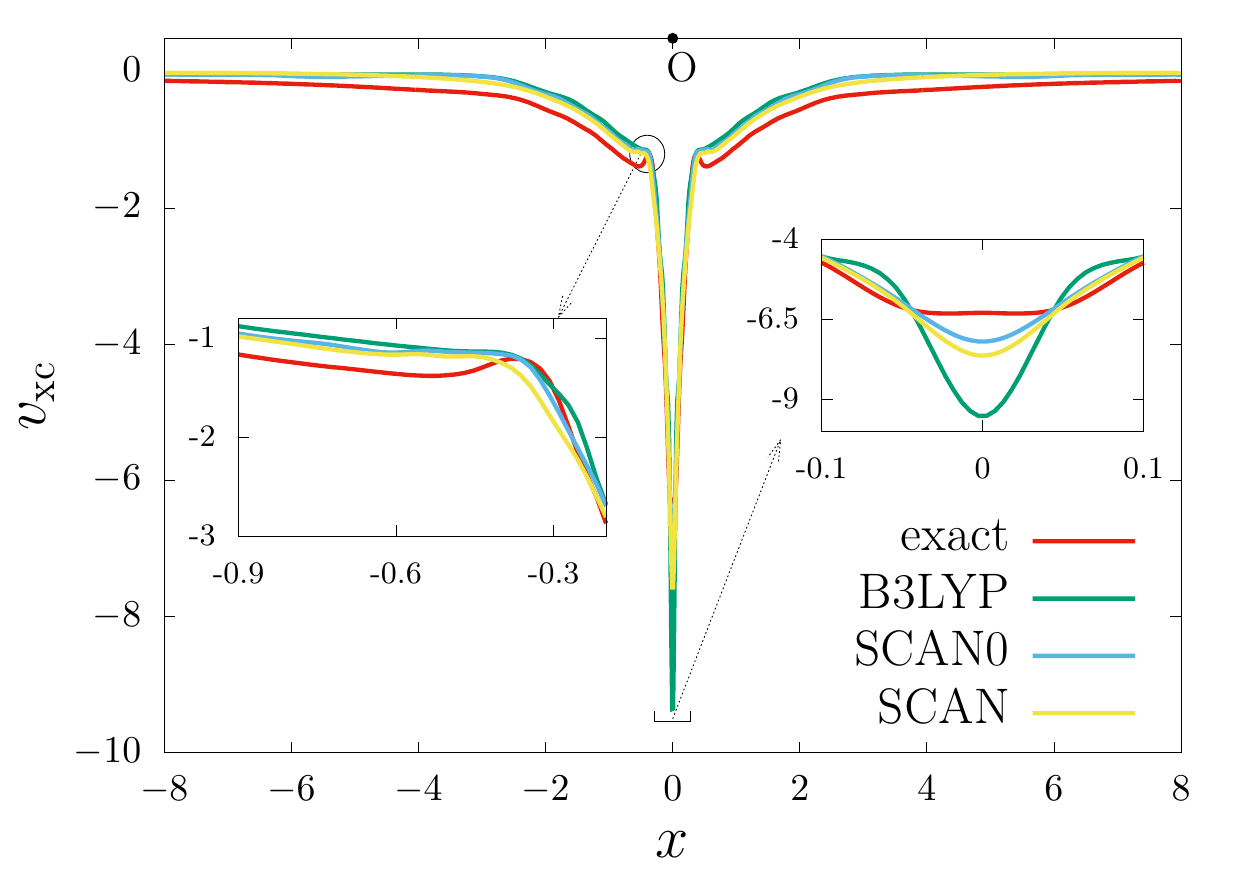}}
  \vspace{-0.3cm}
   \caption{Comparison of exact and model XC potentials for C (top) and O (bottom) atoms for the majority-spin. The x-axis corresponds to the dominant principal axis of the moment of inertia tensor of their densities.}
   \label{fig:C_O_Up}
\end{figure}

Fig.~\ref{fig:C_O_Up} compares the exact and model XC potentials corresponding to B3LYP, SCAN0, and SCAN densities (all majority-spin), for C and O. The model potentials differ significantly in the low density region, owing to their incorrect far-field asymptotics. Importantly, the model potentials differ qualitatively even in the high density region. All the model XC potentials are deeper at the atoms compared to the exact one (see insets in Fig.~\ref{fig:C_O_Up}), with the B3LYP potential being substantially deeper. The SCAN0 and SCAN based model potentials exhibit the atomic inter-shell structure---a distinctive feature of the exact potential marked by a local maxima to minima transition---otherwise absent in the B3LYP potentials (see left insets in Fig.~\ref{fig:C_O_Up}). Quantitatively, SCAN0 based potentials provide better agreement with the exact one (see Table~\ref{tab:errWeightedUp}) than other model potentials. While the relative $L_2$ errors in densities for the model functionals are of $\mathcal{O}(10^{-3}-10^{-2})$, the relative errors in the XC potentials are of $\mathcal{O}(10^{-1}-10^0)$, manifesting in significant differences in the KS eigenvalues, especially for the minority-spin (cf. SI). In other words, for assessing the XC functionals, the XC potentials are more descriptive than the densities. This presents a strong case for using the exact XC potentials in the design and modeling of XC functionals.
%

\begin{figure}[htbp!]
  \centering
  \includegraphics[scale=0.75]{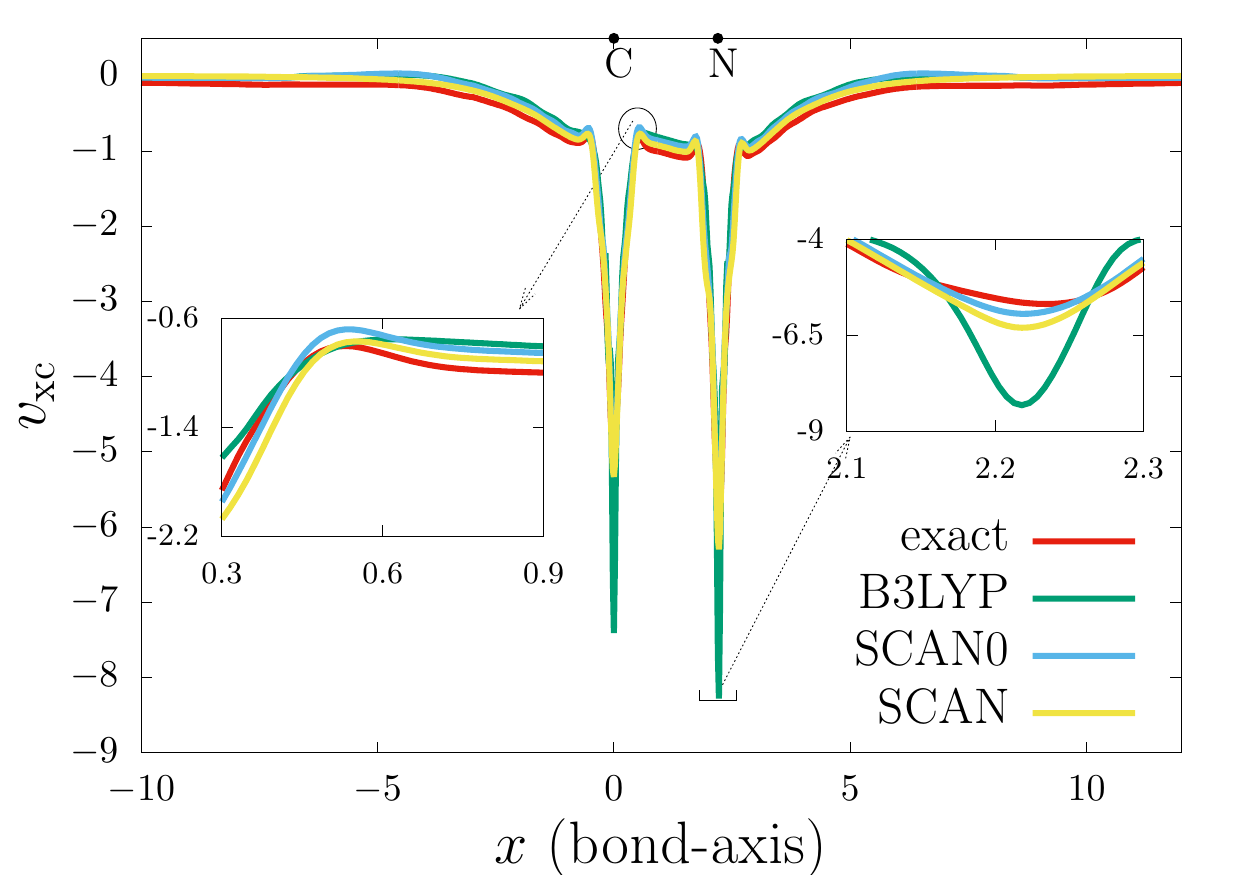}
  \vspace{-0.3cm}
   \caption{Comparison of exact and model XC potentials for CN along along the bond for the majority-spin.}
   \label{fig:CNUp}
\end{figure}

\begin{figure}[htbp!]
  \vspace{-0.3cm}
  \centering
  \subfigure{\includegraphics[scale=0.75]{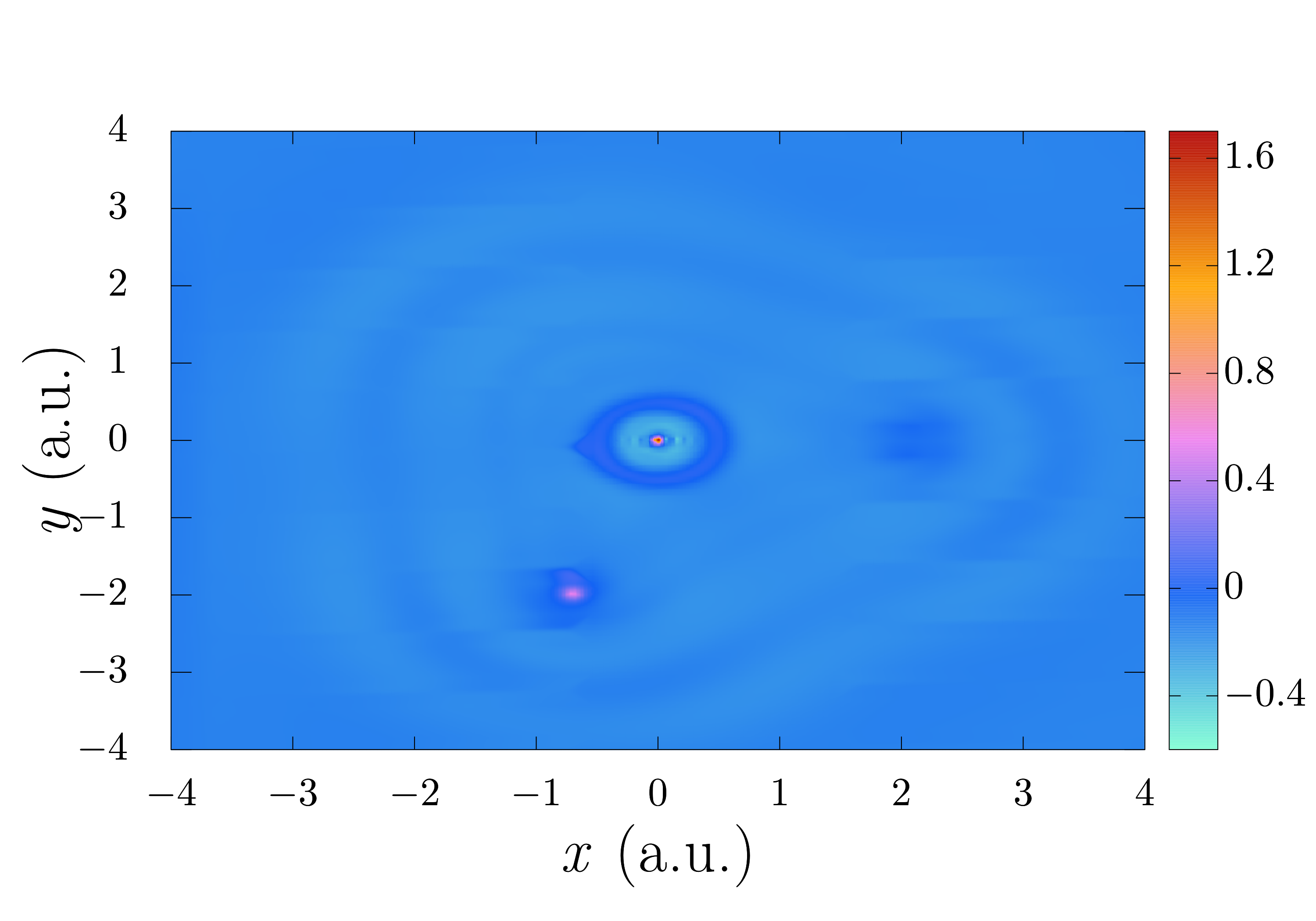}} \\[-2ex]
  \vspace{-0.3cm}
  \subfigure{\includegraphics[scale=0.75]{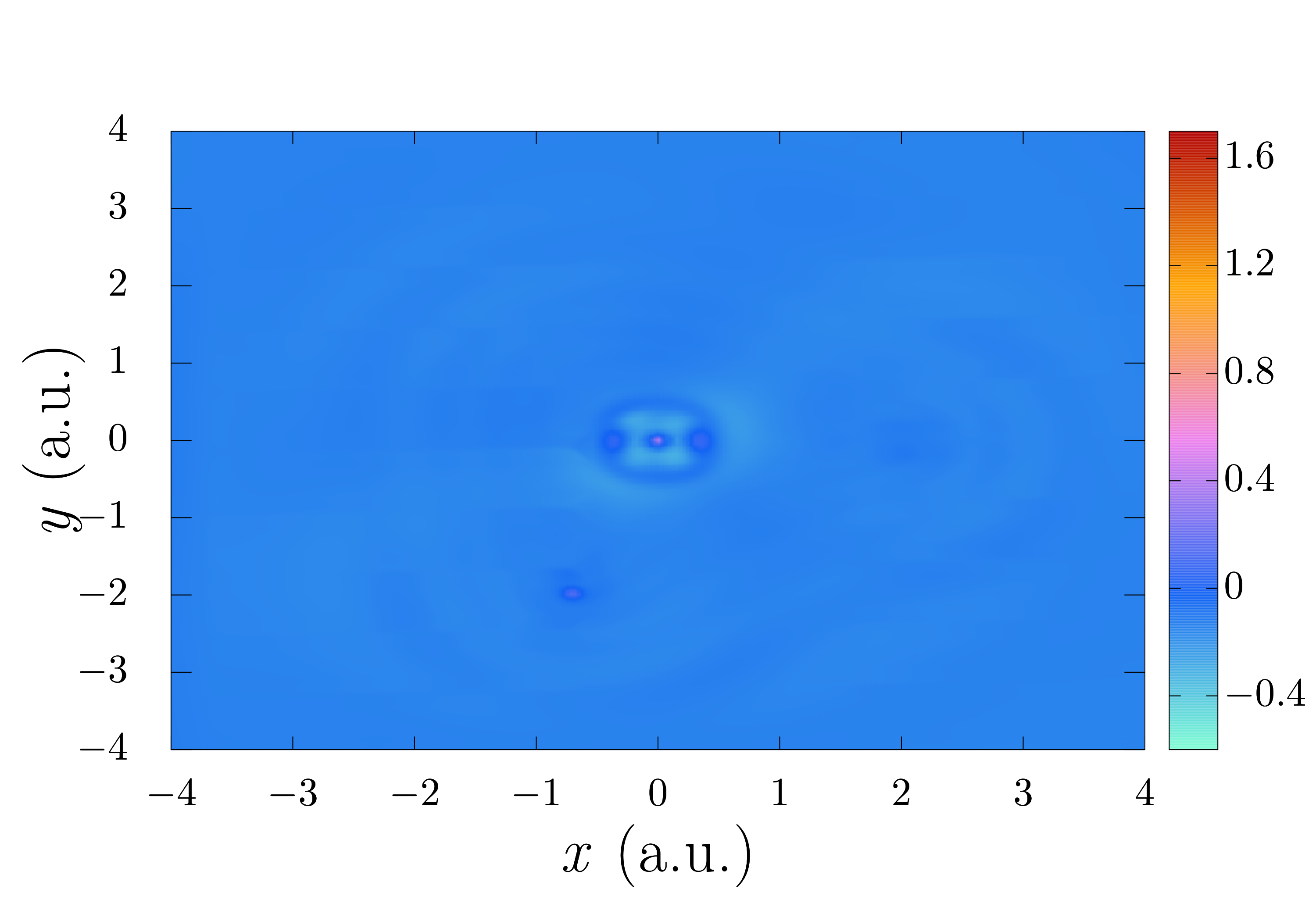}}
  \vspace{-0.3cm}
   \caption{Error in the B3LYP (top) and SCAN0 (bottom) based model XC potentials for CH$_2$ on the plane of the molecule.}
   \label{fig:CH2_B3LYP_SCAN0_Err_Up}
\end{figure}

Turning to the molecules, Fig.~\ref{fig:CNUp} compares the exact and model XC potentials for CN (majority-spin). For CH$_2$, Fig.~\ref{fig:CH2_B3LYP_SCAN0_Err_Up} presents the error ($\vxcsig^{\text{exact}}-\vxcsig^{\text{model}}$) in the B3LYP and SCAN0 based XC potentials (cf. SI for a similar comparison for SCAN and the individual $\vxcsig$). Similar to the atoms, the model potentials are deeper at the atoms. Once again, SCAN0 and SCAN based potentials offer better qualitative and quantitative agreement than the rest, including the presence of atomic intershell structure around both C and N atom in CN and around the C atom in CH$_2$ (cf. yellow rings around the C atom in the $\vxcsig$ plots for CH$_2$ in the SI). 
Results for minority-spin counterparts of Figs.~\ref{fig:C_O_Up}-\ref{fig:CH2_B3LYP_SCAN0_Err_Up} as well as Li and N can be found in the SI.
\begin{table*}
\caption{\small Comparison of the model XC potentials in terms of the error metrics $e_{1,\sigma}$ and $e_{2,\sigma}$ (defined in Eq.~\ref{eq:vxcErrors}), for the majority-spin. See the SI for the error metrics for minority-spin.}
\begin{tabular}{|p{1.4cm} | M{0.9cm} | M{0.8cm} | M{0.8cm} | M{0.8cm} | M{0.8cm} | M{0.8cm} | M{0.8cm} | M{0.8cm} | M{0.8cm} | M{0.8cm} | M{0.8cm}| M{0.8cm}|}
\hline
\multirow{2}{2cm}{Model} & \multicolumn{2}{M{1.6cm}|}{Li} & \multicolumn{2}{M{1.6cm}|} {C} & \multicolumn{2}{M{1.6cm}|}{N} & \multicolumn{2}{M{1.6cm}|}{O} & \multicolumn{2}{M{1.6cm}|}{CN} & \multicolumn{2}{M{1.6cm}|}{CH$_2$} \\
\cline{2-13}
& $e_{1,\sigma}$ & $e_{2,\sigma}$ & $e_{1,\sigma}$ & $e_{2,\sigma}$ & $e_{1,\sigma}$ & $e_{2,\sigma}$ &  $e_{1,\sigma}$ & $e_{2,\sigma}$ &  $e_{1,\sigma}$ & $e_{2,\sigma}$ & $e_{1,\sigma}$ & $e_{2,\sigma}$\\
\hline 
\hline 
B3LYP & 0.100 & 1.870 & 0.114 & 1.694 & 0.106 & 1.315 & 0.106 & 1.392 & 0.109 & 1.511 & 0.113 & 1.700\\ \hline
SCAN0 & 0.070 & 0.526 & 0.064 & 0.493 & 0.060 & 0.434 & 0.051 & 0.429 & 0.058 & 0.372 & 0.061 & 0.399\\ \hline
SCAN & 0.086 & 0.890 & 0.062 & 0.655 & 0.060 & 0.583 & 0.054 & 0.576 & 0.055 & 0.499 & 0.058 & 0.546\\ \hline
PBE & 0.116 & 2.200 & 0.124 & 1.938 & 0.116 & 1.502 & 0.118 & 1.581 & 0.120 & 1.738 & 0.117 & 1.731\\ \hline
PW92 & 0.137 & 0.358 & 0.128 & 0.714 & 0.123 & 0.591 & 0.120 & 0.606 & 0.117 & 0.639 & 0.122 & 0.636\\ \hline
\end{tabular}
\label{tab:errWeightedUp}
\end{table*}


By generalizing the inverse DFT approach to degenerate KS eigenvalues and ensemble-v-representable densities, new insights into the XC potential for open-shell electronic states can be gained. First, this allowed comparisons to be made between exact and model XC potentials, providing a quantitative measure of the quality of model XC potentials for open-shell states. 
The availability of exact XC potentials for the open-shell case can serve as an important guide for the creation of new, accurate models of XC within DFT.



\section*{Supporting Information}
\noindent Proofs, accuracy tests, and additional data for exact and model potentials (SI.pdf)\\
Density, exact potentials, and Kohn-Sham orbitals for lithium and nitrogen (Li\_N.xlsx)

\section*{Acknowledgements}
We gratefully acknowledge DOE grant DE-SC0022241 which supported this study. V.G. also acknowledges support from AFOSR grant FA9550-21-1-0302 that supported the analysis of degenerate eigenvalue problems. This research used resources of the NERSC Center, a DOE Office of Science User Facility supported by the Office of Science of the U.S. Department of Energy under Contract No. DE-AC02-05CH11231. We acknowledge the support of DURIP grant W911NF1810242, which also provided computational resources for this work.

%

\clearpage
\pagebreak
\widetext
\begin{center}
\textbf{\Large Supplemental Material}
\end{center}
\setcounter{equation}{0}
\setcounter{figure}{0}
\setcounter{table}{0}
\setcounter{page}{1}
\setcounter{section}{0}
\makeatletter
\renewcommand{\theequation}{S\arabic{equation}}
\renewcommand{\thefigure}{S\arabic{figure}}
\renewcommand{\bibnumfmt}[1]{[S#1]}
\renewcommand{\citenumfont}[1]{S#1}
\renewcommand{\thesection}{S\arabic{section}}
\renewcommand{\thepage}{S-\Roman{page}}

\section{Optimality conditions}
In this section we derive the optimality conditions (Eqs. 7, 8, and 9 in the main manuscript) and provide the details of their solution procedure. First, we begin with a relation that will be useful subsequently. 
\begin{itemize}

    \item Given the matrix function $f(\bE)=\left(\bI+e^{-(\bE-\mu\bI)/k_BT}\right)^{-1}$, 
\begin{equation} \label{eq:SI_dfdE}
    \frac{\partial f(\bE)}{\partial \Eij} = \frac{1}{k_BT}\left(\bI+e^{-(\bE-\mu\bI)/k_BT}\right)^{-1}e^{-(\bE-\mu\bI)/k_BT}\left(\bLambdaij + \frac{1}{2k_BT}[\bE,\bLambdaij]\right)\left(\bI+e^{-(\bE-\mu\bI)/k_BT}\right)^{-1}\,,
\end{equation}
where $[\bA,\bB]=\bA\bB-\bB\bA$ is the commutator of two matrices and $\bLambdaij$ is a matrix which is zero except in the $(i,j)-$th entry, which is equal to 1. In other words, $\Lambdaijkl=\delta_{ik}\delta_{jl}$.\\
\textit{Proof}.
The proof of the above follows from the following definition of $\frac{\partial f(\bE)}{\partial \Eij}$ in terms of a directional derivative along $\bLambdaij$,
\begin{equation} \label{eq:SI_directionalDer}
    \frac{\partial f(\bE)}{\partial \Eij} = \lim_{h \to 0}\frac{f(\bE+h\bLambdaij)-f(\bE)}{h}\,.
\end{equation}
Using the defintion of $f(\bE)$,
\begin{equation}
\begin{split}
    f(\bE + h\bLambdaij) &= \left(\bI + e^{-(\bE+h\bLambdaij-\mu\bI)/k_BT}\right)^{-1} =  \left(\bI + e^{\mu/k_BT} e^{-(\bE+h\bLambdaij)/k_BT}\right)^{-1}\,,
\end{split}
\end{equation}
where the last equality follows from the fact that $\mu\bI$ commutes with $(\bE+h\bLambdaij)$ (Note: If $[\bA,\bB]=0$, $e^{\bA+\bB}=e^{\bA}e^{\bB}$). Letting, $s=e^{\mu/k_BT}$ and $t=1/k_BT$, 
\begin{equation}
\begin{split}
    f(\bE + h\bLambdaij) &= \left(\bI + s e^{-t(\bE+h\bLambdaij)}\right)^{-1}\\
    &=  \left(\bI + s e^{-t\bE}e^{-ht\bLambdaij}e^{-\frac{ht^2}{2}[\bE,\bLambdaij]}e^{-\frac{h^2t^3}{6}\left(2[\bLambdaij,[\bE,\bLambdaij]]+[\bE,[\bE,\bLambdaij]]\right)}\ldots\right)^{-1}\\
    &= \left(\bI + s e^{-t\bE}\left(\bI-ht\left(\bLambdaij+\frac{t}{2}[\bE,\bLambdaij]\right)+\mathcal{O}(h^2)\right)\right)^{-1}\\
    &= \left(\bI + se^{-t\bE}-htse^{-t\bE}\left(\bLambdaij+\frac{t}{2}[\bE,\bLambdaij]\right)+\mathcal{O}(h^2)\right)^{-1}
\end{split}
\end{equation}
where in the second equality we used the Zassenhaus formula~\cite{Magnus1954}. For two square matrices $\bA$ and $\bB$ of same dimensions, $\left(\bA-\bB\right)^{-1}=\sum_{l=0}^{\infty}\left(\bA^{-1}\bB\right)^l\bA^{-1}$. Taking $\bA=\bI+se^{-t\bE}$ and $\bB=htse^{-t\bE}\left(\bLambdaij+\frac{t}{2}[\bE,\bLambdaij]\right)+\mathcal{O}(h^2)$, the above equation yields,
\begin{equation}
\begin{split}
    f(\bE + h\bLambdaij) &= \left(\bI+se^{-t\bE}\right)^{-1} + \left(\bI+se^{-t\bE}\right)^{-1}\left(htse^{-t\bE}\left(\bLambdaij+\frac{t}{2}[\bE,\bLambdaij]\right)\right)\left(\bI+se^{-t\bE}\right)^{-1}\\
    & + \mathcal{O}(h^2)\,.
\end{split}
\end{equation}
Using the above relation in Eq.~\ref{eq:SI_directionalDer} gives,
\begin{equation}
    \frac{\partial f(\bE)}{\partial \Eij} = ts\left(\bI + s e^{-t\bE}\right)^{-1}e^{-t\bE}\left(\bLambdaij+\frac{t}{2}[\bE,\bLambdaij]\right)\left(\bI + s e^{-t\bE}\right)^{-1}\,.
\end{equation}
This concludes the derivation of Eq.~\ref{eq:SI_dfdE}.
\end{itemize}

\subsection{Derivation of Eq. 7 in the main manuscript}
Taking the variation of a $\lag$ with respect to a KS orbtial $\psi_{k,\sigma}^{(i)}$, we have
\begin{equation} \label{eq:SI_varPsi}
    \frac{\delta \lag}{\delta \psi_{k,\sigma}^{(i)}(\br)} = -2w_\sigma(\br)\left(\rhodsig(\br)-\rhosig(\br)\right))\frac{\partial\rhosig(\br)}{\partial \psi_{k,\sigma}^{(i)}(\br)} + \Hopsig p_{k,\sigma}^{(i)}(\br)  - \sum_j^{m_{k,\sigma}} p_{j,\sigma}^{(j)}(\br)E_{ij} + \sum_{j=1}^{m_{k,\sigma}} (D_{k,\sigma,ij}+D_{k,\sigma,ji})\psi_{k,\sigma}^{(j)}\,.
\end{equation}
Using $\rhosig(\br)=\sum_{k=1}^{M_\sigma}\tr\left(f(\bEksig)\bPsiksig\trans(\br)\bPsiksig(\br)\right)$, we have
\begin{equation}
    \frac{\partial\rhosig(\br)}{\partial \psi_{k,\sigma}^{(i)}(\br)} = \sum_j^{m_{k,\sigma}} \left(f(\bEksig)_{ij}+f(\bEksig)_{ji}\right)\psi_{k,\sigma}^{(j)}(\br)=2 \sum_j^{m_{k,\sigma}}f(\bEksig)_{ij}\psi_{k,\sigma}^{(j)}(\br)\,,
\end{equation}
where the last equality uses the fact that $f(\bEksig)$ is a symmetric matrix. Now, using the above relation in Eq.~\ref{eq:SI_varPsi} and setting $\frac{\delta \lag}{\delta \psi_{k,\sigma}^{(i)}(\br)}$ to zero gives
\begin{equation}
    \Hopsig p_{k,\sigma}^{(i)}(\br) - \sum_j^{m_{k,\sigma}} p_{j,\sigma}^{(j)}(\br)E_{ij} = 4w_\sigma(\br)\left(\rhodsig(\br)-\rhosig(\br)\right)\sum_j^{m_{k,\sigma}} f(\bEksig)_{ij}\psi_{k,\sigma}^{(j)}(\br) -  \sum_{j=1}^{m_{k,\sigma}} (D_{k,\sigma,ij}+D_{k,\sigma,ji})\psi_{k,\sigma}^{(j)}\,.
\end{equation}
Combining the above for all the degenerate $\psi_{k,\sigma}^{(i)}$'s results in 
\begin{equation} \label{eq:SI_adjoint}
    \Hopsig\bPksig(\br)-\bPksig\bEksig=4w_\sigma(\br)\left(\rhodsig(\br)-\rhosig(\br)\right)\bPsiksig(\br)f(\bEksig)-\bPsiksig(\br)\left(\bDksig+\bDksig\trans\right)\,,
\end{equation}
same as Eq. 7 of the main manuscript.

\subsection{Derivation of Eq. 8 in the main manuscript}
Let $\Eksigij$ be the $(i,j)-$th entry of $\bEksig$. Taking the partial derivative of $\lag$ with $\Eksigij$, we have
\begin{equation} \label{eq:SI_varEij}
\begin{split}
    \frac{\partial \lag}{\partial \Eksigij} =& -2\int w_\sigma(\br)\left(\rhodsig(\br)-\rhosig(\br)\right)\tr\left(\frac{\partial f(\bEksig)}{\partial \Eksigij}\bPsiksig\trans(\br)\bPsiksig(\br)\right)\dr  \\
    & + \eta_\sigma \tr\left(\frac{\partial f(\bEksig)}{\partial \Eksigij}\right) - \int \psi_{k,\sigma}^{(i)}(\br) p_{k,\sigma}^{(j)}(\br)\dr\,.
    \end{split}
\end{equation}
We now use Eq.~\ref{eq:SI_dfdE} to simplify the above equation. Substituting $\bE=\bEksig$ in Eq.~\ref{eq:SI_dfdE},
\begin{equation}
\begin{split}
    \frac{\partial f(\bEksig)}{\partial \Eksigij} &= \frac{1}{k_BT}\left(\bI_{m_{k,\sigma}}+e^{\frac{-\left(\bEksig-\mu_\sigma\bI_{m_{k,\sigma}}\right)}{k_BT}}\right)^{-1} e^{\frac{-\left(\bEksig-\mu_\sigma\bI_{m_{k,\sigma}}\right)}{k_BT}}\left(\bLambdaij + \frac{[\bEksig,\bLambdaij]}{2k_BT}\right) \\
    & \left(\bI_{m_{k,\sigma}}+e^{\frac{-\left(\bEksig-\mu_\sigma\bI_{m_{k,\sigma}}\right)}{k_BT}}\right)^{-1}\\
    &=\frac{\partial f_{k,\sigma}^{\mu_\sigma}}{\partial \epsilonksig}\bLambdaij\,,
\end{split}
\end{equation}
where the last line uses the definition of $\bEksig=\epsilonksig\bI_{m_k,\sigma}$ and $f_{k,\sigma}^{\mu_\sigma} = \left(1+e^{-(\epsilonksig-\mu_\sigma)/k_BT}\right)^{-1}$ along with the fact that $[\bEksig,\bLambdaij]=\epsilonksig[\bI_{m_k,\sigma},\bLambdaij]=0$. Using the above relation, we have
\begin{equation}
\begin{split}
    \tr\left(\frac{\partial f(\bEksig)}{\partial \Eksigij}\bPsiksig\trans(\br)\bPsiksig(\br)\right) &=  \frac{\partial f_{k,\sigma}^{\mu_\sigma}}{\partial \epsilonksig}\sum_{\alpha=1}^{m_{k,\sigma}}\sum_{\beta=1}^{m_{k,\sigma}} \Lambdaijab\psi_{k,\sigma}^{(\beta)}(\br)\psi_{k,\sigma}^{(\alpha)}(\br)\,.
\end{split}    
\end{equation}
where $\Lambdaijab$ is the $(\alpha,\beta)-$the entry of $\bLambdaij$. Since $\Lambdaijab=\delta_{i\alpha}\delta_{j\beta}$, the above equation simplifies to
\begin{equation} \label{eq:SI_dTrRho_dE}
    \begin{split}
        \tr\left(\frac{\partial f(\bEksig)}{\partial \Eksigij}\bPsiksig\trans(\br)\bPsiksig(\br)\right) &= \frac{\partial f_{k,\sigma}^{\mu_\sigma}}{\partial \epsilonksig}\psi_{k,\sigma}^{(i)}(\br)\psi_{k,\sigma}^{(j)}(\br)\,.
    \end{split}
\end{equation}
Similarly, 
\begin{equation}\label{eq:SI_dTrE_dE}
    \tr\left(\frac{\partial f(\bEksig)}{\partial \Eksigij}\right)=\frac{\partial f_{k,\sigma}^{\mu_\sigma}}{\partial \epsilonksig}\sum_{\alpha=1}^{m_{k,\sigma}}\Lambdaijaa=\frac{\partial f_{k,\sigma}^{\mu_\sigma}}{\partial \epsilonksig}\delta_{ij}\,.
\end{equation}
Finally, using Eq.~\ref{eq:SI_dTrRho_dE} and ~\ref{eq:SI_dTrE_dE} in Eq.~\ref{eq:SI_varEij} as well as setting $\frac{\partial \lag}{\partial \Eksigij}$ to zero, we get
\begin{equation}
    \int\psi_{k,\sigma}^{(i)}(\br)p_{k,\sigma}^{(j)}(\br)\dr = \frac{\partial f_{k,\sigma}^{\mu_\sigma}}{\partial \epsilonksig}\left[-2\int w_\sigma(\br)\left(\rhodsig(\br)-\rhosig(\br)\right)\psi_{k,\sigma}^{(i)}(\br)\psi_{k,\sigma}^{(j)}(\br)\dr+\eta_\sigma\delta_{ij}\right]\,.
\end{equation}
Thus, the above relation in matrix form can be written as
\begin{equation} \label{eq:SI_adjointOverlap}
    \int \bPsiksig\trans(\br)\bPksig(\br)\dr= \frac{\partial f_{k,\sigma}^{\mu_\sigma}}{\partial \epsilonksig}\left[-2\int w_\sigma(\br) \left(\rhodsig(\br)-\rhosig(\br)\right)\bPsiksig\trans(\br)\bPsiksig(\br)\dr+\eta_\sigma\bI_{m_{k,\sigma}}\right]\,.
\end{equation}
This concludes the derivation of Eq. 8 in the main manuscript. 

\subsection{Derivation of Eq. 9 in the main manuscript}
The partial derivative of $\lag$ with respect to $\mu_\sigma$ is given by
\begin{equation}
    \frac{\partial \lag}{\partial \mu_\sigma} = -2\int w_\sigma(\br) \left(\rhodsig(\br)-\rhosig(\br)\right)\sum_{k=1}^{M_\sigma}\tr\left(\frac{\partial f(\bEksig)}{\partial \mu_\sigma} \bPsiksig\trans(\br)\bPsiksig(\br)\right)\dr + \eta_\sigma \sum_{k=1}^{M_\sigma}\tr\left(\frac{\partial f(\bEksig)}{\partial \mu_\sigma}\right)\,.
\end{equation}
Using $f(\bEksig)=f_{k,\sigma}^{\mu_\sigma}\bI_{m_{k,\sigma}}$ in the above and setting $\frac{\partial \lag}{\partial \mu_\sigma}$ to zero, leads to 
\begin{equation} \label{eq:SI_eta}
    \eta_\sigma \sum_{k=1}^{M_\sigma}m_{k,\sigma}\frac{\partial f_{k,\sigma}^{\mu_\sigma}}{\partial \mu_\sigma} =2\sum_{k=1}^{M_\sigma}\frac{\partial f_{k,\sigma}^{\mu_\sigma}}{\partial \mu_\sigma} \int w_\sigma(\br)\left(\rhodsig(\br)-\rhosig(\br)\right)\tr\left(\bPsiksig\trans(\br)\bPsiksig(\br)\right)\dr\,,
\end{equation}
same as Eq. 9 in the main manuscript.



\section{Uniqueness of ${\delta \lag}/{\delta \vxcsig(\br)}$}
To show the uniqueness of ${\delta \lag}/{\delta \vxcsig(\br)}$ for a given $\vxcsig(\br)$,  we begin with adjoint equation (Eq. 7 in the main manuscript), 
\begin{equation}\label{eq:SI_adjoint_repeat}
    \Hopsig(\br)\bPksig-\bPksig(\br)\bEksig = 4w_\sigma(\br)\left(\rhodsig(\br)-\rhosig(\br)\right)\bPsiksig(\br)f(\bEksig) 
    - \bPsiksig(\br)\left(\bDksig+\bDksig\trans\right)\,.
\end{equation}
Left multiplying the above equation with $\bPsiksig\trans(\br)$ and integrating over the domain, yields 
\begin{equation}\label{eq:SI_D}
    \bDksig+\bDksig\trans=4\left[\int w_\sigma(\br)\bPsiksig\trans(\br)\left(\rhodsig(\br)-\rhosig(\br)\right)\bPsiksig\dr\right]f(\bEksig)\,,
\end{equation}
where we have used the fact that $\bPsiksig$ are eigenfunctions of $\Hopsig$ (i.e., $\Hopsig\bPsiksig=\bPsiksig\bEksig$). In the case of a degenerate eigenvalue, $\bPsiksig$ cannot be determined uniquely. That is, given an orthogonal matrix $\bQksig$ (i.e,. $\bQksig\trans\bQksig=\bQksig\bQksig\trans=\bI_{m_{k,\sigma}}$), $\bPsitildeksig=\bPsiksig\bQksig$ will satisfy the KS eigenvalue problem and the orthonormality condition. Further, $\bPsitildeksig$ will also preserve the density $\rhosig$ (see Eq. 5 of the main manuscript). Denoting the corresponding $\bPksig(\br)$ and $\bDksig$ for $\bPsitildeksig$ as $\bPtildeksig(\br)$ and $\bDtildeksig$, respectively, Eq.~\ref{eq:SI_adjoint} and Eq.~\ref{eq:SI_D} can be rewritten as
\begin{equation}\label{eq:SI_Adjointtilde}
\begin{split}
    \Hopsig(\br)\bPtildeksig(\br)-\bPtildeksig(\br)\bEksig = 4w_\sigma(\br)\left(\rhodsig(\br)-\rhosig(\br)\right)\bPsitildeksig(\br)f(\bEksig) -  \bPsitildeksig(\br)\left(\bDtildeksig+\bDtildeksig\trans\right)\, 
\end{split}
\end{equation}
\begin{equation}\label{eq:SI_Dtilde}
    \bDtildeksig+\bDtildeksig\trans=4\left[\int w_\sigma(\br)\bPsitildeksig\trans(\br)\left(\rhodsig(\br)-\rhosig(\br)\right)\bPsitildeksig\dr\right]f(\bEksig)\,.
\end{equation}
Substituting $\bPsitildeksig=\bPsiksig\bQksig$ in the above two equations leads to
\begin{equation}\label{eq:SI_Adjointtilde2}
\begin{split}
    \Hopsig(\br)\bPtildeksig-\bPtildeksig(\br)\bEksig =4 w_\sigma(\br)\left(\rhodsig(\br)-\rhosig(\br)\right)\bPsiksig(\br)\bQksig f(\bEksig) - \bPsiksig(\br)\bQksig\left(\bDtildeksig+\bDtildeksig\trans\right)\,, 
\end{split}
\end{equation}
\begin{equation}\label{eq:SI_Dtilde2}
    \bDtildeksig+\bDtildeksig\trans=4\left[\int\bQksig\trans\bPsiksig\trans(\br)w_\sigma(\br)\left(\rhodsig(\br)-\rhosig(\br)\right)\bPsiksig \bQksig \dr\right]f(\bEksig)\,.
\end{equation}
Multiplying the above equation with $\bQksig$ from the left yields,
\begin{equation}
\begin{split}
    \bQksig\left(\bDtildeksig+\bDtildeksig\trans\right)&=4\left[\int\bQksig\bQksig\trans\bPsiksig\trans(\br)w_\sigma(\br)\left(\rhodsig(\br)-\rhosig(\br)\right)\bPsiksig \bQksig \dr\right]f(\bEksig)\\
    &=\left(\bDksig+\bDksig\trans\right)\bQksig\,.  
\end{split}
\end{equation}
Now, using the above relation in Eq.~\ref{eq:SI_Adjointtilde2} results in
\begin{equation}\label{eq:SI_adjoint2}
     \Hopsig(\br)\bPtildeksig(\br)-\bPtildeksig(\br)\bEksig=4w_\sigma(\br)\left(\rhodsig(\br)-\rhosig(\br)\right)\bPsiksig(\br)\bQksig f(\bEksig) - \bPsiksig(\br)\left(\bDksig+\bDksig\trans\right)\bQksig\,.
\end{equation}
Comparing the above equation with Eq.~\ref{eq:SI_adjoint}, it is straightforward to note that $\bPtildeksig=\bPksig\bQksig$. That, is for an orthogonal transformation of $\bPsiksig$, its corresponding adjoint function is also transformed similarly. Finally, rewriting $\frac{\delta \lag}{\delta \vxcsig(\br)} = \sum_{k=1}^{M_{\sigma}}\tr\left(\bPksig\trans(\br) \bPsiksig(\br)\right)$ (see Eq. 10 in main manuscript) in terms of $\bPsitildeksig$ and $\bPtildeksig$, we have
\begin{equation}
        \frac{\delta \lag}{\delta \vxcsig(\br)} = \sum_{k=1}^{M_{\sigma}}\tr\left(\bPtildeksig\trans(\br) \bPsitildeksig(\br)\right) =  \sum_{k=1}^{M_{\sigma}}\tr\left(\bQksig\trans\bPksig\trans(\br) \bPsiksig(\br)\bQksig\right)\,.
\end{equation}
Using the fact that the trace of products of matrices is invariant with respect to cyclic permutation (i.e., $\tr(\bA\bB\bC)=\tr(\bB\bC\bA)$), the above equation simplifies to
   \begin{equation}\label{eq:SI_vxc2}
        \frac{\delta \lag}{\delta \vxcsig(\br)} = \sum_{k=1}^{M_{\sigma}}\tr\left(\bPksig\trans(\br) \bPsiksig(\br)\bQksig\bQksig\trans\right) = \sum_{k=1}^{M_{\sigma}}\tr\left(\bPksig\trans(\br) \bPsiksig(\br)\right)\,.
\end{equation}
This shows the uniqueness of $\delta \lag/\delta \vxcsig(\br)$ for a given $\vxcsig(\br)$.

\section{Solution Procedure}
Below we provide the overall solution procedure to solve the inverse DFT problem for a given spin densities ($\rhodsig(\br)$).
\begin{enumerate}
    \item For the current iterate of $\vxcsig(\br)$, solve the KS eigenvalue problem (Eq. 2 in the main manuscript) to find $\bPsiksig(\br)$ and $\bEksig$. Subsequently, evaluate $\rhosig(\br)$ (Eq. 4 in the main manuscript) and $\mu_\sigma$ (Eq. 5 in the main manuscript).
    \item Using $\bPsiksig$, $\bEksig$ and $\mu_\sigma$, solve for $\eta_\sigma$ (Eq. 9 in the main manuscript)
    \item Using $\bPsiksig$, $\bEksig$, $\mu_\sigma$, and $\eta_\sigma$, solve for the overlap between the KS orbitals and their adjoint functions (Eq. 8 in the main manuscript)
    \item Using Eq.~\ref{eq:SI_D}, evaluate $\bDksig+\bDksig\trans$ and substitute it in the adjoint equation (Eq. 7 in the main manuscript).
    \item Solve the adjoint equation (Eq. 7 in the main manuscript) to find the adjoint functions ($\bPksig$). Note that the adjoint equation, by itself, does not provide a unique solution for $\bPksig$. That is, if $\bPksig$ is a solution to the adjoint equation, it can be trivially shown that for any $m_{k,\sigma}\times m_{k,\sigma}$ matrix $\bA$, $\bPksig+\bPsiksig\bA$ is also a solution. This is owing to the fact that $\bPsiksig$ are the eigenfunctions of $\Hopsig$. Nevertheless, we have an additional condition (Eq. 8 in the main manuscript) that provides the overlap between the KS orbitals and their adjoint functions, and hence, uniquely determines the $\bPksig$. To do so, we first solve the adjoint equation in a space orthogonal to $\bPsiksig$ to find $\bPksig^{\perp}$ that is orthogonal to $\bPsiksig$. Subsequently, we find $\bPksig=\bPksig^{\perp}+\bPsiksig\bB$, where $\bB=\frac{\partial f_{k,\sigma}^{\mu_\sigma}}{\partial\epsilonksig}\left[-2w_\sigma(\br)\int\left(\rhodsig(\br)-\rhosig(\br)\right)\bPsiksig\trans(\br)\bPsiksig(\br)\dr+\eta_\sigma\bI_{m_{k,\sigma}}\right]$ (i.e.,the right hand side of Eq. 8 in the main manuscript). 
    \item Update $\vxcsig(\br)$ using Eq. 10 in the main manuscript 
    \item Go to step 1 and repeat until convergence in $\rhosig$ (i.e., $||\rhodsig-\rhosig||$ is below a tolerance).
\end{enumerate}

\section{Non-interacting ensemble-v-representable density}
The KS density matrix ($\hat{\mathcal{D}}_\sigma$), in general, can be non-interacting ensemble-v-representable (e-$v_s$) as it can be expressed as an ensemble of $L$ degenerate KS Slater determinants,
\begin{equation} \label{eq:SI_DM_ensemble}
  \hat{\mathcal{D}}_\sigma=\sum_{j=1}^{L} d_j \ket{\bPhi_{j,\sigma}}\bra{\bPhi_{j,\sigma}}\,,\quad d_j \geq 0\,,\quad \sum_{j=1}^L d_j = 1\,.  
\end{equation}
In the above equation, $\left\{\ket{\bPhi_{j,\sigma}}\right\}$ denote the $L$ degenerate KS determinants. The corresponding density is given by,
\begin{equation}
    \rhosig(\br) = \tr\left(\hat{\mathcal{D}}_\sigma\hat{\rho}_\sigma(\br)\right) = \sum_{j=1}^L d_j \rho_{j,\sigma}(\br)\,,
\end{equation}
where $\hat{\rho}_\sigma$ is the density operator and $\rho_{j,\sigma}(\br)$ is the density corresponding to $\bPhi_{j,\sigma}$. The non-interacting pure-v-representable (pure-$v_s$) density is a special case with $L=1$. Typically, for an e-$v_s$ density, the different KS Slater determinants ($\bPhi_{j,\sigma}$) differ only in their highest occupied molecular orbital (HOMO), all of which are degenerate with their KS eigenvalues equal to the Fermi level (chemical potential). In other words, a typical e-$v_s$ density can be written as
\begin{equation}
    \rhosig(\br) = \sum_{i:\epsilon_{i,\sigma}<\mu} |\psi_{i,\sigma}(\br)|^2 + \sum_{i:\epsilon_{i,\sigma}=\mu} f |\psi_{i,\sigma}(\br)|^2\,,
\end{equation}
where $f$ is the fractional occupancy of the HOMO level (typically equal to $1/L$).

\begin{figure}[htbp!]
    \centering
    \includegraphics[scale=1]{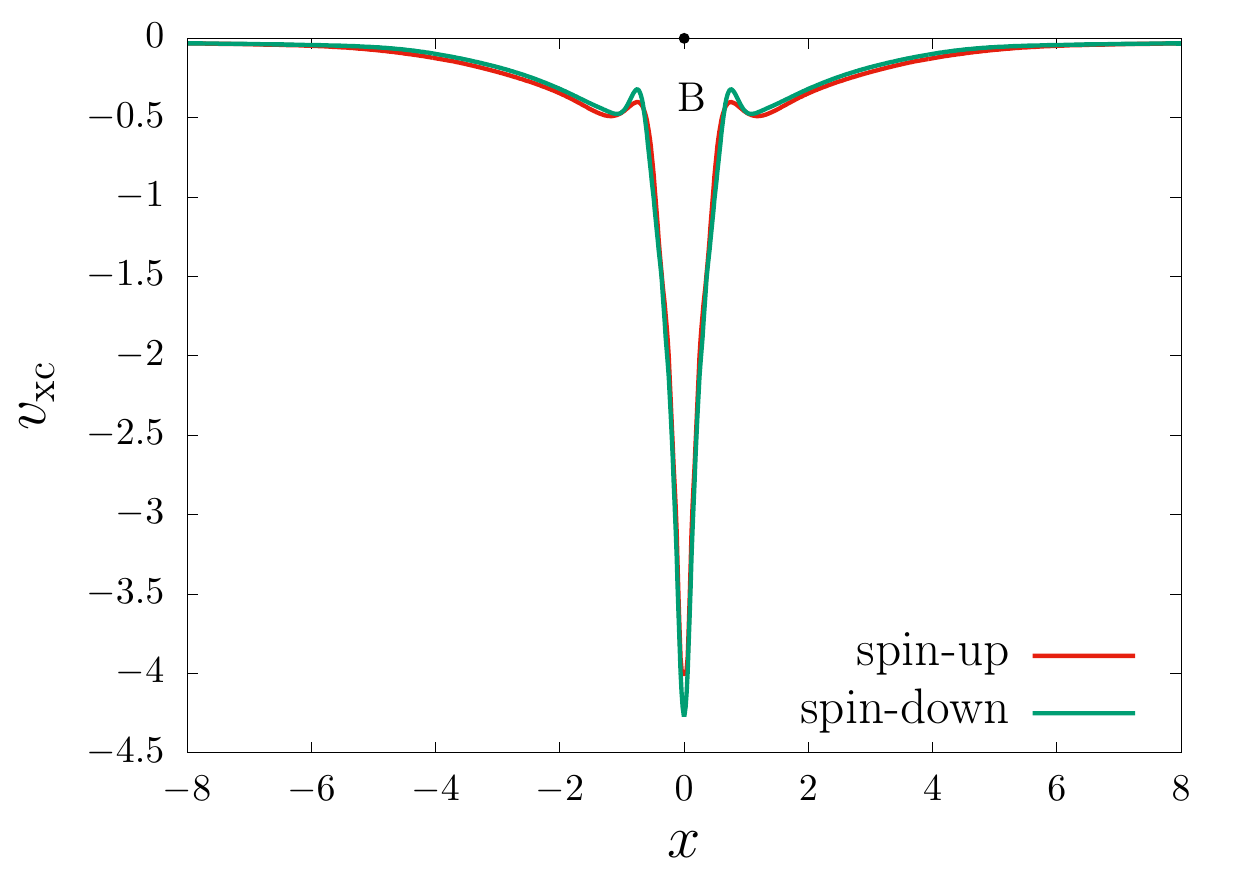}
    \caption{SCAN0 based XC potentials for B. The majority-spin density is an e-$v_s$ density.}
    \label{fig:SI_B_SCAN0}
\end{figure}

For a given density, it is \textit{apriori} difficult to ascertain if it is a pure-$v_s$ or an e-$v_S$ density. However, any robust approach to the inverse DFT problem should be flexible enough to admit both kinds of densities. All the densities for the benchmark systems considered (Li, C, N, O, CN, and CH$_2$) turn out to be pure-$v_s$. In this example, we demonstrate the efficacy of the proposed inverse DFT approach for e-$v_s$ density by using the SCAN0 based density for boron (B), obtained using a finite temperature Fermi-Dirac smearing in the groundstate calculation. Upon inversion, we obtain an ensemble of three KS Slater determinants for the majority-spin. To elaborate, we obtain three degenerate KS orbitals at the Fermi level ($\mu$). The density for the minority-spin turns out to be pure-$v_s$. Fig.~\ref{fig:SI_B_SCAN0} presents the SCAN0 based XC potentials for B.

\section{Accuracy tests}
In this section, we provide various tests to measure the accuracy of the proposed approach to inverse DFT as well as quantify the uncertainties with respect to the choice of basis set used, both for generating the target densities as well as for performing the inverse DFT calculations.

\subsection{Verification with LDA densities}
\begin{figure}[htbp!]
  \centering
  \subfigure[]{\includegraphics[scale=0.65]{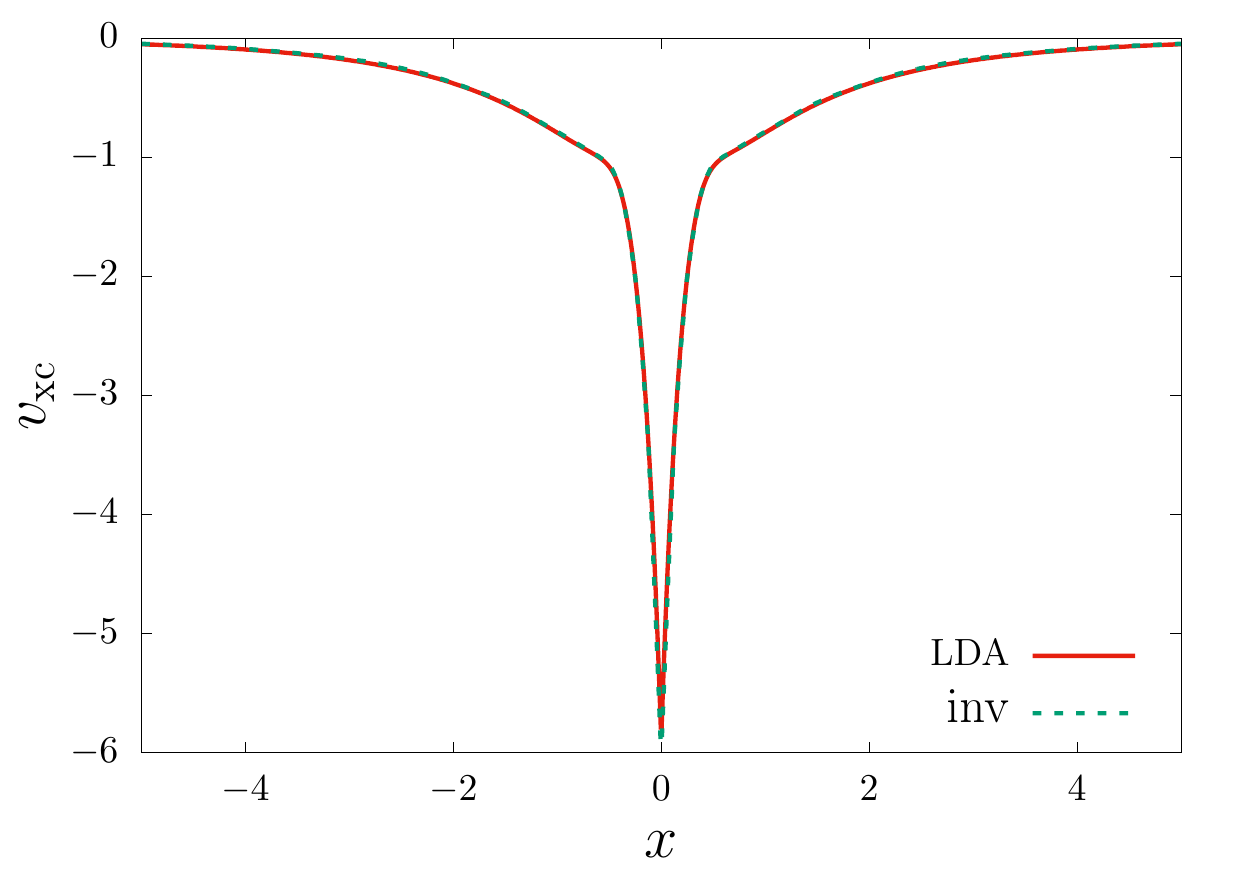}}
  \subfigure[]{\includegraphics[scale=0.65]{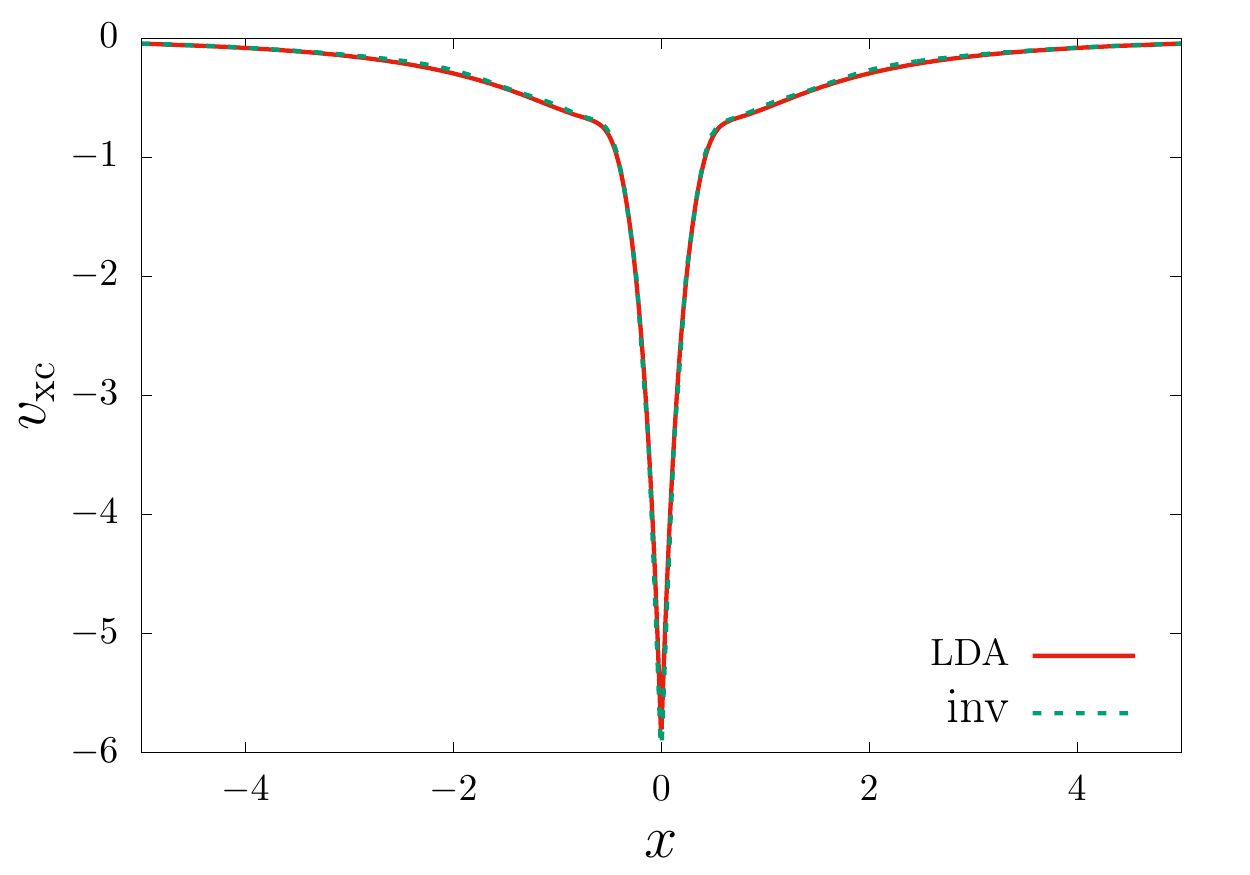}}
   \caption{LDA density based verification study on N: (a) majority-spin, and (b) minority-spin. The solid line corresponds to the LDA XC potential directly evaluated using $\rhosig^{\text{LDA}}$. The dashed line corresponds to the XC potential obtained from the inverse DFT calculation using $\rhosig^{\text{LDA}}$ as the target density.}
   \label{fig:SI_N_LDA_test}
\end{figure}

We, first, assess the accuracy and robustness of our inverse DFT method using LDA~\cite{Perdew1992} spin-densities, $\rhosig^{\text{LDA}}$, as our target densities. We use nitrogen as our benchmark system. Given that the corresponding XC potential, $\vxcsig^{\text{LDA}}[\rhosig^{\text{LDA}}](\br)$ is exactly known, this test allows for a direct assessment of the quality of the XC potential obtained from an inverse DFT calculation. As evident from Fig.~\ref{fig:SI_N_LDA_test}, the XC potential obtained from the inverse DFT calculation is virtually indistinguishable from $\vxcsig^{\text{LDA}}[\rhosig^{\text{LDA}}](\br)$. The $L_2$ norm in the density, $||\rhosig^{\text{LDA}} -\rhosig||$, is driven below $10^{-4}$.  Additionally, the Kohn–Sham eigenvalues computed using the XC potential from inverse DFT agrees to the exact LDA Kohn-Sham eigenvalues to within 1 mHa. 

\subsection{Sensitivity to Gaussian and finite element basis}
\begin{figure}[htbp!]
  \centering
  \subfigure[]{\includegraphics[scale=0.65]{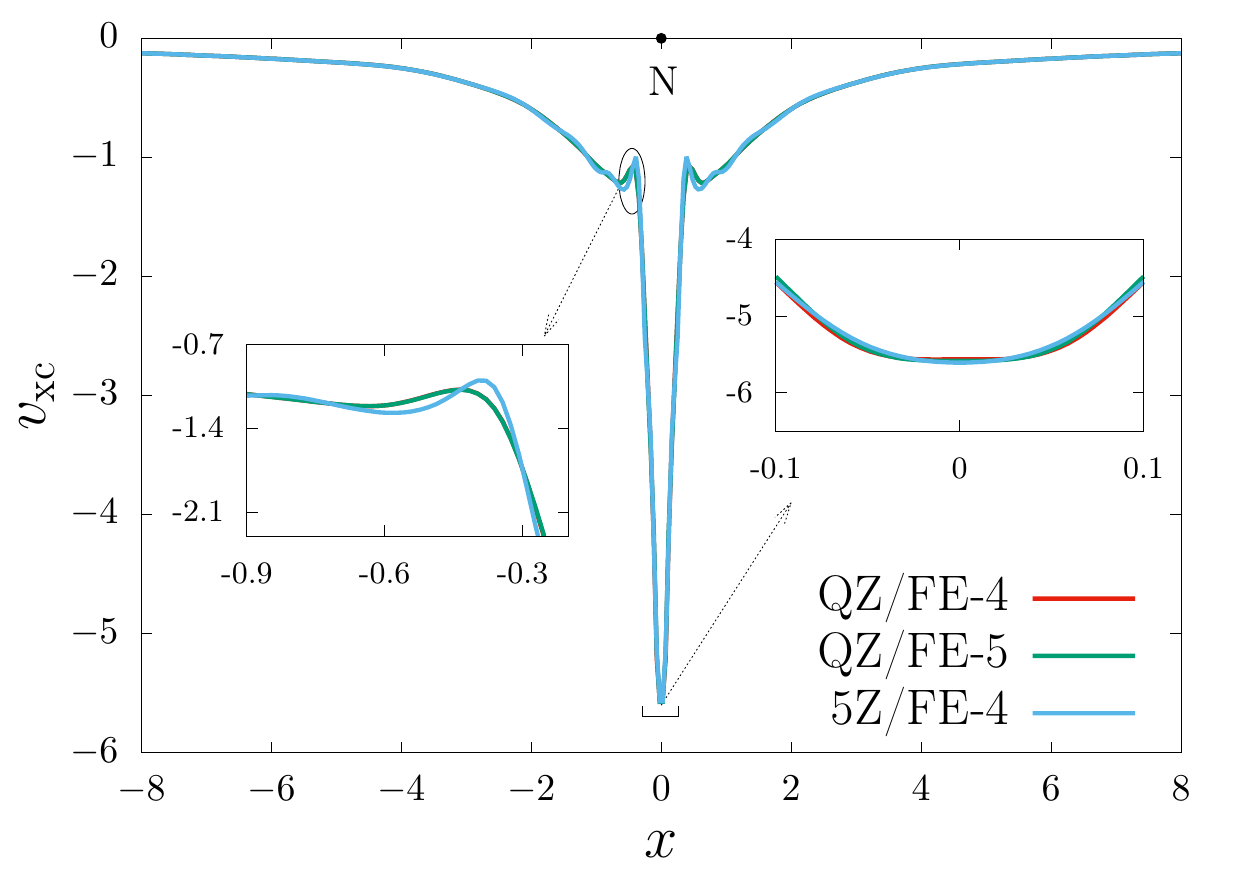}}
  \subfigure[]{\includegraphics[scale=0.65]{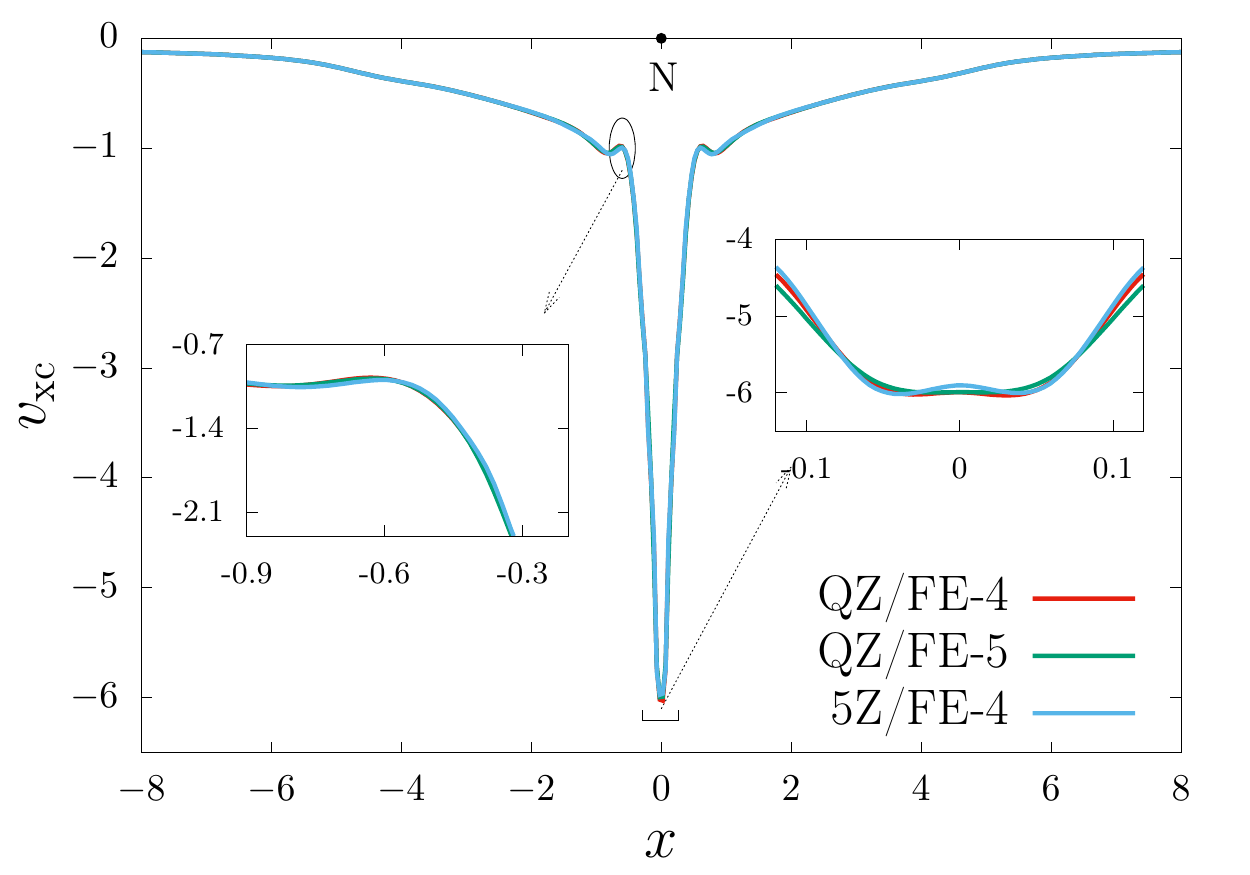}}
   \caption{Sensitivity of the exact XC potential for N with respect to Gaussian and finite element basis: (a) majority-spin, and (b) minority-spin.}
   \label{fig:SI_N_basis_test}
\end{figure}

In this exercise, we measure the sensitivity of the XC potentials obtained in this work with respect to the Gaussian basis used to generate the target densities as well as the finite element basis used to discretize the inverse DFT problem. We, once again, use the nitrogen atom as our benchmark system. We use two Gaussian basis---cc-pCVQZ and cc-pCV5Z---to generate the CI densities. Further, we use two different finite element discretization---one with fourth-order and the other with fifth-order finite elements (i.e., fourth- and fifth-order Lagrange polynomial in each finite element). In the following discussion the notation QZ/FE-4 refers to an inverse calculation where the CI density is obtained using cc-pCVQZ Gaussian basis and the inverse problem is discretized using fourth-order finite element basis. We adopt similar definitions for QZ/FE-5 and 5Z/FE-4. Fig.~\ref{fig:SI_N_basis_test} compares the exact XC potential for N, obtained using three different combination of Gaussian and finite element basis. As evident, except minor differences near the nuclei and the intershell structure, the potentials from different combinations of Gaussian and finite element basis are virtually identical.  Table~\ref{tab:SI_N_basis_test} compares the correlation part of the kinetic energy ($T_\text{c}$) and the mean-absolute error (MAE) in the Kohn-Sham eigenvalues from the three different combinations of Gaussian and finite element basis. $T_\text{c}=T-T_s$ is the difference between the kinetic energy ($T$) and the Kohn-Sham non-interacting kinetic energy ($T_S$).  We use the QZ/FE-4 XC potentials as the reference potentials to define the above error metrics. As evident, the uncertainty in both $T_\text{c}$ and the Kohn-Sham eigenvalues are $\sim$2 mHa. This exercise ascertains that the sensitivity of the XC potentials, beyond the QZ/FE-4, is negligible. Thus, in all the other calculations reported in this work we use the QZ/FE-4 combination.

\begin{table*}
\centering
\caption{\small Sensitivity of $T_\text{c}$ and Kohn-Sham eigenvalues from exact XC potential for N with respect to Gaussian and finite element basis. $T_\text{c}=T-T_s$ denotes the correlation part of the kinetic energy. MAE$-\epsilon_\sigma$ denotes the mean absolute error in the Kohn-Sham eigenvalues, using the QZ/FE-4 values as the reference.}
\begin{tabular}{|p{2.6cm} | M{2cm} | M{2cm} | M{2cm}|} 
\hline
& QZ/FE-4 & QZ/FE-5 & 5Z/FE-4 \\ \hline
$T_\text{c}$ (Ha) & 0.156 & 0.157 & 0.155 \\ \hline
MAE-$\epsilon_{\sigma_1}$ (Ha) & 0 & 0.0001 & 0.0003 \\  \hline
MAE-$\epsilon_{\sigma_2}$ (Ha) & 0 & 0.0026 & 0.0018 \\ \hline
\end{tabular}
\label{tab:SI_N_basis_test}
\end{table*}


\subsection{Agreement with Koopmans' theorem}
The Koopmans' theorem in DFT~\cite{Perdew1997} mandates that for the exact XC potential the eigenvalue for the Kohn-Sham highest occupied molecular orbital (HOMO), $\epsilon_{\text{H}}$, should be equal to the negative of the ionization potential ($I_{\text{P}}$). Thus, the agreement between $\epsilon_{\text{H}}$ and $-I_{\text{P}}$ is a important indicator of the accuracy of the XC potential. We compare the two for all the benchmark systems in Table~\ref{tab:SI_Koopmans}. As evident, we attain good agreement between $\epsilon_{\text{H}}$ and $-I_{\text{P}}$, with the largest deviation being 12 mHa. We remark that this is a stringent test and is crucially dependent on the accuracy of the target density in the low density region, which may not be adequately captured through the Gaussian basis. We expect even better agreement with more accurate densities and better initial guess/boundary conditions to the XC potential in the far-field (e.g., Slater XC-hole potential). 

\begin{table*}
\centering
\caption{\small Comparison of the Kohn-Sham HOMO level ($\epsilon_{\text{H}}$) and the negative of the ionization potential ($I_{\text{P}}$). All values in Ha.}
\begin{tabular}{|p{2.6cm} | M{2cm} | M{2cm} | M{2cm}| M{2cm} | M{2cm} | M{2cm}|} 
\hline
& Li & C & N & O & CN & CH2 \\ \hline
$\epsilon_{\text{H}}$ & -0.197 & -0.406 & -0.521 & -0.494 & -0.500 & -0.388\\ \hline
$-I_{\text{P}}$ & -0.198 & -0.413 & -0.533 & -0.496 & -0.509 & -0.386 \\ \hline 
\end{tabular}
\label{tab:SI_Koopmans}
\end{table*}

\subsection{Virial of the potentials}
The virial ($t_{\text{xc}}$) of the XC potential , $\vxcsig$, corresponding to the density $\rhodsig$ is defined as,
\begin{equation} \label{eq:SI_txc}
    t_{\text{xc}}= -\sum_{\sigma} \int \rhodsig(\br) \br \cdot \nabla \vxcsig(\br) d\br\,.
\end{equation}
It is known from virial relations~\cite{Levy1985} that for the exact XC potential,
\begin{equation} \label{eq:SI_virial}
    E_{\text{xc}} + T_\text{c} = t_{\text{xc}}^{\text{exact}}\,,
\end{equation}
where the XC energy $E_{\text{xc}}= E - T_s - E_\text{H}[\rhod] -\sum_\sigma\int \rhodsig(\br)\vN(\br)\dr$, with $E_\text{H}[\rhod]$ being the Hartree energy corresponding to $\rhod(\br)=\sum_\sigma\rhodsig(\br)$. Thus, an agreement between the left and the right side of the above equation forms an important indicator of the accuracy of the XC potential. 
For atoms, since total energy is same as the negative of the kinetic energy (i.e., $E=-T$), $E_{\text{xc}} + T_\text{c} =-2T_s - E_\text{H}[\rhod] -\sum_\sigma \int\rhodsig(\br)\vN(\br)\dr$. Table~\ref{tab:SI_virial_atoms} provides a comparison of $E_{\text{xc}} + T_\text{c}$ and $t_{\text{xc}}^{\text{exact}}$ for all the atoms considered in this work. As, evident we obtain good agreement between the two, with the largest deviation being $\sim$ 4 mHa. 

We note that for molecules the above comparison requires converged values not just for the total energy ($E$) but also for the kinetic energy ($T$) from CI. Given that in CI, the total energy is minimized, it attains much faster convergence with respect to the basis set than the individual components in it (e.g., kinetic energy). For instance, in our numerical studies using cc-pCVQZ and cc-pCV5Z basis, while the total energy for molecules converged to within a few mHa/atom, the difference in the kinetic energy is $\sim$50 mHa. Thus, a meaningful comparison of the above relation for molecules cannot be performed at this stage, for want of accurate kinetic energies. Nevertheless, the good agreement of the virial relation for atoms underlines the accuracy of the proposed inverse DFT method.      

\begin{table*}
\centering
\caption{\small Comparison of $E_{\text{xc}} + T_\text{c}$ and the virial of the XC potental, $t_{\text{xc}}^{\text{exact}}$. All values in Ha.}
\begin{tabular}{|p{2.6cm} | M{2cm} | M{2cm} | M{2cm}| M{2cm}|} 
\hline
& Li & C & N & O\\ \hline
$t_{\text{xc}}^{\text{exact}}$ & -1.7816 & -5.0856 & -6.6132 & -8.2225\\ \hline
$E_{\text{xc}} + T_\text{c}$ & -1.7821 & -5.0886 & -6.6178 & -8.2251 \\ \hline 
\end{tabular}
\label{tab:SI_virial_atoms}
\end{table*}

\subsection{Inverse DFT with Hartree-Fock density}
In this exercise, we perform inversion on Hartree-Fock density (say $\rho_\sigma^{\text{HF}}$), using N as a benchmark system. Corresponding to the exact exchange potential ($\bar{v}_{\text{x},\sigma}[\rho_\sigma^\text{HF}]$), it is known that the exchange energy evaluated from the KS orbitals ($\{\bar{\psi}_{i,\sigma}\}$) should be equal to the virial, i.e.,
\begin{equation} \label{eq:SI_exchange_virial}
-\frac{1}{2}\sum_{\sigma} \sum_i \sum_j \int \int \frac{\bar{\psi}_{i,\sigma}^{\dagger}(\br)\bar{\psi}_{j,\sigma}^{\dagger}(\br') \bar{\psi}_{j,\sigma}(\br) \bar{\psi}_{i,\sigma}(\br')}{|\br-\br'|}\dr\dr' = -\sum_\sigma \int \rho_\sigma^{\text{HF}}(\br) \br \cdot \nabla \bar{v}_{\text{x},\sigma}[\rho_\sigma^\text{HF}](\br) \dr\,.
\end{equation}
The evaluation of the exact exchange potential ($\bar{v}_\text{x}[\rho_\sigma^\text{HF}]$) requires an optimized effective potential (OEP) procedure~\cite{Kummel2008}, which remains beyond the scope of this work. However, the XC potential ($\vxcsig[\rho_{\sigma}^{\text{HF}}]$) corresponding to $\rho_{\sigma}^{\text{HF}}$ is expected to be close to $\bar{v}_{\text{x},\sigma}[\rho_\sigma^\text{HF}]$. As a result, the above relation should hold for $\vxcsig[\rho_{\sigma}^{\text{HF}}]$ and its corresponding KS orbitals, and hence, is an instructive measure of the accuracy of any inverse DFT method. We note that the evaluation of the exchange energy in a finite-element basis is computationally expensive, owing to the large number of basis and quadrature points involved. While the evaluation can be accelerated through the use of reduced order basis (e.g., tensor decomposition of the KS orbitals), we currently lack such capabilities in our implementation of the proposed inverse DFT method. Nevertheless,  we can use accurate data from past efforts at the same exercise~\cite{Ryabinkin2013}. To be consistent with the density used in ~\cite{Ryabinkin2013}, we use the Hartree-Fock density for N, obtained using the UGBS basis~\cite{De1998}, as the target density. Using the virial of the resulting XC potential as the exchange energy, we obtain a total energy of -54.4033 Ha, which agrees to Ref.~\cite{Ryabinkin2013} (see Table I in the reference) to within 0.1 mHa.             

\subsection{Spherical harmonic expansion for atoms}
For atoms, it is known that the density as well as the corresponding potential should have a finite expansion in terms of the spherical harmonics~\cite{Fertig2000}. Particularly, the density should only have even harmonics. We note that evaluating the spherical harmonics expansion of the XC potential is, in general, not feasible within the finite element framework we developed for KS inversion. To elaborate, our finite element implementation is restricted to a finite parallelepiped domain. Thus, for a field that does not have a finite compact support (e.g., XC potential), evaluation of spherical harmonics coefficients is not compatible with the geometry of the parallelepiped domain. However, given that the electron density has a compact support, we can are able to evaluate its expansion coefficients. For all the atoms considered in the work, as expected, we obtain a finite even harmonic expansion.

\section{Energies from exact potentials}
We report various energies related to the exact XC potentials in Table~\ref{tab:SI_exact_energy}.
For the atoms, since from virial theorem the kinetic energy ($T$) is negative of the total energy $E$, we evaluate the correlation part of the kinetic energy $T_\text{c} = T-T_s$ (which simplifies to $-E-T_s$ for atoms). Alternatively, for both atoms and molecules, we provide the $T_\text{c}$ evaluated from the virial of the potential: $T_\text{c}=t_{\text{xc}}^{\text{exact}}-E_{\text{xc}}$ (see Eq.~\ref{eq:SI_virial}).

\begin{table*}
\centering
\caption{\small Various energy components corresponding to the exact XC potentials. $E$ refers to the groundstate energy obtained from CI, $T_s$ is the non-interacting kinetic energy corresponding to the exact XC potentials, $E_\text{xc}$ is the exchange correlation energy, and $T_\text{c}$ is the correlation part of the kinetic energy. We provide two different values of $T_\text{c}$: $T_\text{c,1}=T-T_s$ and $T_{\text{c},2}=t_\text{xc}^{\text{exact}}-E_\text{xc}$ (see Eq.~\ref{eq:SI_virial}). All values in Ha.}
\begin{tabular}{|p{2cm} | M{2cm} | M{2cm} | M{2cm}| M{2cm} | M{2cm} | M{2cm}|} 
\hline
& Li & C & N & O & CN & CH2 \\ \hline
$E$	&	-7.476	&	-37.840	&	-54.582	&	-75.055	&	-92.692	&	-39.133	\\ \hline
$T_s$	&	7.434	&	37.711	&	54.425	&	74.845	&	92.346	&	38.898	\\ \hline
$E_\text{xc}$	&	-1.824	&	-5.218	&	-6.776	&	-8.435	&	-12.102	&	-6.049	\\ \hline
$T_\text{c,1}$ & 0.042 & 0.129 & 0.157 & 0.210 & - & - \\ \hline
$T_\text{c,2}$ & 0.042 & 0.132 & 0.161 & 0.213 & 0.336 & 0.175 \\ \hline

\end{tabular}
\label{tab:SI_exact_energy}
\end{table*}

\section{Errors in model densities and potentials}
We report the error in the densities obtained from self-consistently solved calculations with approximate XC functionals (denoted as $\rho_{\sigma,\text{model}}^{\text{data}}$), relative to the ground-state density from heat-bath configuration interaction (HBCI) calculations (denoted as $\rho_{\sigma,\text{exact}}^{\text{data}}$). We quantify the errors in the density using two metrics
\begin{equation}  \label{eq:SI_rhoErrors}
f_{1,\sigma} = \frac{\norm{\rho_{\sigma,\text{exact}}^{\text{data}}-\rho_{\sigma,\text{model}}^{\text{data}}}_{L_2}}{\norm{\rho_{\sigma,\text{exact}}^{\text{data}}}_{L_2}}\,,
\quad f_{2,\sigma} = \frac{\norm{\modulus{\nabla(\rho_{\sigma,\text{exact}}^{\text{data}}-\rho_{\sigma,\text{model}}^{\text{data}})}}_{L_2}}{\norm{\modulus{\nabla\rho_{\sigma,\text{exact}}^{\text{data}}}}_{L_2}}\,.
\end{equation}
We also report four additional error metrics for the model XC potentials, given by 
\begin{equation} \label{eq:SI_vxcErrors2}
e_{3,\sigma} = \frac{\norm{\delta\vxcsig}_{L_2}}{\norm{\vxcsig^{\text{exact}}}_{L_2}}\,,
\quad e_{4,\sigma} = \frac{\norm{\modulus{\nabla\delta\vxcsig}}_{L_2}}{\norm{\modulus{\nabla\vxcsig^{\text{exact}}}}_{L_2}}\,,
\end{equation}
\begin{equation} \label{eq:SI_vxcErrors3}
e_{5,\sigma} = \frac{\norm{\delta\vxcsig-\Delta\mu^{\text{model}}}_{L_2}}{\norm{\vxcsig^{\text{exact}}}_{L_2}}\,,
\quad e_{6,\sigma} = \frac{\norm{\rhodsig(\delta\vxcsig - \Delta\mu^{\text{model}})}_{L_2}}{\norm{\rhodsig\vxcsig^{\text{exact}}}_{L_2}}\,,
\end{equation}
where $\delta\vxcsig=\vxcsig^{\text{exact}}-\vxcsig^{\text{model}}$ and $\Delta\mu^{\text{model}}=\epsilon_{\text{H}}^{\text{exact}}- \epsilon_{\text{H}}^{\text{model}}$ is the difference between the Kohn-Sham HOMO level for the exact and model potentials. We note that while $e_{1,\sigma}$ and $e_{2,\sigma}$ (presented in the main manuscript) are $\rhodsig-$weighted error metrics, $e_{3,\sigma}$ and $e_{4,\sigma}$ are their unweighted counterparts, respectively. Further, we use the $e_{5,\sigma}$ and $e_{6,\sigma}$ metrics to eliminate any constant shift in the potentials by aligning their HOMO levels. 

Tables~\ref{tab:SI_densityErrUp} and ~\ref{tab:SI_densityErrDown} list the $f_1$ and $f_2$ values for the majority- and minority-spin, for all the benchmark systems considered in this study. Table~\ref{tab:SI_errWeightedDown} lists the $e_{1,\sigma}$ and $e_{2,\sigma}$ errors (defined in Eq.12 of the main manuscript) for the minority-spin. Tables~\ref{tab:SI_errUnweightedUp} and ~\ref{tab:SI_errUnweightedDown} list the $e_{3,\sigma}$ and $e_{4,\sigma}$ errors for the majority- and minority-spin, respectively. Tables~\ref{tab:SI_errShiftedUp} and ~\ref{tab:SI_errShiftedDown} list the $e_{5,\sigma}$ and $e_{6,\sigma}$ errors for the majority- and minority-spin, respectively.

Comparing Tables~\ref{tab:SI_densityErrUp} and ~\ref{tab:SI_densityErrDown} with Table I from the main manuscript (and Tables~\ref{tab:SI_errWeightedDown}, \ref{tab:SI_errUnweightedUp} and \ref{tab:SI_errUnweightedDown} here), it is evident that while the relative errors in the density are of $\mathcal{O}(10^{-3}-10^{-2})$, the relative errors in the XC potentials are two-orders higher (i.e., $\mathcal{O}(10^{-1}-10^0)$).  This establishes the XC potential to be a quantity of greater sensitivity than the density, and hence, can be instrumental in development of future XC functionals. 

\begin{table}
\caption{\small Comparing the exact and the model density in terms of $f_{1,\sigma}$ and $f_{2,\sigma}$  values (Eq.~\ref{eq:SI_rhoErrors}) for the majority-spin.} 
\begin{tabular}{|p{1.6cm} | M{0.8cm} | M{0.8cm} | M{0.8cm} | M{0.8cm} | M{0.8cm}| M{0.8cm} | M{0.8cm} | M{0.8cm} | M{0.8cm} | M{0.8cm}|M{0.8cm}| M{0.8cm}|}
\hline
\multirow{1}{1.6cm}{Model} & \multicolumn{2}{M{1.6cm}|}{Li} & \multicolumn{2}{M{1.6cm}|} {C} & \multicolumn{2}{M{1.6cm}|}{N} & \multicolumn{2}{M{1.6cm}|}{O} & \multicolumn{2}{M{1.6cm}|}{CN} &
\multicolumn{2}{M{1.6cm}|}{CH$_2$}\\
\cline{2-13}
& $f_1$ & $f_2 $ & $f_1$ & $f_2$ &  $f_1$ & $f_2$ &  $f_1$ & $f_2$ &  $f_1$ & $f_2$ & $f_1$ & $f_2$ \\
\hline 
\hline
B3LYP &  0.007  &  0.013 &  0.004  &  0.006 &  0.004  &  0.005 &  0.003  &  0.004 &  0.003  &  0.005 &  0.004  &  0.005 \\ \hline
SCAN0 &  0.003  &  0.006 &  0.003  &  0.003 &  0.002  &  0.002 &  0.002  &  0.002 &  0.005  &  0.003 &  0.003  &  0.003 \\ \hline
SCAN &  0.004  &  0.008 &  0.004  &  0.004 &  0.003  &  0.003 &  0.002  &  0.002 &  0.002  &  0.003 &  0.003  &  0.004 \\ \hline
PBE &  0.007  &  0.011 &  0.004  &  0.006 &  0.004  &  0.005 &  0.003  &  0.004 &  0.004  &  0.006 &  0.014  &  0.015 \\ \hline
PW92 &  0.021  &  0.022 &  0.012  &  0.013 &  0.012  &  0.013 &  0.01  &  0.012 &  0.012  &  0.013 &  0.014  &  0.015 \\ \hline
\end{tabular}
\label{tab:SI_densityErrUp}
\end{table}

\begin{table} 
\caption{\small Comparing the exact and the model density in terms of $f_{1,\sigma}$ and $f_{2,\sigma}$ values (Eq.~\ref{eq:SI_rhoErrors}) for the minority-spin.} 
\begin{tabular}{|p{1.6cm} | M{0.8cm} | M{0.8cm} | M{0.8cm} | M{0.8cm} | M{0.8cm}| M{0.8cm} | M{0.8cm} | M{0.8cm} | M{0.8cm} | M{0.8cm}|M{0.8cm}| M{0.8cm}|}
\hline
\multirow{1}{1.6cm}{Model} & \multicolumn{2}{M{1.6cm}|}{Li} & \multicolumn{2}{M{1.6cm}|} {C} & \multicolumn{2}{M{1.6cm}|}{N} & \multicolumn{2}{M{1.6cm}|}{O} & \multicolumn{2}{M{1.6cm}|}{CN} &
\multicolumn{2}{M{1.6cm}|}{CH$_2$}\\
\cline{2-13}
& $f_1$ & $f_2 $ & $f_1$ & $f_2$ &  $f_1$ & $f_2$ &  $f_1$ & $f_2$ &  $f_1$ & $f_2$ & $f_1$ & $f_2$ \\
\hline 
\hline
B3LYP &  0.007  &  0.011 &  0.003  &  0.006 &  0.003  &  0.005 &  0.003  &  0.004 &  0.003  &  0.005 &  0.003  &  0.005 \\ \hline
SCAN0 &  0.003  &  0.005 &  0.002  &  0.003 &  0.002  &  0.003 &  0.002  &  0.002 &  0.006  &  0.003 &  0.002  &  0.002 \\ \hline
SCAN &  0.004  &  0.008 &  0.003  &  0.004 &  0.002  &  0.003 &  0.002  &  0.002 &  0.003  &  0.003 &  0.003  &  0.003 \\ \hline
PBE &  0.007  &  0.012 &  0.004  &  0.007 &  0.003  &  0.006 &  0.003  &  0.005 &  0.004  &  0.006 &  0.012  &  0.013 \\ \hline
PW92 &  0.021  &  0.022 &  0.011  &  0.012 &  0.01  &  0.011 &  0.009  &  0.01 &  0.012  &  0.013 &  0.012  &  0.013 \\ \hline
\end{tabular}
\label{tab:SI_densityErrDown}
\end{table}

\begin{table*}
\caption{\small Comparison of the model XC potentials in terms of the error metrics $e_{1,\sigma}$ and $e_{2,\sigma}$ (defined in Eq.12 of the main manuscript), for the minority-spin. See the main manuscript for the error metrics for majority-spin.}
\begin{tabular}{|p{1.6cm} | M{0.8cm} | M{0.8cm} | M{0.8cm} | M{0.8cm} | M{0.8cm}| M{0.8cm} | M{0.8cm} | M{0.8cm} | M{0.8cm} | M{0.8cm}|M{0.8cm}| M{0.8cm}|}
\hline
\multirow{1}{1.6cm}{Model} & \multicolumn{2}{M{1.6cm}|}{Li} & \multicolumn{2}{M{1.6cm}|} {C} & \multicolumn{2}{M{1.6cm}|}{N} & \multicolumn{2}{M{1.6cm}|}{O} & \multicolumn{2}{M{1.6cm}|}{CN} &
\multicolumn{2}{M{1.6cm}|}{CH$_2$}\\
\cline{2-13}
& $e_{1,\sigma}$ & $e_{2,\sigma}$ & $e_{1,\sigma}$ & $e_{2,\sigma}$ & $e_{1,\sigma}$ & $e_{2,\sigma}$ &  $e_{1,\sigma}$ & $e_{2,\sigma}$ &  $e_{1,\sigma}$ & $e_{2,\sigma}$ & $e_{1,\sigma}$ & $e_{2,\sigma}$\\
\hline 
\hline 
B3LYP & 0.200 & 1.859 & 0.146 & 1.774 & 0.155 & 1.358 & 0.114 & 1.399 & 0.113 & 1.516 & 0.135 & 1.646\\ \hline
SCAN0 & 0.187 & 0.462 & 0.112 & 0.472 & 0.125 & 0.363 & 0.073 & 0.395 & 0.067 & 0.357 & 0.104 & 0.478\\ \hline
SCAN & 0.189 & 0.885 & 0.102 & 0.637 & 0.118 & 0.475 & 0.071 & 0.530 & 0.058 & 0.476 & 0.090 & 0.617\\ \hline
PBE & 0.209 & 2.501 & 0.147 & 1.983 & 0.158 & 1.577 & 0.124 & 1.613 & 0.120 & 1.724 & 0.132 & 1.764\\ \hline
PW92 & 0.258 & 0.576 & 0.183 & 0.742 & 0.207 & 0.652 & 0.143 & 0.642 & 0.126 & 0.627 & 0.164 & 0.696\\ \hline
\end{tabular}
\label{tab:SI_errWeightedDown}
\end{table*}

\begin{table*}
\caption{\small Comparison of the model XC potentials in terms of the error metrics $e_{3,\sigma}$ and $e_{4,\sigma}$ (defined in Eq.~\ref{eq:SI_vxcErrors2}) for the majority-spin.}

\begin{tabular}{|p{1.6cm} | M{0.8cm} | M{0.8cm} | M{0.8cm} | M{0.8cm} | M{0.8cm}| M{0.8cm} | M{0.8cm} | M{0.8cm} | M{0.8cm} | M{0.8cm}|M{0.8cm}| M{0.8cm}|}
\hline
\multirow{1}{1.6cm}{Model} & \multicolumn{2}{M{1.6cm}|}{Li} & \multicolumn{2}{M{1.6cm}|} {C} & \multicolumn{2}{M{1.6cm}|}{N} & \multicolumn{2}{M{1.6cm}|}{O} & \multicolumn{2}{M{1.6cm}|}{CN} &
\multicolumn{2}{M{1.6cm}|}{CH$_2$}\\
\cline{2-13}
& $e_{3,\sigma}$ & $e_{4,\sigma}$ & $e_{3,\sigma}$ & $e_{4,\sigma}$ & $e_{3,\sigma}$ & $e_{4,\sigma}$ &  $e_{3,\sigma}$ & $e_{4,\sigma}$ &  $e_{3,\sigma}$ & $e_{4,\sigma}$ & $e_{3,\sigma}$ & $e_{4,\sigma}$\\
\hline 
\hline 
B3LYP & 0.723 & 0.401 & 0.789 & 0.448 & 0.777 & 0.513 & 0.775 & 0.397 & 0.773 & 0.388 & 0.781 & 0.493\\ \hline
SCAN0 & 0.701 & 0.300 & 0.746 & 0.396 & 0.733 & 0.510 & 0.728 & 0.309 & 0.730 & 0.322 & 0.739 & 0.392\\ \hline
SCAN & 0.839 & 0.282 & 0.899 & 0.339 & 0.905 & 0.529 & 0.915 & 0.289 & 0.898 & 0.320 & 0.895 & 0.317\\ \hline
PBE & 0.819 & 0.931 & 0.887 & 0.478 & 0.896 & 0.422 & 0.911 & 0.409 & 0.893 & 0.459 & 0.887 & 0.476\\ \hline
PW92 & 0.834 & 0.408 & 0.896 & 0.396 & 0.904 & 0.397 & 0.915 & 0.403 & 0.897 & 0.389 & 0.893 & 0.398\\ \hline
\end{tabular}
\label{tab:SI_errUnweightedUp}
\end{table*}

\begin{table*}
\caption{\small Comparison of the model XC potentials in terms of the error metrics $e_{3,\sigma}$ and $e_{4,\sigma}$ (defined in Eq.~\ref{eq:SI_vxcErrors2}) for the minority-spin.}
\begin{tabular}{|p{1.6cm} | M{0.8cm} | M{0.8cm} | M{0.8cm} | M{0.8cm} | M{0.8cm}| M{0.8cm} | M{0.8cm} | M{0.8cm} | M{0.8cm} | M{0.8cm}|M{0.8cm}| M{0.8cm}|}
\hline
\multirow{1}{1.6cm}{Model} & \multicolumn{2}{M{1.6cm}|}{Li} & \multicolumn{2}{M{1.6cm}|} {C} & \multicolumn{2}{M{1.6cm}|}{N} & \multicolumn{2}{M{1.6cm}|}{O} & \multicolumn{2}{M{1.6cm}|}{CN} &
\multicolumn{2}{M{1.6cm}|}{CH$_2$}\\
\cline{2-13}
& $e_{3,\sigma}$ & $e_{4,\sigma}$ & $e_{3,\sigma}$ & $e_{4,\sigma}$ & $e_{3,\sigma}$ & $e_{4,\sigma}$ &  $e_{3,\sigma}$ & $e_{4,\sigma}$ &  $e_{3,\sigma}$ & $e_{4,\sigma}$ & $e_{3,\sigma}$ & $e_{4,\sigma}$\\
\hline 
\hline 
B3LYP & 0.746 & 0.417 & 0.792 & 0.496 & 0.803 & 0.543 & 0.784 & 0.418 & 0.776 & 0.397 & 0.795 & 0.669\\ \hline
SCAN0 & 0.739 & 0.392 & 0.745 & 0.434 & 0.759 & 0.527 & 0.735 & 0.349 & 0.734 & 0.334 & 0.754 & 0.654\\ \hline
SCAN & 0.854 & 0.375 & 0.913 & 0.343 & 0.910 & 0.455 & 0.922 & 0.299 & 0.900 & 0.301 & 0.896 & 0.528\\ \hline
PBE & 1.035 & 1.024 & 0.927 & 0.525 & 0.925 & 0.538 & 0.923 & 0.491 & 0.898 & 0.462 & 0.908 & 0.663\\ \hline
PW92 & 0.852 & 0.371 & 0.912 & 0.408 & 0.909 & 0.462 & 0.922 & 0.415 & 0.898 & 0.387 & 0.894 & 0.596\\ \hline
\end{tabular}
\label{tab:SI_errUnweightedDown}
\end{table*}

\begin{table*}
\caption{\small Comparison of the model XC potentials in terms of the error metrics $e_{5,\sigma}$ and $e_{6,\sigma}$ (defined in Eq.~\ref{eq:SI_vxcErrors3}) for the majority-spin.}
\begin{tabular}{|p{1.6cm} | M{0.8cm} | M{0.8cm} | M{0.8cm} | M{0.8cm} | M{0.8cm}| M{0.8cm} | M{0.8cm} | M{0.8cm} | M{0.8cm} | M{0.8cm}|M{0.8cm}| M{0.8cm}|}
\hline
\multirow{1}{1.6cm}{Model} & \multicolumn{2}{M{1.6cm}|}{Li} & \multicolumn{2}{M{1.6cm}|} {C} & \multicolumn{2}{M{1.6cm}|}{N} & \multicolumn{2}{M{1.6cm}|}{O} & \multicolumn{2}{M{1.6cm}|}{CN} &
\multicolumn{2}{M{1.6cm}|}{CH$_2$}\\
\cline{2-13}
& $e_{5,\sigma}$ & $e_{6,\sigma}$ & $e_{5,\sigma}$ & $e_{6,\sigma}$ & $e_{5,\sigma}$ & $e_{6,\sigma}$ &  $e_{5,\sigma}$ & $e_{6,\sigma}$ &  $e_{5,\sigma}$ & $e_{6,\sigma}$ & $e_{5,\sigma}$ & $e_{6,\sigma}$\\
\hline 
\hline 
B3LYP & 0.564 & 0.094 & 3.008 & 0.108 & 3.252 & 0.103 & 3.855 & 0.105 & 2.821 & 0.105 & 3.723 & 0.108\\ \hline
SCAN0 & 0.710 & 0.038 & 2.665 & 0.041 & 2.997 & 0.044 & 3.399 & 0.042 & 2.569 & 0.038 & 3.206 & 0.039\\ \hline
SCAN & 0.444 & 0.060 & 1.739 & 0.056 & 2.039 & 0.061 & 2.423 & 0.057 & 1.420 & 0.053 & 2.051 & 0.055\\ \hline
PBE & 0.421 & 0.102 & 1.969 & 0.122 & 2.314 & 0.117 & 2.590 & 0.120 & 1.571 & 0.120 & 2.366 & 0.118\\ \hline
PW92 & 0.458 & 0.094 & 1.949 & 0.083 & 2.231 & 0.079 & 2.678 & 0.080 & 1.500 & 0.086 & 2.358 & 0.079\\ \hline
\end{tabular}
\label{tab:SI_errShiftedUp}
\end{table*}

\begin{table*}
\caption{\small Comparison of the model XC potentials in terms of the error metrics $e_{5,\sigma}$ and $e_{6,\sigma}$ (defined in Eq.~\ref{eq:SI_vxcErrors3}) for the minority-spin.}
\begin{tabular}{|p{1.6cm} | M{0.8cm} | M{0.8cm} | M{0.8cm} | M{0.8cm} | M{0.8cm}| M{0.8cm} | M{0.8cm} | M{0.8cm} | M{0.8cm} | M{0.8cm}|M{0.8cm}| M{0.8cm}|}
\hline
\multirow{1}{1.6cm}{Model} & \multicolumn{2}{M{1.6cm}|}{Li} & \multicolumn{2}{M{1.6cm}|} {C} & \multicolumn{2}{M{1.6cm}|}{N} & \multicolumn{2}{M{1.6cm}|}{O} & \multicolumn{2}{M{1.6cm}|}{CN} &
\multicolumn{2}{M{1.6cm}|}{CH$_2$}\\
\cline{2-13}
& $e_{5,\sigma}$ & $e_{6,\sigma}$ & $e_{5,\sigma}$ & $e_{6,\sigma}$ & $e_{5,\sigma}$ & $e_{6,\sigma}$ &  $e_{5,\sigma}$ & $e_{6,\sigma}$ &  $e_{5,\sigma}$ & $e_{6,\sigma}$ & $e_{5,\sigma}$ & $e_{6,\sigma}$\\
\hline 
\hline 
B3LYP & 0.566 & 0.166 & 3.038 & 0.110 & 3.107 & 0.114 & 3.861 & 0.099 & 2.816 & 0.102 & 3.568 & 0.104\\ \hline
SCAN0 & 0.712 & 0.146 & 2.691 & 0.062 & 2.865 & 0.075 & 3.405 & 0.039 & 2.565 & 0.038 & 3.075 & 0.056\\ \hline
SCAN & 0.451 & 0.152 & 1.754 & 0.068 & 1.952 & 0.081 & 2.427 & 0.050 & 1.417 & 0.051 & 1.969 & 0.060\\ \hline
PBE & 0.576 & 0.177 & 1.981 & 0.124 & 2.210 & 0.128 & 2.593 & 0.116 & 1.568 & 0.118 & 2.266 & 0.115\\ \hline
PW92 & 0.463 & 0.220 & 1.966 & 0.139 & 2.133 & 0.163 & 2.683 & 0.102 & 1.498 & 0.094 & 2.260 & 0.120\\ \hline
\end{tabular}
\label{tab:SI_errShiftedDown}
\end{table*}

\pagebreak

\section{Comparison of energies, virial, and Kohn-Sham eigenvalues}
In this section, we compare the groundstate energies ($E$), non-interacting kinetic energy ($T_s$), exchange-correlation energy ($E_\text{xc}$), and  the virial ($t_{\text{xc}}$) corresponding to the exact and the model XC potentials. Additionally, we also compare the Kohn-Sham eigenvalues for the exact and the model XC potentials. The groundstate energy for the exact potentials refers to the CI groundstate energies. For the model XC potentials, the energies ($E$, $T_s$, $E_\text{xc}$) refer to  self-consistently solved energies of their XC approximations.  Table~\ref{tab:SI_energy_comparison}, Table~\ref{tab:SI_Ts_comparison} and Table~\ref{tab:SI_Exc_comparison} compares the $E$, $T_s$, and $E_\text{xc}$, respectively.   Table~\ref{tab:SI_virial_comparison} compares the virial of the exact and model XC potentials.  Additionally, Tables~\ref{tab:SI_eigs_Li}$-$~\ref{tab:SI_eigs_CH2} compares the Kohn-Sham eigenvalues for Li, C, N, O, CN, and CH$_2$, respectively.  For each system, we provide the error in the eigenvalues of a model XC, defined as 
\begin{equation}
  \Delta \epsilon_{\sigma}^{\text{model}} = \frac{1}{N_\sigma}\sum_i^{N_\sigma} |\epsilon_{i,\sigma}^{\text{exact}}-(\epsilon_{i,\sigma}^{\text{model}}+\Delta \mu^{\text{model}})|\,, 
\end{equation}
where $\epsilon_{i,\sigma}^{\text{exact}}$ and  $\epsilon_{i,\sigma}^{\text{model}}$ are the $i^{\text{th}}$ Kohn-Sham eigenvalue for spin-index $\sigma$ corresponding to the exact and model XC potentials, respectively; and $\Delta \mu^{\text{model}}=\mu^{\text{exact}}-\mu^{\text{model}}$ is the difference between the Kohn-Sham HOMO level for the exact and model XC potentials. The $\Delta \mu^{\text{model}}$ helps to remove any constant shift in the eigenvalues.  

\begin{table*}
\centering
\caption{\small Comparison of the groundstate energies. All values in Ha.}
\begin{tabular}{|p{2.6cm} | M{2cm} | M{2cm} | M{2cm}| M{2cm} | M{2cm} | M{2cm}|} 
\hline
& Li & C & N & O & CN & CH2 \\ \hline
exact & -7.476 & -37.840 & -54.582 & -75.055 & -92.692 & -39.125\\ \hline
B3LYP & -7.493 & -37.861 & -54.606 & -75.098 & -92.754 &-39.162\\ \hline
SCAN0 & -7.479 & -37.839 & -54.588 & -75.065 & -92.689 & -39.141\\ \hline
SCAN & -7.480 & -37.840 & -54.591 & -75.070 & -92.715 & -39.144\\ \hline
PBE & -7.462 & -37.798 & -54.534 & -75.013 & -92.648 & -39.098\\ \hline
PW92 & -7.343 & -37.468 & -54.133 & -74.525 & -91.952 & -38.754\\ \hline
\end{tabular}
\label{tab:SI_energy_comparison}
\end{table*}

\begin{table*}
\centering
\caption{\small Comparison of the non-interacting kinetic energies ($T_s$). All values in Ha.}
\begin{tabular}{|p{2.6cm} | M{2cm} | M{2cm} | M{2cm}| M{2cm} | M{2cm} | M{2cm}|} 
\hline
& Li & C & N & O & CN & CH2 \\ \hline
exact	&	7.434	&	37.711	&	54.425	&	74.845	&	92.346	&	38.898	\\ \hline
B3LYP	&	7.434	&	37.719	&	54.434	&	74.870	&	92.302	&	38.920	\\ \hline
SCAN0	&	7.432	&	37.728	&	54.464	&	74.873	&	92.237	&	38.904	\\ \hline
SCAN	&	7.428	&	37.731	&	54.478	&	74.880	&	92.300	&	38.923	\\ \hline
PBE	&	7.415	&	37.680	&	54.389	&	74.824	&	92.296	&	38.894	\\ \hline
PW92	&	7.249	&	37.246	&	53.865	&	74.193	&	91.343	&	38.446	\\ \hline
\end{tabular}
\label{tab:SI_Ts_comparison}
\end{table*}

\begin{table*}
\centering
\caption{\small Comparison of the exchange-correlation energies ($E_{\text{xc}}$). All values in Ha.}
\begin{tabular}{|p{2.6cm} | M{2cm} | M{2cm} | M{2cm}| M{2cm} | M{2cm} | M{2cm}|} 
\hline
& Li & C & N & O & CN & CH2 \\ \hline
exact	&	-1.824	&	-5.218	&	-6.776	&	-8.435	&	-12.102	&	-6.049	\\ \hline
B3LYP	&	-1.836	&	-5.233	&	-6.792	&	-8.477	&	-12.155	&	-6.073	\\ \hline
SCAN0	&	-1.824	&	-5.219	&	-6.789	&	-8.458	&	-12.101	&	-6.062	\\ \hline
SCAN	&	-1.823	&	-5.219	&	-6.791	&	-8.459	&	-12.121	&	-6.065	\\ \hline
PBE	&	-1.802	&	-5.162	&	-6.712	&	-8.382	&	-12.046	&	-6.005	\\ \hline
PW92	&	-1.664	&	-4.806	&	-6.285	&	-7.864	&	-11.301	&	-5.629	\\ \hline
\end{tabular}
\label{tab:SI_Exc_comparison}
\end{table*}

\begin{table*}
\centering
\caption{\small Comparison of the virial of the XC potentials (see Eq.~\ref{eq:SI_txc} for definition). All values in Ha.}
\begin{tabular}{|p{2.6cm} | M{2cm} | M{2cm} | M{2cm}| M{2cm} | M{2cm} | M{2cm}|} 
\hline
& Li & C & N & O & CN & CH2 \\ \hline
exact	&	-1.782	&	-5.086	&	-6.616	&	-8.228	&	-11.766	&	-5.874	\\ \hline
B3LYP	&	-1.778	&	-5.089	&	-6.614	&	-8.241	&	-11.752	&	-5.881	\\ \hline
SCAN0	&	-1.777	&	-5.105	&	-6.658	&	-8.258	&	-11.764	&	-5.900	\\ \hline
SCAN	&	-1.771	&	-5.105	&	-6.667	&	-8.258	&	-11.799	&	-5.904	\\ \hline
PBE	&	-1.756	&	-5.039	&	-6.559	&	-8.181	&	-11.685	&	-5.819	\\ \hline
PW92	&	-1.571	&	-4.581	&	-6.013	&	-7.521	&	-10.748	&	-5.338	\\ \hline
\end{tabular}
\label{tab:SI_virial_comparison}
\end{table*}

\begin{table*}
\caption{\small Comparison of Kohn-Sham eigenvalues corresponding to the exact and model XC potentials for Li. $\sigma_1$ and $\sigma_2$ denote majority and minority spins, respectively. The ones in boldface denote the KS highest occupied molecular orbital (HOMO). Ionization potential ($I_\text{P}$) = 0.198. All values in Ha.}
\begin{tabular}
{|l|l|l|l||l|l|l|l|}
\hline
\multicolumn{4}{|M{7.2cm}||}{$\sigma_1$} & \multicolumn{4}{ M{7.2cm}|}{$\sigma_2$}\\
\hline
\multicolumn{1}{|M{1.8cm}|}{Exact} & \multicolumn{1}{M{1.8cm}|} {B3LYP} & \multicolumn{1}{M{1.8cm}|}{SCAN0} & \multicolumn{1}{M{1.8cm}||}{SCAN} & \multicolumn{1}{M{1.7cm}|}{Exact} & \multicolumn{1}{M{1.8cm}|} {B3LYP} & \multicolumn{1}{M{1.8cm}|}{SCAN0} & \multicolumn{1}{M{1.8cm}|}{SCAN}\\
\cline{1-8}
-2.054	& -1.931 &	-1.916 &	-1.896 & -2.346 &	-1.899 &	-1.932 &	-1.925\\
-\textbf{0.197} 	& \textbf{-0.117} &	\textbf{-0.109} &	\textbf{-0.117} & & & &\\
\hline \hline
$\Delta \epsilon_{\sigma_1}^{\text{model}}$ & 0.021 & 0.025 & 0.039 & $\Delta \epsilon_{\sigma_2}^{\text{model}}$ & 0.367 & 0.326 & 0.341 \\
\hline
\end{tabular}
\label{tab:SI_eigs_Li}
\end{table*}

\begin{table*}
\caption{\small Comparison of Kohn-Sham eigenvalues corresponding to the exact and model XC potentials for C. $\sigma_1$ and $\sigma_2$ denote majority and minority spins, respectively. The degenerate ones are maked with their multiplicity in parenthesis. The ones in boldface  denote the KS highest occupied molecular orbital (HOMO). $I_\text{P}$ = 0.413. All values in Ha.}
\begin{tabular}
{|l|l|l|l||l|l|l|l|}
\hline
\multicolumn{4}{|M{7.2cm}||}{$\sigma_1$} & \multicolumn{4}{ M{7.2cm}|}{$\sigma_2$}\\
\hline
\multicolumn{1}{|M{1.8cm}|}{Exact} & \multicolumn{1}{M{1.8cm}|} {B3LYP} & \multicolumn{1}{M{1.8cm}|}{SCAN0} & \multicolumn{1}{M{1.8cm}||}{SCAN} & \multicolumn{1}{M{1.8cm}|}{Exact} & \multicolumn{1}{M{1.8cm}|} {B3LYP} & \multicolumn{1}{M{1.8cm}|}{SCAN0} & \multicolumn{1}{M{1.8cm}|}{SCAN}\\
\cline{1-8}
-10.250	&	-9.983	&	-10.010	&	-10.070	&	-10.514	&	-9.962	&	-10.007	&	-10.051\\
-0.721	&	-0.463	&	-0.489	&	-0.539	&	-0.625	&	-0.360	&	-0.377	&	-0.442\\
\textbf{-0.406}	(2) &	\textbf{-0.160}	(2) &	\textbf{-0.186}	(2) &	\textbf{-0.238} (2)	&		&		&		&	\\
\hline \hline
$\Delta \epsilon_{\sigma_1}^{\text{model}}$ & 0.008 & 0.008 & 0.006 & $\Delta \epsilon_{\sigma_2}^{\text{model}}$ & 0.163 & 0.158 & 0.155 \\
\hline
\end{tabular}
\label{tab:SI_eigs_C}
\end{table*}

\begin{table*}
\caption{\small Comparison of Kohn-Sham eigenvalues corresponding to the exact and model XC potentials for N. $\sigma_1$ and $\sigma_2$ denote majority and minority spins, respectively. The degenerate ones are marked with their multiplicity in parenthesis. The ones in boldface  denote the KS highest occupied molecular orbital (HOMO). $I_\text{P}$ = 0.533. All values in Ha.}
\begin{tabular}
{|l|l|l|l||l|l|l|l|}
\hline
\multicolumn{4}{|M{7.2cm}||}{$\sigma_1$} & \multicolumn{4}{ M{7.2cm}|}{$\sigma_2$}\\
\hline
\multicolumn{1}{|M{1.8cm}|}{Exact} & \multicolumn{1}{M{1.8cm}|} {B3LYP} & \multicolumn{1}{M{1.8cm}|}{SCAN0} & \multicolumn{1}{M{1.8cm}||}{SCAN} & \multicolumn{1}{M{1.8cm}|}{Exact} & \multicolumn{1}{M{1.8cm}|} {B3LYP} & \multicolumn{1}{M{1.8cm}|}{SCAN0} & \multicolumn{1}{M{1.8cm}|}{SCAN}\\
\cline{1-8}
-14.319	&	-14.056	&	-14.084	&	-14.164	&	-14.780	&	-14.032	&	-14.100	&	-14.129 \\
-0.955	&	-0.667	&	-0.695	&	-0.750	&	-0.980	&	-0.505	&	-0.508	&	-0.565 \\
\textbf{-0.521} (3)	&	\textbf{-0.243} (3)	&	\textbf{-0.264} (3)	&	\textbf{-0.322} (3)	&		&		&		&	\\
\hline \hline
$\Delta \epsilon_{\sigma_1}^{\text{model}}$ & 0.005 & 0.005 & 0.010 & $\Delta \epsilon_{\sigma_2}^{\text{model}}$ & 0.309 & 0.294 & 0.309 \\
\hline
\end{tabular}
\label{tab:SI_eigs_N}
\end{table*}

\begin{table*}
\caption{\small Comparison of Kohn-Sham eigenvalues corresponding to the exact and model XC potentials for O. $\sigma_1$ and $\sigma_2$ denote majority and minority spins, respectively. The degenerate ones are marked with their multiplicity in parenthesis. The ones in boldface  denote the KS highest occupied molecular orbital (HOMO). $I_\text{P}$ = 0.496. All values in Ha.}
\begin{tabular}
{|l|l|l|l||l|l|l|l|}
\hline
\multicolumn{4}{|M{7.2cm}||}{$\sigma_1$} & \multicolumn{4}{ M{7.2cm}|}{$\sigma_2$}\\
\hline
\multicolumn{1}{|M{1.8cm}|}{Exact} & \multicolumn{1}{M{1.8cm}|} {B3LYP} & \multicolumn{1}{M{1.8cm}|}{SCAN0} & \multicolumn{1}{M{1.8cm}||}{SCAN} & \multicolumn{1}{M{1.8cm}|}{Exact} & \multicolumn{1}{M{1.8cm}|} {B3LYP} & \multicolumn{1}{M{1.8cm}|}{SCAN0} & \multicolumn{1}{M{1.8cm}|}{SCAN}\\
\cline{1-8}
-19.114	&	-18.849	&	-18.891	&	-18.951	&	-19.237	&	-18.805	&	-18.831	&	-18.871\\
-1.143	&	-0.866	&	-0.913	&	-0.953	&	-1.008	&	-0.719	&	-0.738	&	-0.787\\
-0.606 (2)	&	-0.340 (2)	&	-0.381 (2)	&	-0.533 (2)	&	\textbf{-0.494}	&	\textbf{-0.208}	&	\textbf{-0.240}	&	\textbf{-0.290}\\
-0.534	&	-0.269	&	-0.301	&	-0.344	&		&		&		&	\\
\hline \hline
$\Delta \epsilon_{\sigma_1}^{\text{model}}$ & 0.018 & 0.027 & 0.023 & $\Delta \epsilon_{\sigma_2}^{\text{model}}$ & 0.050 & 0.056 & 0.060 \\
\hline
\end{tabular}
\label{tab:SI_eigs_O}
\end{table*}

\begin{table*}
\caption{\small Comparison of Kohn-Sham eigenvalues corresponding to the exact and model XC potentials for CN. $\sigma_1$ and $\sigma_2$ denote majority and minority spins, respectively. The degenerate ones are marked with their multiplicity in parenthesis. The ones in boldface  denote the KS highest occupied molecular orbital (HOMO). $I_\text{P}$ = 0.509. All values in Ha.}
\begin{tabular}
{|l|l|l|l||l|l|l|l|}
\hline
\multicolumn{4}{|M{7.2cm}||}{$\sigma_1$} & \multicolumn{4}{ M{7.2cm}|}{$\sigma_2$}\\
\hline
\multicolumn{1}{|M{1.8cm}|}{Exact} & \multicolumn{1}{M{1.8cm}|} {B3LYP} & \multicolumn{1}{M{1.8cm}|}{SCAN0} & \multicolumn{1}{M{1.8cm}||}{SCAN} & \multicolumn{1}{M{1.8cm}|}{Exact} & \multicolumn{1}{M{1.8cm}|} {B3LYP} & \multicolumn{1}{M{1.8cm}|}{SCAN0} & \multicolumn{1}{M{1.8cm}|}{SCAN}\\
\cline{1-8}
-14.249	&	-13.970	&	-13.948	&	-14.025	&	-14.280	&	-13.959	&	-13.947	&	-14.024	\\ 
-10.195	&	-9.935	&	-9.933	&	-9.997	&	-10.247	&	-9.914	&	-9.914	&	-9.989	\\ 
-1.027	&	-0.810	&	-0.833	&	-0.894	&	-1.033	&	-0.783	&	-0.813	&	-0.877	\\ 
-0.601	&	-0.376	&	-0.408	&	-0.468	&	-0.575	&	-0.323	&	-0.345	&	-0.411	\\ 
-0.503	&	-0.289	&	-0.309	&	-0.375	&	-0.511 (2)	&	-0.269 (2)	&	-0.295 (2)	&	-0.361 (2)	\\ 
\textbf{-0.500} (2)	&	\textbf{-0.281} (2)	&	\textbf{-0.300} (2)	&	\textbf{-0.364} (2)	&	&		&		&		\\ \hline
\hline 
$\Delta \epsilon_{\sigma_1}^{\text{model}}$ & 0.019 & 0.027 & 0.024 & $\Delta \epsilon_{\sigma_2}^{\text{model}}$ & 0.043 & 0.054 & 0.079 \\
\hline
\end{tabular}
\label{tab:SI_eigs_CN}
\end{table*}

\begin{table*}
\caption{\small Comparison of Kohn-Sham eigenvalues corresponding to the exact and model XC potentials for CH$_2$. $\sigma_1$ and $\sigma_2$ denote majority and minority spins, respectively. The ones in boldface  denote the KS highest occupied molecular orbital (HOMO). $I_\text{P}$ = 0.386. All values in Ha.}
\begin{tabular}
{|M{1.6cm} | M{1.6cm} | M{1.6cm} | M{1.6cm} || M{1.6cm} | M{1.6cm} |  M{1.6cm} | M{1.6cm}|}
\hline
\multicolumn{4}{|M{6.4cm}||}{$\sigma_1$} & \multicolumn{4}{ M{6.4cm}|}{$\sigma_2$}\\
\hline
\multicolumn{1}{|M{1.6cm}|}{Exact} & \multicolumn{1}{M{1.6cm}|} {B3LYP} & \multicolumn{1}{M{1.6cm}|}{SCAN0} & \multicolumn{1}{M{1.6cm}||}{SCAN} & \multicolumn{1}{M{1.6cm}|}{Exact} & \multicolumn{1}{M{1.6cm}|} {B3LYP} & \multicolumn{1}{M{1.6cm}|}{SCAN0} & \multicolumn{1}{M{1.6cm}|}{SCAN}\\
\cline{1-8}
-10.145	&	-9.894	&	-9.922	&	-9.988	&	-10.320	&	-9.859	&	-9.886	&	-9.949	\\
-0.796	&	-0.547	&	-0.575	&	-0.627	&	-0.830	&	-0.476	&	-0.501	&	-0.566	\\
-0.553	&	-0.310	&	-0.328	&	-0.382	&	-0.628	&	-0.276	&	-0.305	&	-0.371	\\
-0.457	&	-0.210	&	-0.240	&	-0.293	&		&		&		&		\\
\textbf{-0.388}	&	\textbf{-0.144} &	\textbf{-0.175}	&	\textbf{-0.229}	&		&		&		&		\\
\hline \hline
$\Delta \epsilon_{\sigma_1}^{\text{model}}$ & 0.003 & 0.007 & 0.006 & $\Delta \epsilon_{\sigma_2}^{\text{model}}$ & 0.150 & 0.149 & 0.138 \\
\hline
\end{tabular}
\label{tab:SI_eigs_CH2}
\end{table*}
\pagebreak

\section{Exact and model XC potentials}
\begin{figure}[htbp!]
  \centering
  \subfigure[]{\includegraphics[scale=0.65]{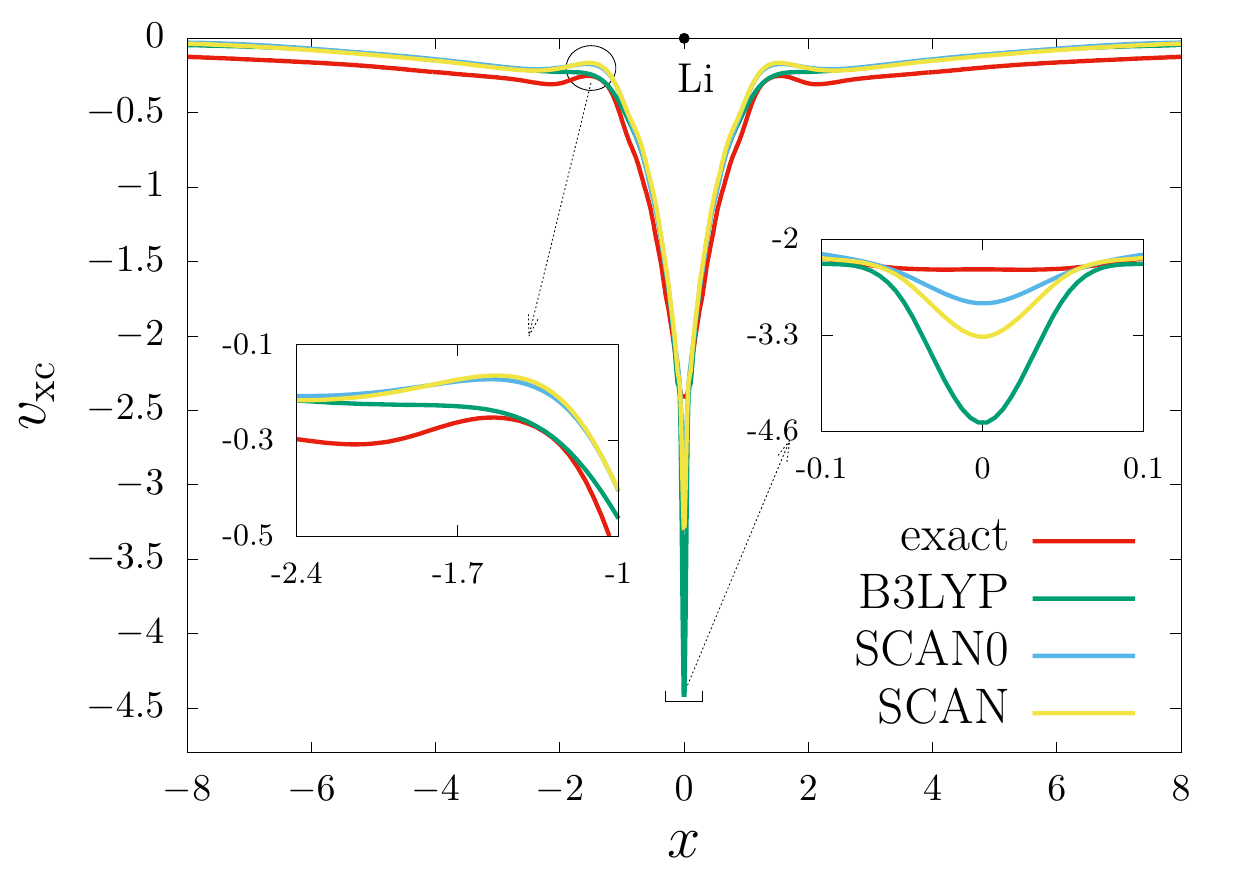}}
  \subfigure[]{\includegraphics[scale=0.65]{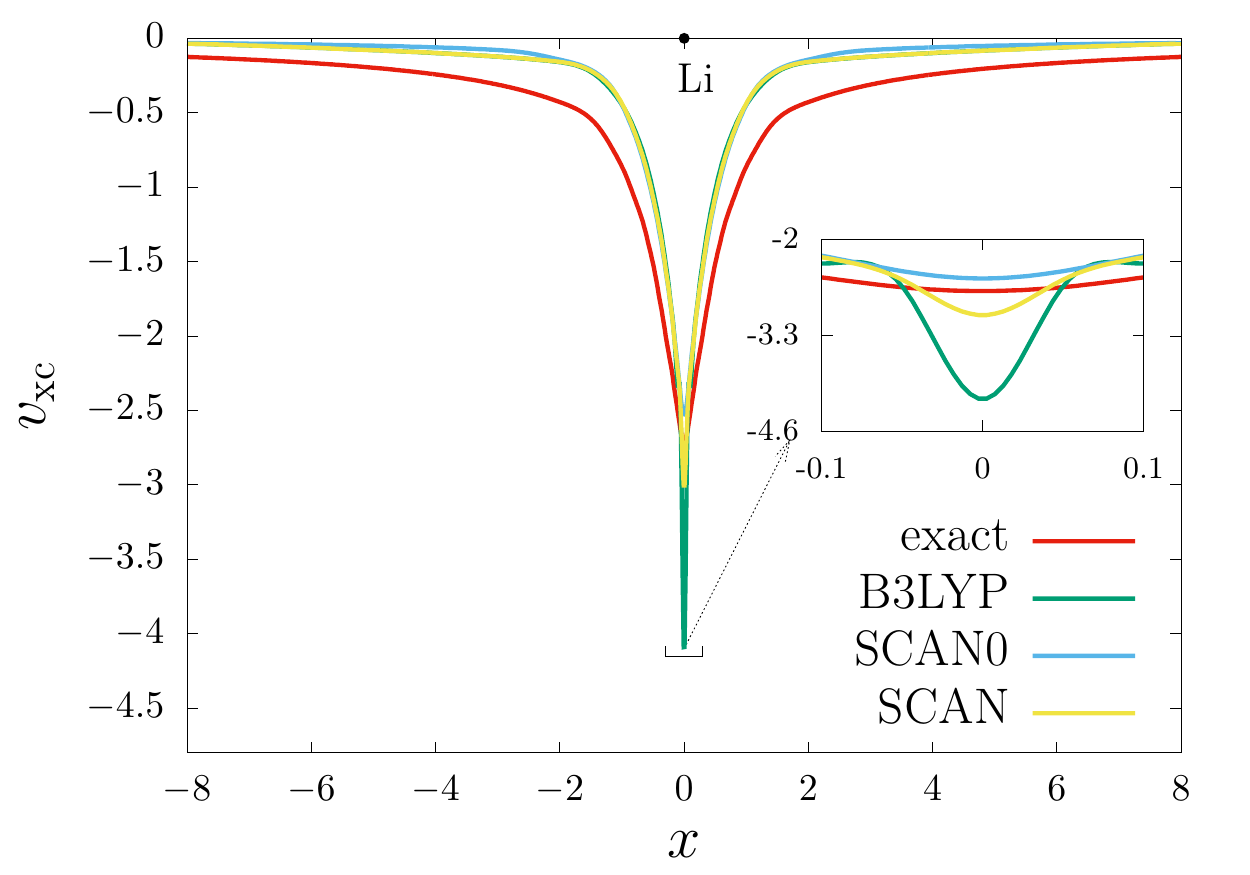}}
   \caption{Comparison of the exact and model XC potentials for Li along the dominant principal axis of the moment of inertia tensor of its density: (a) majority-spin, and (b) minority-spin.}
   \label{fig:SI_Li_Up_Down}
\end{figure}

\begin{figure}[htbp!]
  \centering
  \subfigure[]{\includegraphics[scale=0.65]{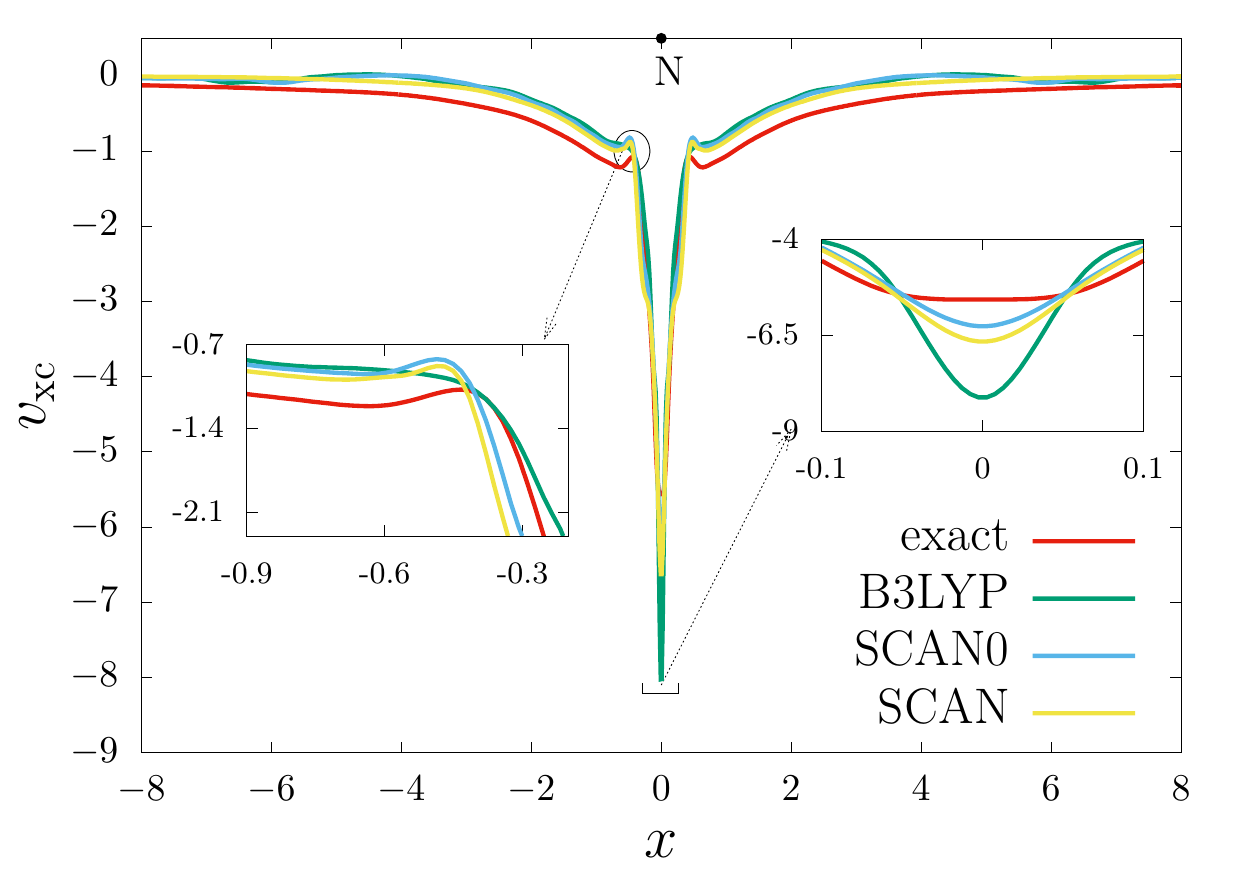}}
  \subfigure[]{\includegraphics[scale=0.65]{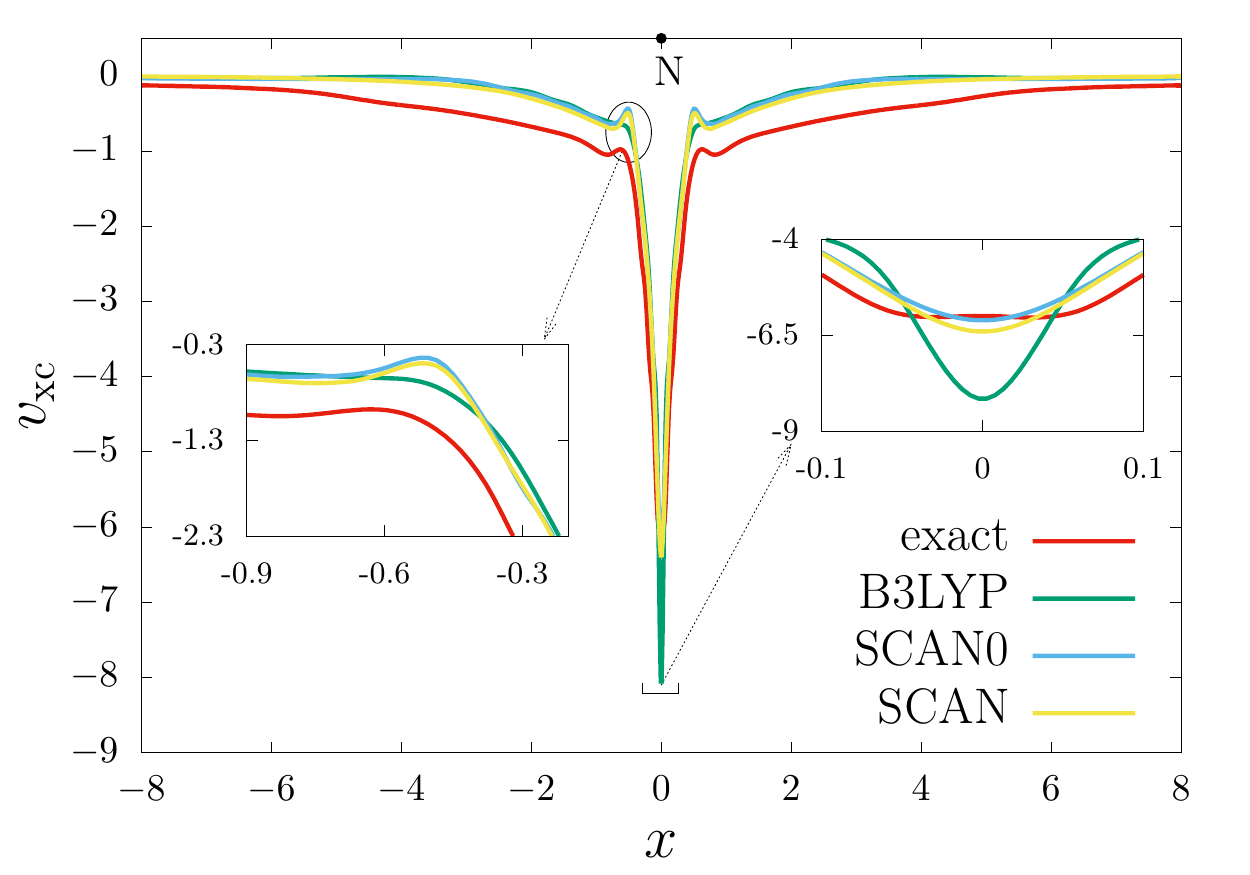}}
   \caption{Comparison of the exact and model XC potentials for N along the dominant principal axis of the moment of inertia tensor of its density: (a) majority-spin, and (b) minority-spin.}
   \label{fig:SI_N_Up_Down}
\end{figure}

\begin{figure}[htbp!]
    \centering
    \includegraphics{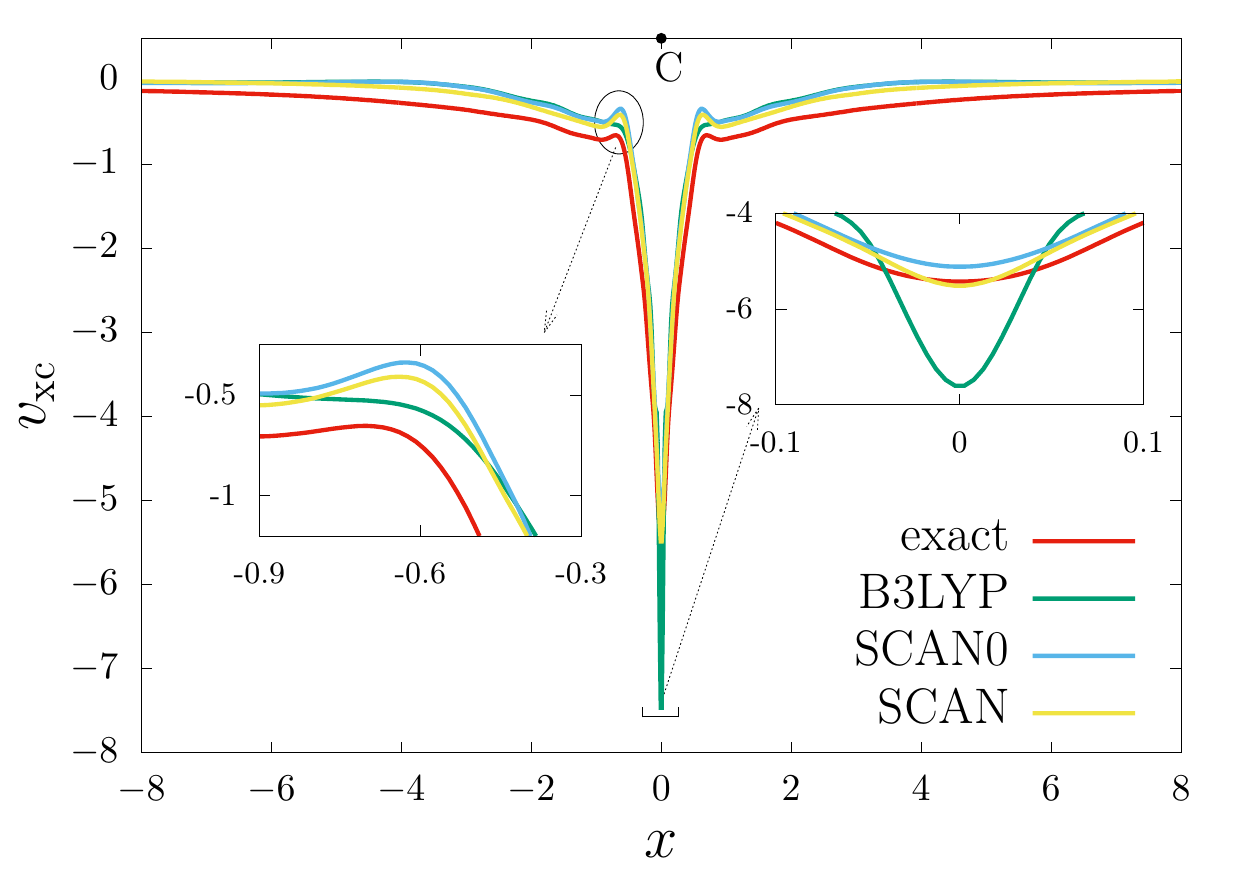}
    \caption{Comparison of the exact and model XC potentials for C for the minority-spin, along the dominant principal axis of the moment of inertia tensor of its density.}
    \label{fig:SI_C_down}
\end{figure}

\begin{figure}[htbp!]
    \centering
    \includegraphics{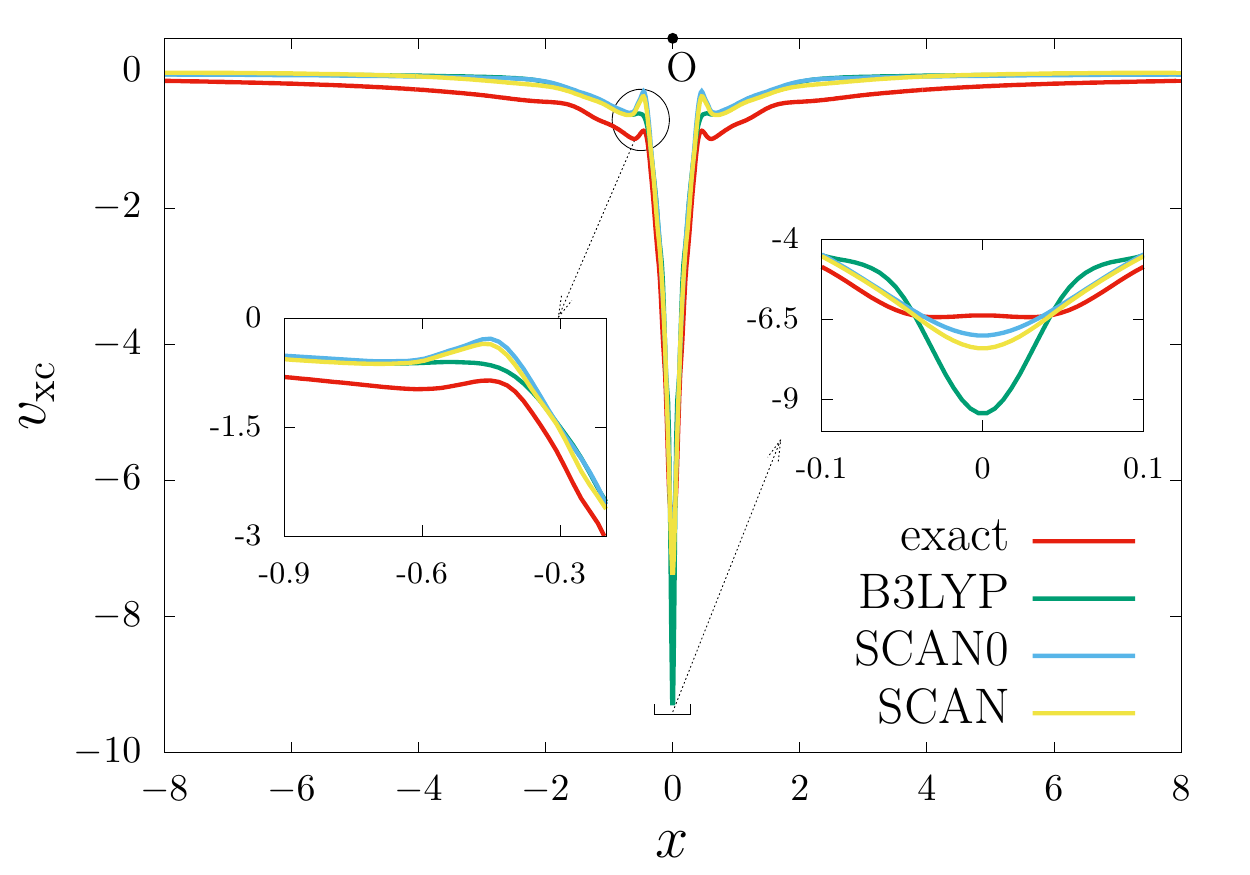}
    \caption{Comparison of the exact and model XC potentials for O for the minority-spin, along the dominant principal axis of the moment of inertia tensor of its density.}
    \label{fig:SI_O_down}
\end{figure}

\begin{figure}[htbp!]
    \centering
    \includegraphics{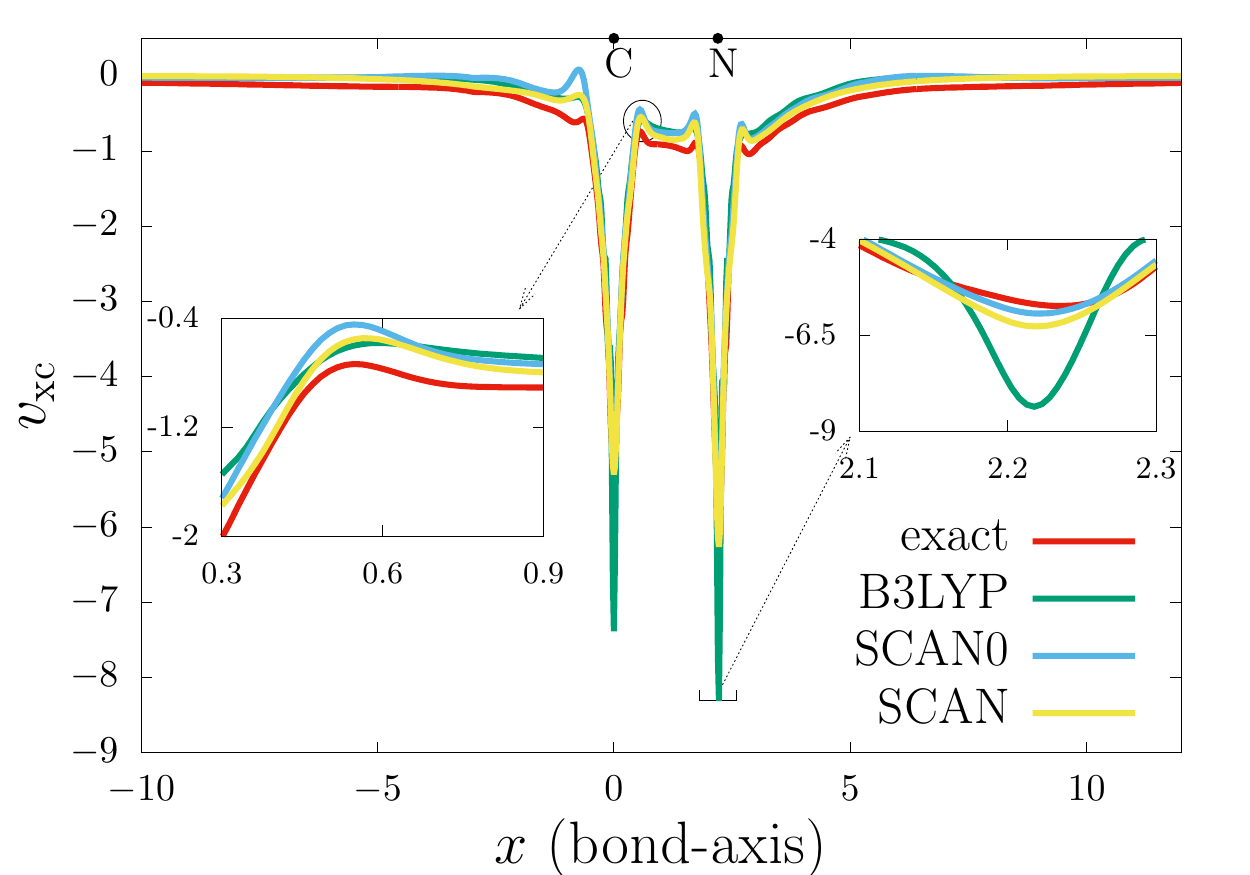}
    \caption{Comparison of the exact and model XC potentials for CN for the minority-spin, along the bond-length.}
    \label{fig:SI_CN_down}
\end{figure}

\begin{figure}[htbp!]
  \centering
  \subfigure[]{\includegraphics[scale=0.65]{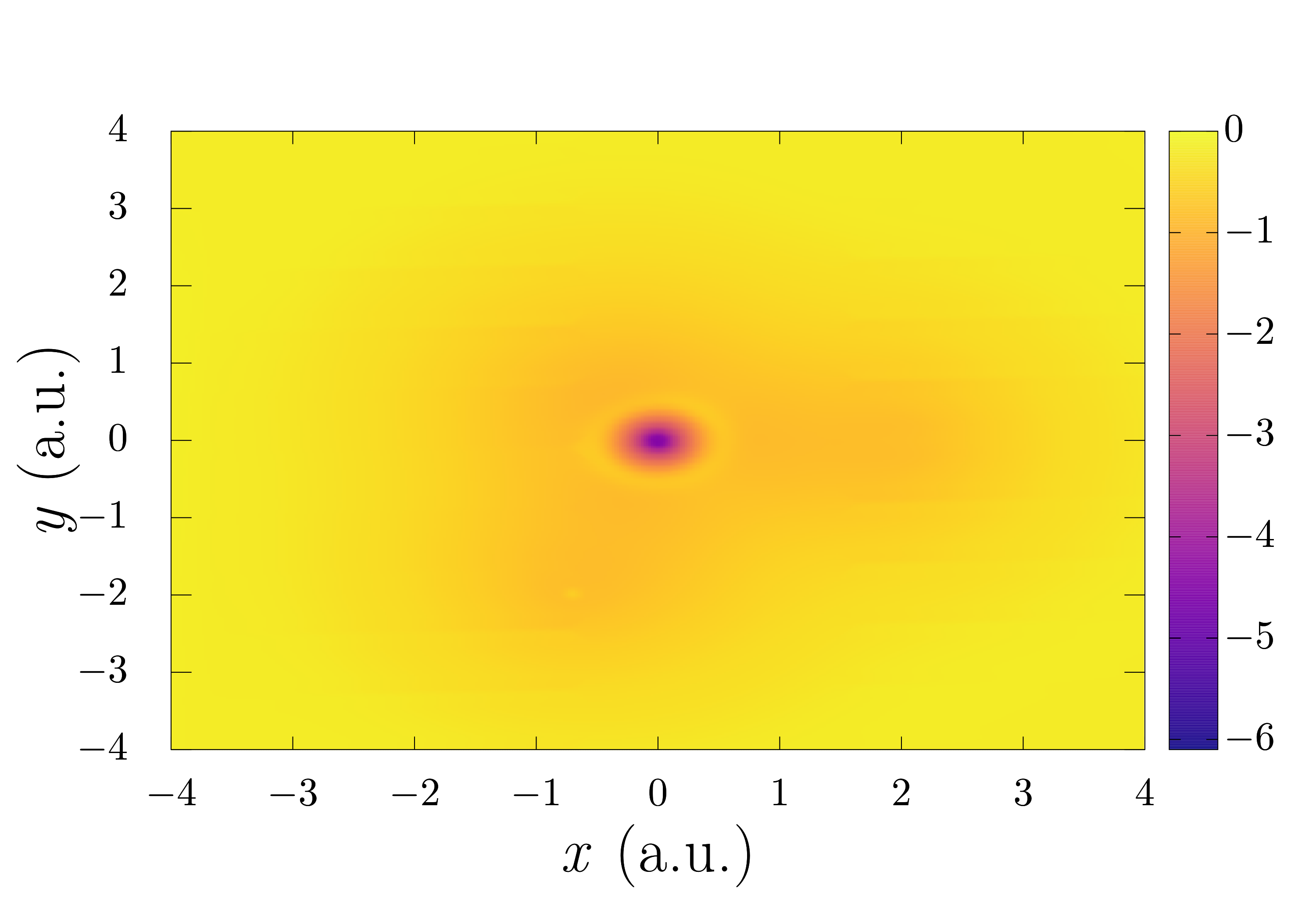}}
  \subfigure[]{\includegraphics[scale=0.65]{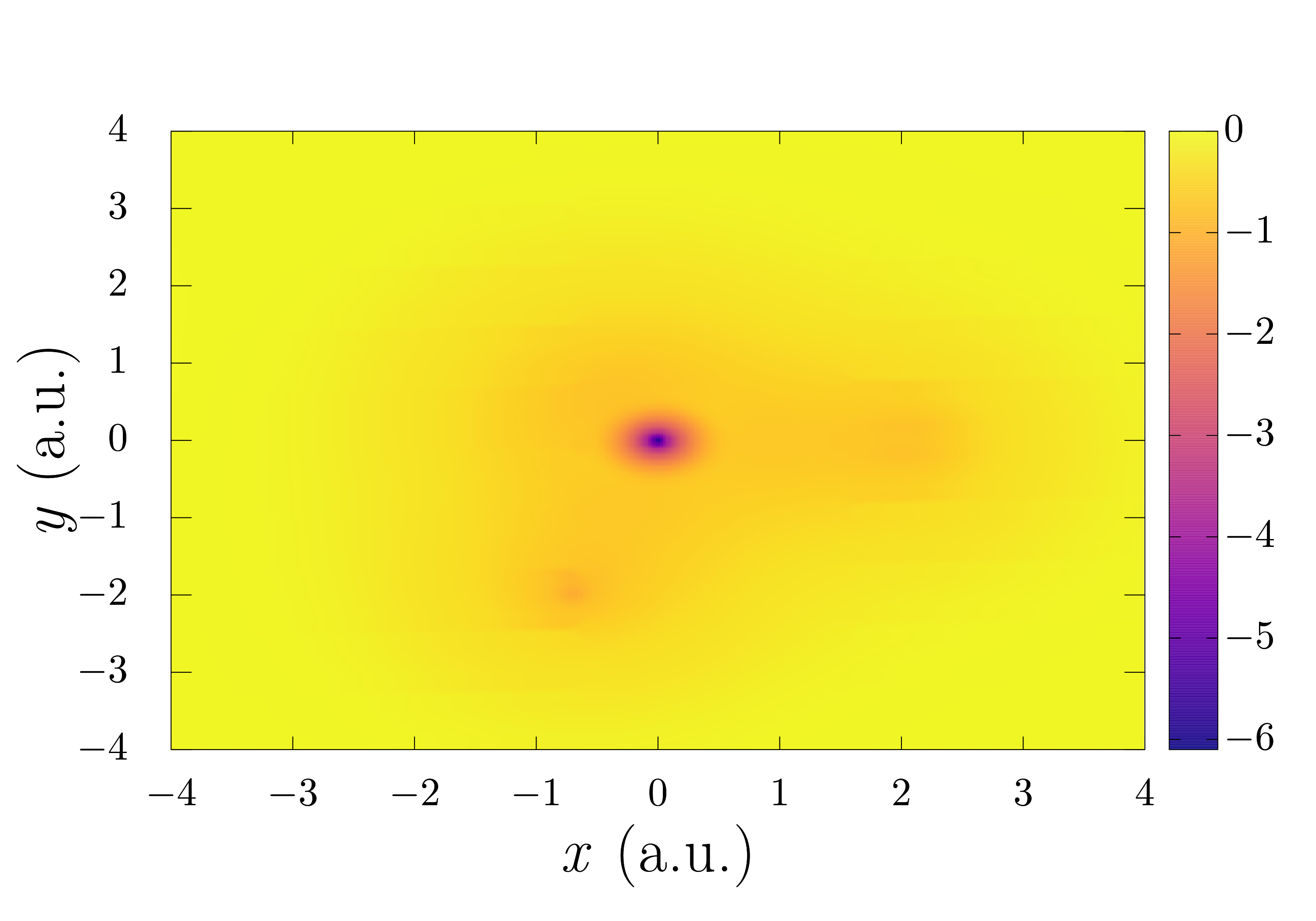}}
  \subfigure[]{\includegraphics[scale=0.65]{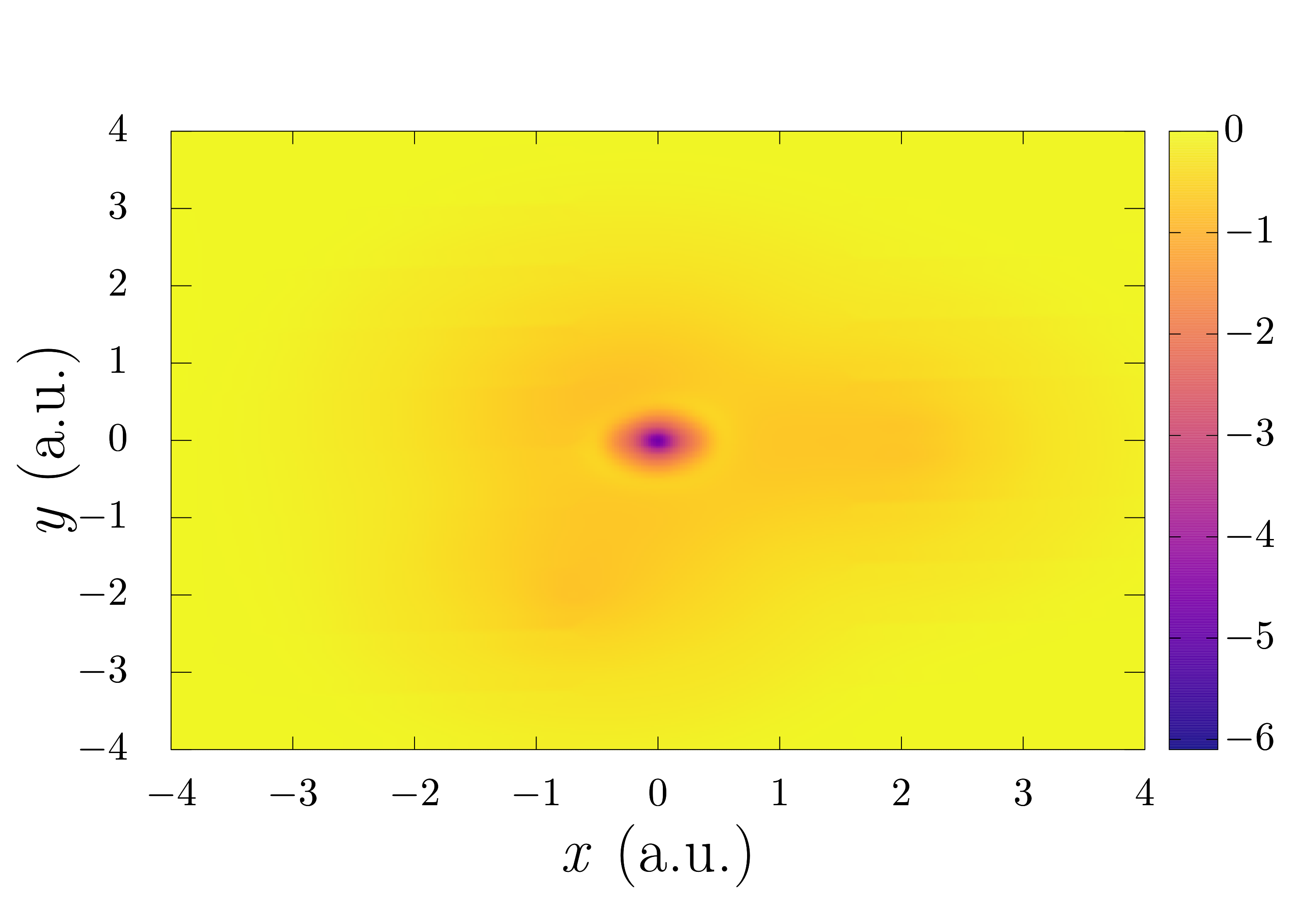}}
  \subfigure[]{\includegraphics[scale=0.65]{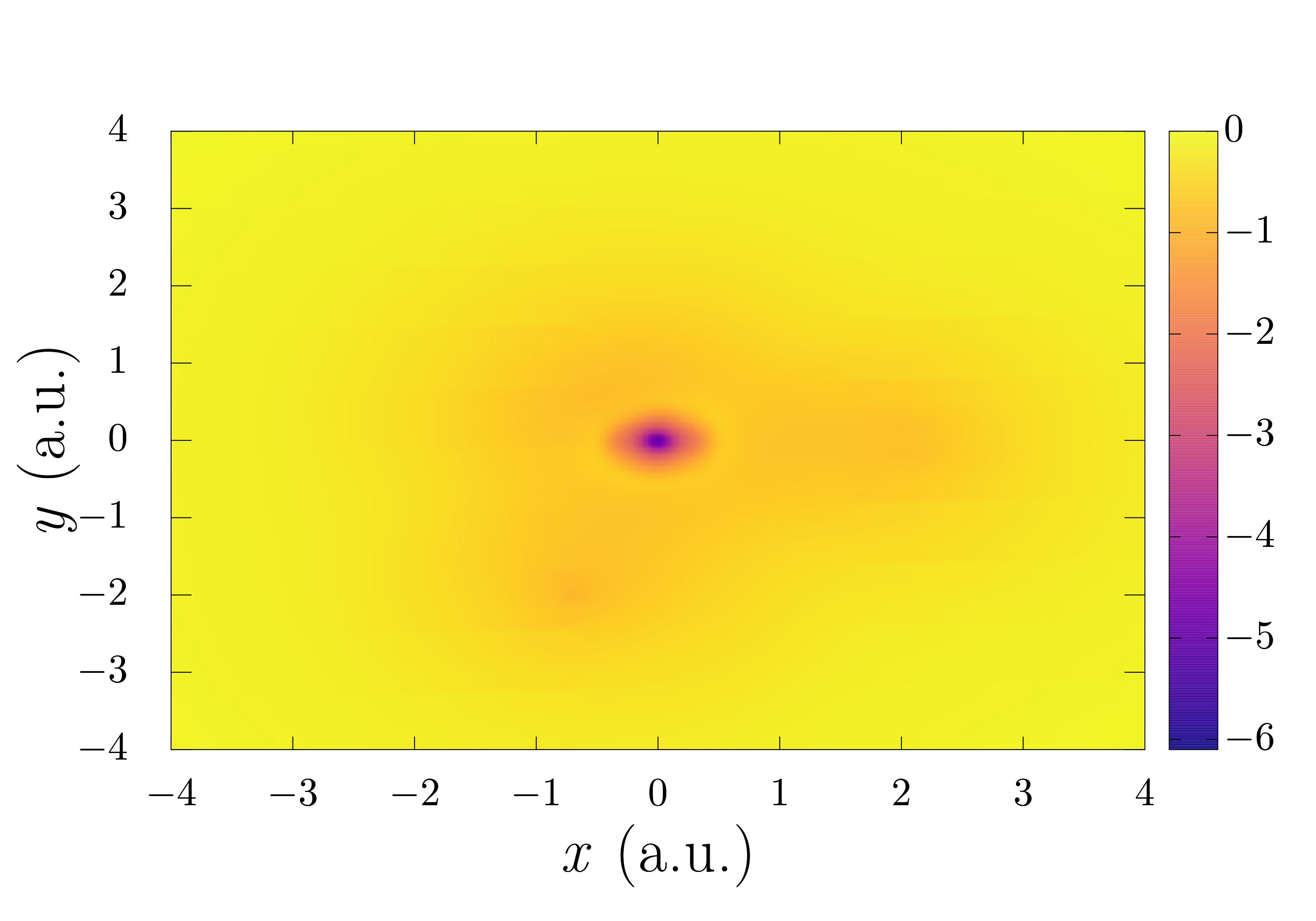}}
   \caption{Exact and model XC potentials for CH$_2$ on the plane of the molecule for the majority-spin: (a) exact potential, (b) B3LYP based model potential, (c) SCAN0 based model potential, and (d) SCAN based model potential. The yellow ring around the C atom in the exact, SCAN0, and SCAN potentials represent the atomic intershell structure, otherwise absent in the B3LYP based potential.}
\label{fig:SI_CH2_exact_b3lyp_scan0_scan_up}
\end{figure}

\begin{figure}[htbp!]
  \centering
  \subfigure[]{\includegraphics[scale=0.65]{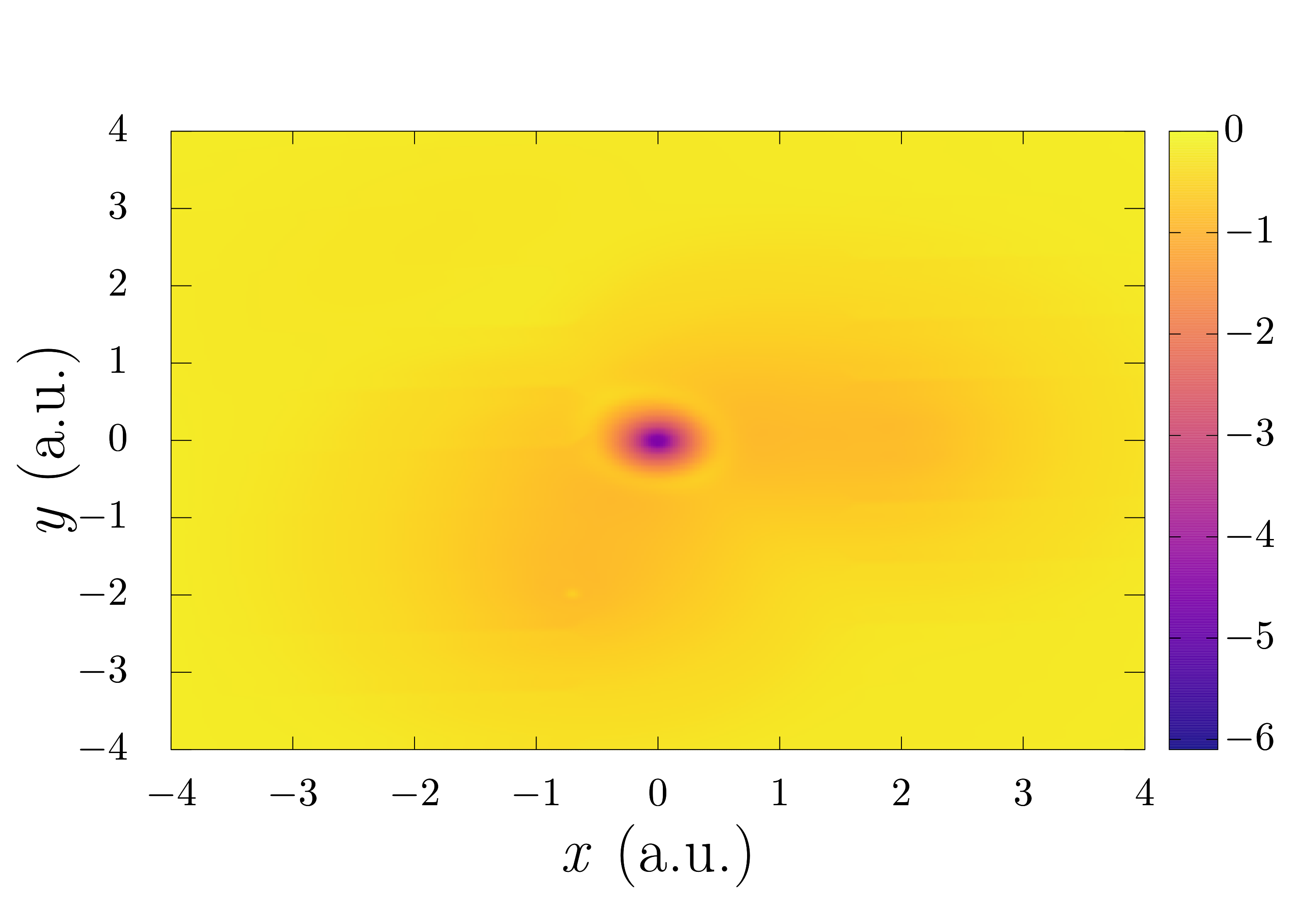}}
  \subfigure[]{\includegraphics[scale=0.65]{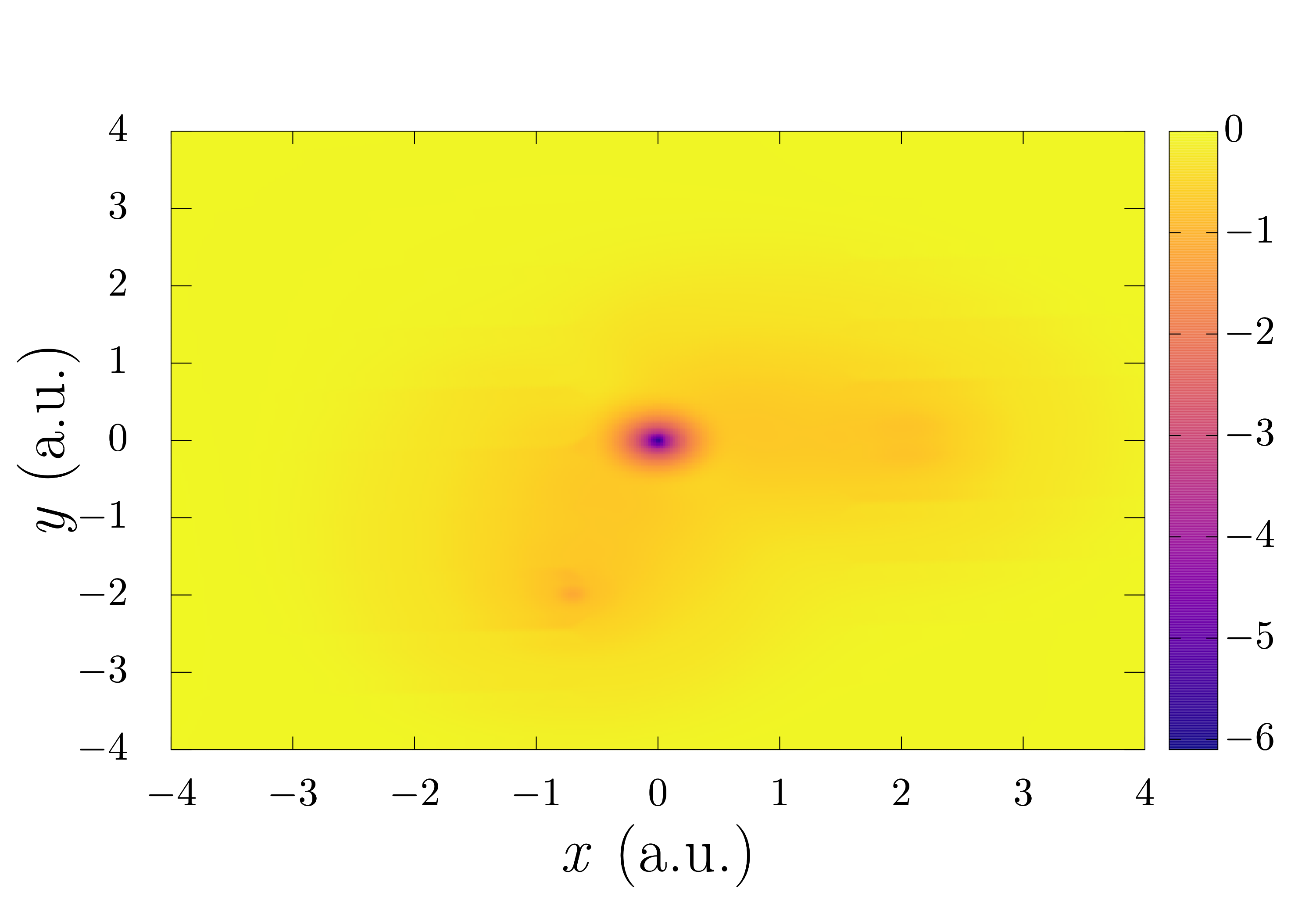}}
  \subfigure[]{\includegraphics[scale=0.65]{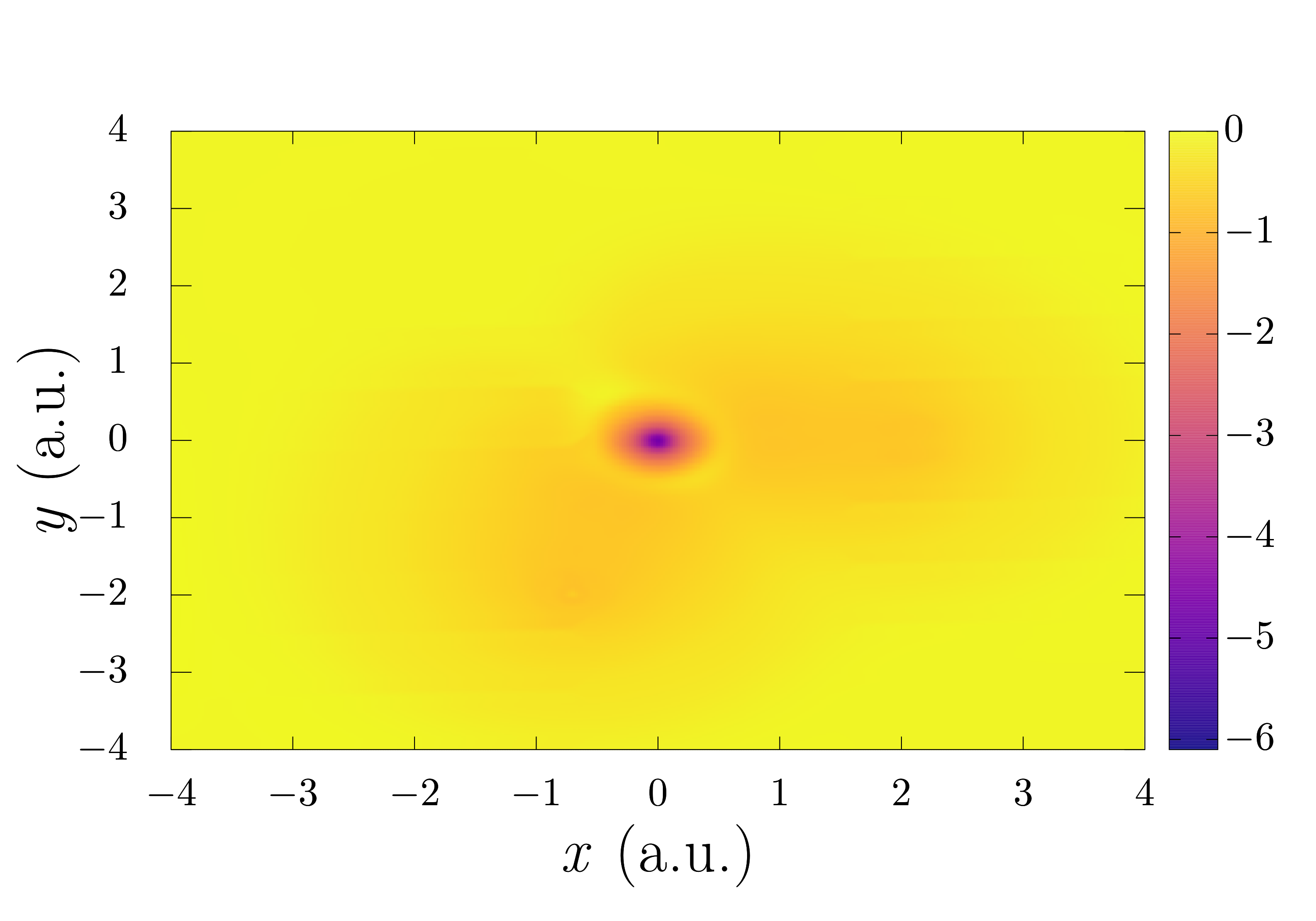}}
  \subfigure[]{\includegraphics[scale=0.65]{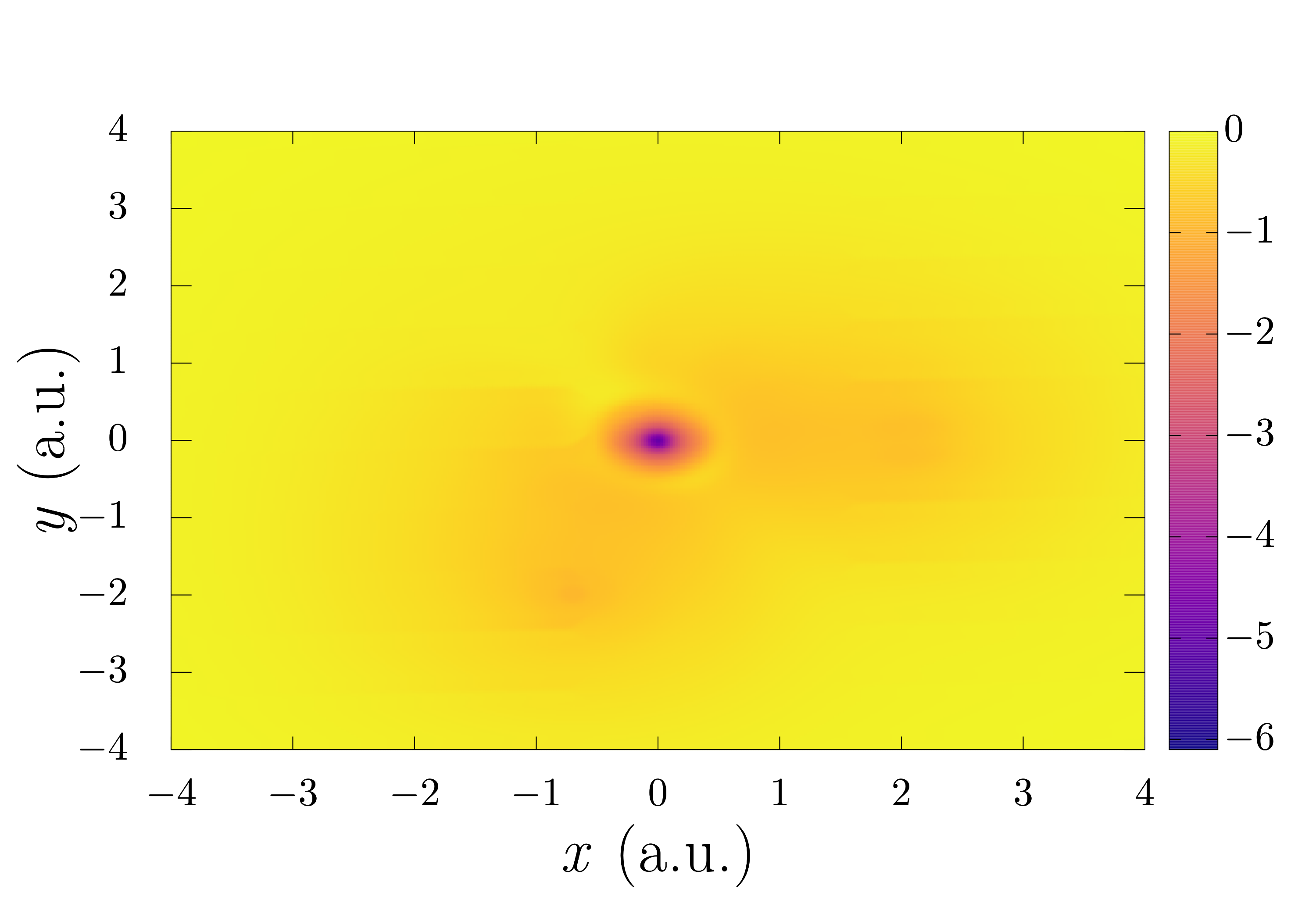}}
   \caption{Exact and model XC potentials for CH$_2$ on the plane of the molecule for the minority-spin: (a) exact, (b) B3LYP based model potential, (c) SCAN0 based model potential, and (d) SCAN based model potential. The yellow ring around the C atom in the exact, SCAN0, and SCAN potentials represent the atomic intershell structure, otherwise absent in the B3LYP based potential.}
\label{fig:SI_CH2_exact_b3lyp_scan0_scan_down}
\end{figure}

\begin{figure}[htbp!]
  \centering
  \subfigure[]{\includegraphics[scale=0.65]{VXC0_exact_b3lyp_diff_CH2.png}}
  \subfigure[]{\includegraphics[scale=0.65]{VXC0_exact_scan0_diff_CH2.png}}
  \subfigure[]{\includegraphics[scale=0.65]{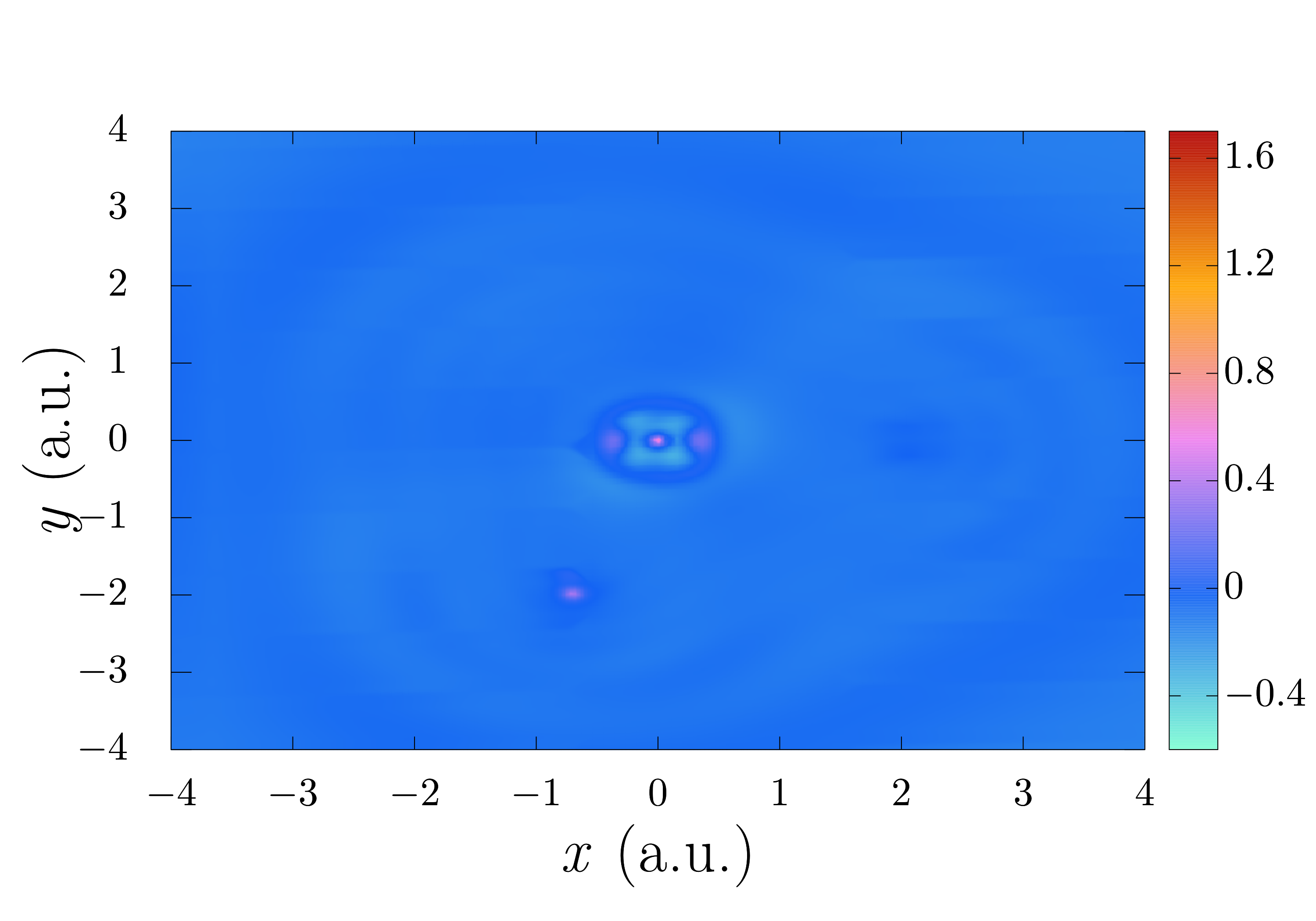}}
   \caption{Error in the model XC potentials (i.e., $\vxcsig^{\text{exact}}-\vxcsig^{\text{model}}$) for CH$_2$ on the plane of the molecule for the majority-spin: (a) B3LYP based model potential, (b) SCAN0 based model potential, and (c) SCAN based model potential. }
\label{fig:SI_CH2_B3LYP_SCAN0_Err_Up}
\end{figure}

\begin{figure}[htbp!]
  \centering
  \subfigure[]{\includegraphics[scale=0.65]{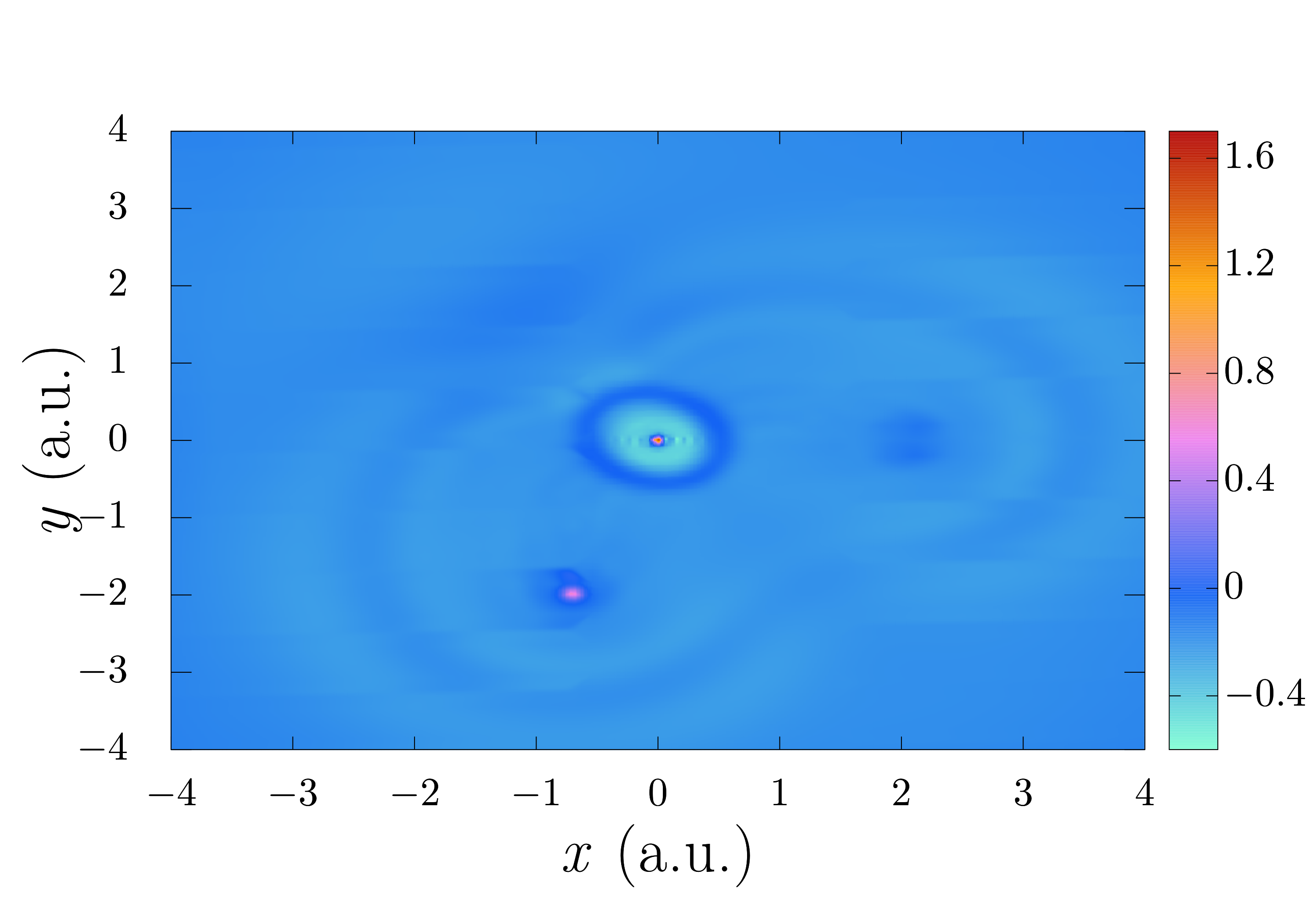}}
  \subfigure[]{\includegraphics[scale=0.65]{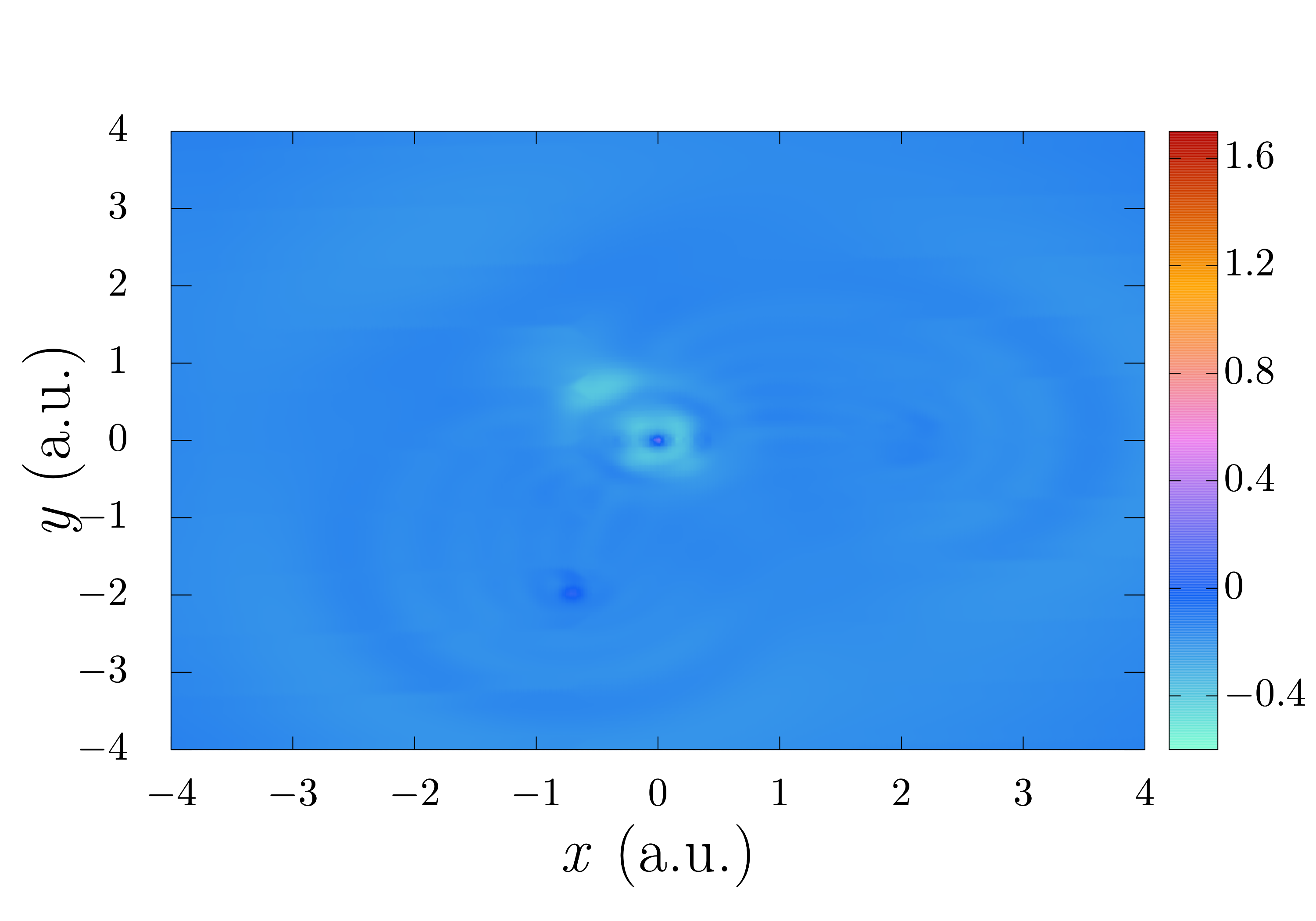}}
  \subfigure[]{\includegraphics[scale=0.65]{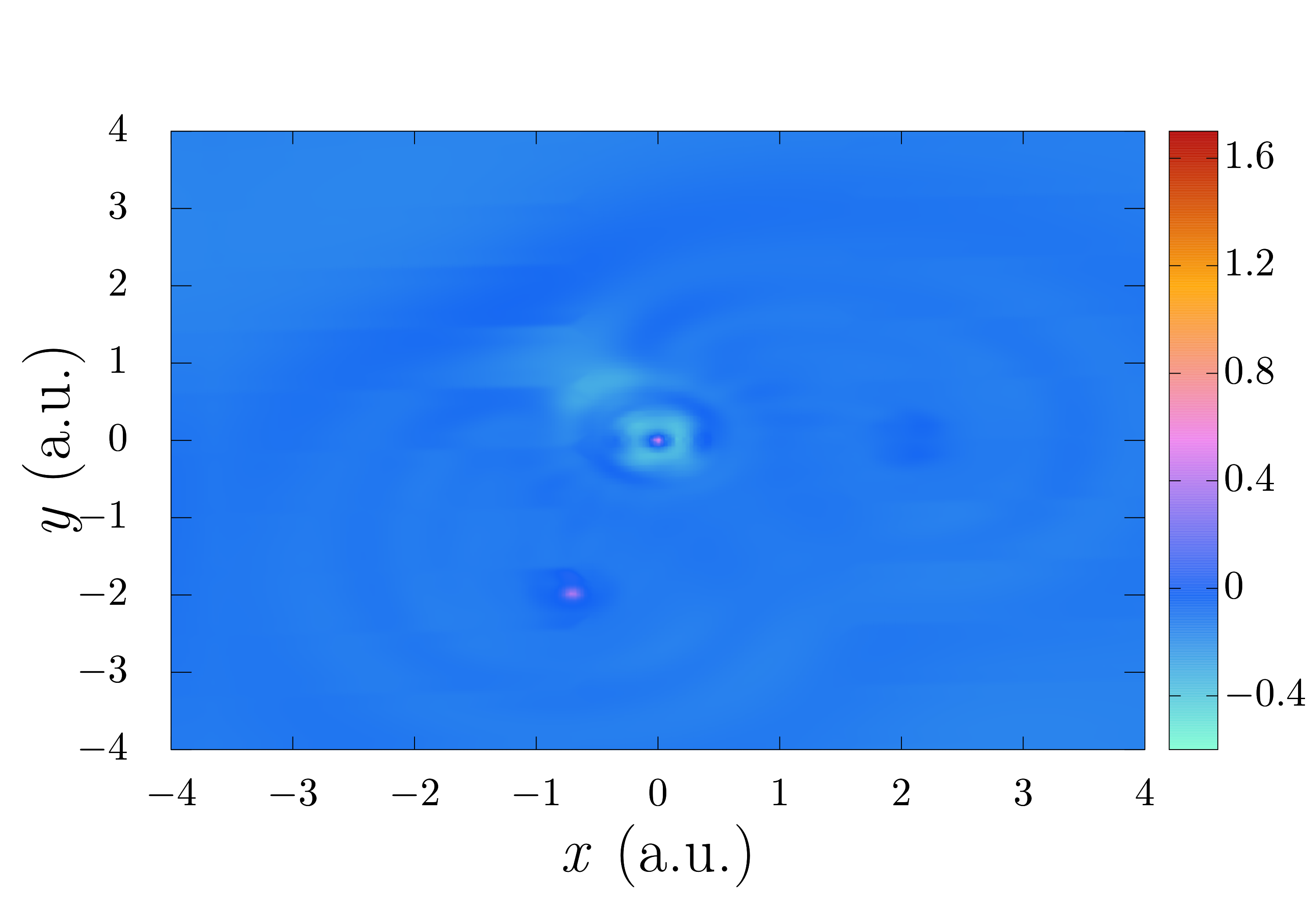}}
   \caption{Error in the model XC potentials (i.e., $\vxcsig^{\text{exact}}-\vxcsig^{\text{model}}$) for CH$_2$ on the plane of the molecule for the minority-spin: (a) B3LYP based model potential, (b) SCAN0 based model potential, and (c) SCAN based model potential.}
\label{fig:SI_CH2_B3LYP_SCAN0_Err_Down}
\end{figure}

In this section, we present additional comparison between the exact and model XC potentials.  Figures~\ref{fig:SI_Li_Up_Down} and ~\ref{fig:SI_N_Up_Down} provide the comparison for Li and N, respectively. Additionally, Figs.~\ref{fig:SI_C_down}-~\ref{fig:SI_CN_down} present the comparison for the minority-spin for C, O, and CN, respectively (see the main manuscript for the comparison of the majority-spin). Figures~\ref{fig:SI_CH2_exact_b3lyp_scan0_scan_up} and ~\ref{fig:SI_CH2_exact_b3lyp_scan0_scan_down} provide the exact and model XC potentials for the CH$_2$ molecule on the plane of the molecule, for the both the spins. For CH$_2$, we also provide the error in  model XC potentials, for both the spins, in Fig.~\ref{fig:SI_CH2_B3LYP_SCAN0_Err_Up} and Fig.~\ref{fig:SI_CH2_B3LYP_SCAN0_Err_Down}. As evident, the model XC potentials differ significantly from the exact ones. In all cases, SCAN0 and SCAN provide better qualitative and quantitative agreement, including the presence of atomic inter-shell structure. Note the minority-spin of Li has no atomic intershell structure, which is expected, as the density is composed of a single KS orbital.    

\clearpage
\section{Wavefunction Energies and Density Matrices}
Heat bath configuration interaction~\cite{sharma2017,holmes2017,Li2018,Brorsen2020,Holmes2016,Yao2021,Chien2018,Dang2022,Dang2023} (HBCI) was used to generate densities and provide total energies, as implemented in a development version of the QChem software package.~\cite{QChem4} HBCI begins from a reference state and systematically expands the wavefunction towards the full CI (FCI) limit. Although many implementations of HBCI start from a single-determinant reference, our implementation uses a complete active space (CAS) reference, up to a CAS (8e,8o) space.~\cite{roos1980,roos1980_2} HBCI consists of two steps: a variational step and a perturbative step. In the variational step, determinants from the FCI space are added if they are highly coupled to a determinant already in the space with the variational wavefunction. The latter is expressed as 
\begin{equation}
\ket{\psi_0}=\sum_{i}c_i\ket{D_i}
\end{equation}
where \(\ket{D_i}\) represent the important determinants. An Epstein-Nesbet perturbation correction is then applied to the remaining determinants, using a tight threshold to determine the importance of each coupling element with the pertubative wavefunction. The perturbative component of the wavefunction is expressed as
\begin{equation}
\ket{\psi_1}=\sum_{k}\frac{\sum_{i}H_{ki}c_i}{E_{var}-H_{kk}}\ket{D_k}
\end{equation}
where $k$ enumerates the determinants not present in the variational wavefunction. New determinants are added using the following selection criteria
\begin{equation}
max(|H_{ki}c_i|)>\epsilon_1
\end{equation}
where the parameter \(\epsilon_1\)  \ controls the addition of new determinants. A similar paramter \(\epsilon_2\) controls the perturbative importance criteria and is naturally smaller in magnitude. The energies were converged using HBCI to tight tolerances of \(\epsilon_1\) and \(\epsilon_2\), (see Table ~\ref{tab:ci_energies}) showing a close approach to the full CI limit.

Correlation from core electrons can significantly contribute to the total correlation energy of atoms and molecules. To provide high accuracy input to inverse DFT, the core electrons were included in the CI space for all species.  The core-valence polarized, quadruple zeta basis set, cc-pCVQZ,~\cite{Woon1995,peterson2002} was therefore applied to all systems. 

All the information needed to calculate the electron density at a given point in space can be found in the one particle density matrix and the one particle, atomic orbital basis set. In closed shell systems, these density matrices are identical for \(\alpha\) and \(\beta\) electrons. For open-shell systems, the \(\alpha\) and \(\beta\) density matrices must be generated separately. These density matrices are generated in the variational step and then corrected during the perturbative step. For example, the \(\alpha\) density matrix can be constructed via
\begin{equation}
D^\alpha = \langle \psi_0 \vert \hat{\rho}_a \vert \psi_0 \rangle +2\langle \psi_0 \vert \hat{\rho}_a \vert \psi_1 \rangle
\end{equation}
with a similar equation for the \(\beta\) component.~\cite{mest} The variational density matrices are normalized such that the trace of each matrix will yield the number of electrons of a given spin.
\begin{equation}
Tr(D^\alpha_{var})=N_\alpha
\end{equation}
where \(N_\alpha\) is the number of alpha electrons and \(D^\alpha_{var}\) is the density matrix for the alpha electrons in the variational step. In the perturbative step, the trace of the perturbative density matrix tends to be on the order of \(10^{-5}\) , as is the case for the oxygen atom at \(3\times 10^{-5}\), indicating a small but significant change in the density due to the perturbation. To preserve the total number of electrons, it is necessary to renormalize these matrices so that the total number of electrons is preserved.
\begin{equation}
Tr(D^\alpha_{var}+D^\alpha_{EN})=N_\alpha
\end{equation}
where \(D^\alpha_{EN}\) represents the density matrix for the alpha electrons in the perturbation step. In order to get the calculation for CN to run with sufficiently tight epsilon values, it was necessary to parallelize it. The perturbative correction to the density matrix is not implemented in this highly parallel version of the code at this time. As a result, the density for CN comes from only the variational step. The perturbative energy correction for CN is still included in this manuscript to show that it is relatively small and therefore, the perturbative density correction would also be small. 

\begin{table*}
\caption{\small Total energy values with a breakdown into contributions from the variational and perturbative step and the values of \(\epsilon_1\) and \(\epsilon_2\).}
\begin{tabular}
{|M{2.4cm} | M{2.4cm} | M{2.4cm} | M{2.4cm} | M{2.4cm} | M{2.4cm} |}
\hline
\multicolumn{1}{|M{2.4cm}|}{Compound} & \multicolumn{1}{M{2.4cm}|} {Variational Energy (Ha)} & \multicolumn{1}{M{2.4cm}|}{Perturbative Energy (Ha)} & \multicolumn{1}{M{2.4cm}|}{Total Energy (Ha)} & \multicolumn{1}{M{2.4cm}|}{$\epsilon_1~(\mu \text{Ha})$} & \multicolumn{1}{M{2.4cm}|} {$\epsilon_2~(\mu \text{Ha})$}\\
\cline{1-6}
Li & -7.4764 & 0.000 & -7.4764 & 0 & -	\\
C & -37.8380 & -0.0002 & -37.8383 & 100 & 0.1 \\
N & -54.5795 & -0.0006 & -54.5801 & 100 & 0.1 \\ 
O & -75.0521 & -0.0007 & -75.0528 & 100 & 0.1 \\ 
$\text{CH}_2$ & -39.1309 & -0.0014 & -39.1323 & 100 & 0.1 \\ 
CN & -92.7006 & -0.0052 & -92.7058 & 50 & 0.05 \\
\hline
\end{tabular}
\label{tab:ci_energies}
\end{table*}

\end{document}